\documentclass{article}
\usepackage[margin=1in]{geometry}
\usepackage{amsmath}
\usepackage{amssymb}
\usepackage{graphicx}
\usepackage{array} 
\usepackage{amsthm}
\usepackage{mathtools}
\usepackage{caption}
\usepackage{subcaption} 
\usepackage{tikz}
\tikzstyle{vertex}=[draw,thick,fill=white,circle,inner sep=2pt]
\tikzstyle{tiny_vertex}=[draw,thick,fill=white,circle,inner sep=1.5pt]
\tikzstyle{full}=[draw,thick,fill=black,circle,inner sep=2pt]
\tikzstyle{empty}=[draw,color=black!40!white,thick,fill=white,circle,inner sep=2pt]

\newtheorem{theorem}{Theorem}
\newtheorem{lemma}{Lemma}
\newtheorem{proposition}{Proposition} 
\newtheorem{definition}{Definition} 
\newcommand{\EE}{\ensuremath{\mathbb{E}}}
\newcommand{\PP}{\ensuremath{\mathbb{P}}}
\DeclareMathOperator{\var}{Var}

\begin{document}

\title{A Comparison of Mutation and Amplification-Driven Resistance Mechanisms and Their Impacts on Tumor Recurrence}
\date{} 
\author{Aaron Li\footnote{School of Mathematics, University of Minnesota, Minneapolis, MN, USA \texttt{lia@umn.edu}} \and  Danika Kibby \footnote{Target Corporation, Minneapolis, MN, USA \texttt{danika.kibby@target.com}} \and  Jasmine Foo \footnote{School of Mathematics, University of Minnesota, Minneapolis, MN, USA \texttt{jyfoo@umn.edu}}}
\maketitle

\begin{abstract}
Tumor recurrence, driven by the evolution of drug resistance is a major barrier to therapeutic success in cancer.  Tumor drug resistance is often caused by genetic alterations such as point mutation, which refers to the modification of a single genomic base pair, or gene amplification, which refers to 
the  duplication of a region of DNA that contains a gene.  These  mechanisms typically confer varying degrees of resistance, and they tend to occur at vastly different frequencies.  Here we  investigate the dependence of tumor recurrence dynamics on these mechanisms of resistance, using stochastic multi-type branching process models. We derive tumor extinction probabilities and deterministic estimates for the tumor recurrence time, defined as the time when an initially drug sensitive tumor surpasses its original size after developing resistance. For models of amplification-driven and mutation-driven resistance, we prove law of large numbers results regarding the convergence of the stochastic recurrence times to their mean. Additionally, we prove sufficient and necessary conditions for a tumor to escape extinction under the gene amplification model, discuss behavior under biologically relevant parameters, and compare the recurrence time and tumor composition in the mutation and amplification models both analytically and using simulations. In comparing these mechanisms, we find that the ratio between recurrence times driven by amplification vs. mutation depends linearly on the number of amplification events required to acquire the same degree of resistance as a mutation event, and we find that the relative frequency of amplification and mutation events plays a key role in determining the mechanism under which recurrence is more rapid for any specific system.  In the amplification-driven resistance model, we also observe that increasing drug concentration leads to a stronger initial reduction in tumor burden, but that the eventual recurrent tumor population is less heterogeneous, more aggressive and harbors higher levels of drug-resistance. 
\end{abstract}

\section{Introduction} 

The emergence of drug-resistance is the principal cause of treatment failure in cancer. Drug resistance and the loss of effective treatment options is responsible for up to 90\% of cancer death \cite{wang_drug_2019}. For example, in non-small cell lung cancer (NSCLC), while many patients initially respond positively to treatment, 30-55\% of those patients eventually experience recurrence \cite{uramoto_recurrence_2014}. Molecularly-targeted therapies are particularly vulnerable to the development of resistance due to their focused action on specific mutable targets \cite{huang_molecularly_2014}. Tumors can acquire drug resistance via a variety of genetic and epigenetic mechanisms, including point mutation, gene amplification, and upregulated drug efflux \cite{housman_drug_2014}.

In this work we examine two common mechanisms of drug-resistance --  point mutations and gene amplification processes. Point mutation refers to a modification, addition or deletion of a single base pair within the genome, while gene amplification refers to the duplication of a region of DNA containing a gene.  In Epidermal Growth Factor Receptor (EGFR)-driven NSCLC, for example, two of the most common mechanisms of resistance to targeted EFGR inhibitors, such as erlotinib and gefitinib, are a T790M point mutation (which alters the binding site of the drug) and amplification of the MET gene (which upregulates an alternate signaling pathway) \cite{tang_erlotinib_2013, bean_met_2007}. Similarly, in chronic myeloid leukemia (CML), resistance to imatinib can arise from either the amplification of oncogene BCR-ABL or a point mutation in the target binding site \cite{komarova_drug_2005}. A review by Albertson \cite{albertson_gene_2006} provides additional examples of drug resistance arising from gene amplification, such as DHFR amplification in response to methotrexate in leukemia patients and TYMS amplification in response to 5-fluorouracil in colorectal cancer patients.  The frequencies of gene amplication and point mutation events can differ by several orders of magnitude, with amplification events often occurring more frequently than point mutation  \cite{tlsty_differences_1989}. On the other hand, individual amplification events may confer weaker drug resistance effects than rare point mutation events which accumulate as copy number increases; these interesting disparities can result in distinct tumor recurrence timing and population dynamics.   Developing an understanding of how resistance mechanisms influence recurrence dynamics  provides critical insights into prognosis for treatment response as well as treatment strategies for recurrent tumors.

To investigate these phenomena, we define and analyze branching process models of drug resistance acquisition occurring through the mechanisms of point mutation and gene amplification. Using these models, we investigate how these mechanisms influence the tumor recurrence time, which refers to the time at which an initially drug-sensitive tumor population rebounds to its initial size due to the outgrowth of resistant clones, as well as the composition of recurrent tumors. Accurate estimates of recurrence times may help physicians schedule surgical interventions and improve understanding of optimal control of combination therapy. Furthermore, understanding how these mechanisms influence the tumor composition at recurrence, i.e.\ the proportion of drug resistant and drug sensitive cells, could help inform physicians about how aggressive the tumor will be when it reaches its original size. 
%As such, developing an understanding of how the evolution of drug resistance influences the time scale and composition of a tumor at recurrence is of clinical importance.
%Since tumors can evolve drug resistance via different mechanisms, being able to compare tumor recurrence under these two regimes will help provide patients with greater specificity of care and better predict recurrence under different scenarios. 

There have been many previous works analyzing mathematical models of point mutation-driven drug resistance in tumors.  For example, 
Tomasetti and Levy \cite{tomasetti_elementary_2010} used ODEs to characterize the average number of resistant cells that arise from point mutations at any given time after treatment and generalize their results to the multi-drug case. Works by Iwasa et al. \cite{iwasa_evolution_2006}, Komarova \cite{komarova_stochastic_2006}, and Komarova and Wodarz \cite{komarova_drug_2005} study the multi-drug point mutation scenario as well, but characterized the probability of developing drug resistance under stochastic branching process models. Hanagal et al.\ \cite{hanagal_large_2022} study recurrence time in a similar gene mutation model but do not examine recurrence in the amplification model. In our previous works, we developed a similar branching process modeling approach to study the stochastic time at which the resistant cell population first begins to dominate the tumor  \cite{foo_dynamics_2013, foo_escape_2014}.  However, these works have not typically explored the tumor recurrence time. On the other hand, mathematical models of gene amplificaton-driven resistance are somewhat less well-explored in the literature.  Works by Kimmel and Axelrod \cite{kimmel_mathematical_1990} and Kimmel et al.\ \cite{kimmel_branching_1992} and others detailed in a review by Swierniak \cite{swierniak_mathematical_2009} use stochastic branching processes to model gene amplification. These works focus on the distribution and dynamics of gene copy numbers under different regimes, rather than on tumor recurrence or extinction; in this work we build upon these existing contributions.

%The gene amplification model presented in this paper along with its corresponding analysis contribution to this body of work which previously primarily focused on the case where resistance arises due to a single point mutation. Works by Kimmel and Axelrod \citet{kimmel_mathematical_1990} and Kimmel, Axelrod, and Wahl \citet{kimmel_branching_1992} and others detailed in a review by Swierniak, Kimmel, and Smieja \citet{swierniak_mathematical_2009} do use stochastic branching processes to model gene amplification, but focus on the distribution and dynamics of gene copy numbers under different regimes, rather than on tumor recurrence or extinction.

The remainder of the paper is organized as follows.  In Section~\ref{models}, we introduce the models for both point mutation and gene amplification. In Section~\ref{results}, we explore sufficient and necessary conditions for non-extinction of the process and analyze the tumor recurrence time in both models.  Theorem~\ref{LLNAmp}  gives a law of large numbers result regarding the convergence of the stochastic recurrence time in the gene amplification model, and Theorem~\ref{LLNMut} gives the analogous result in the point mutation model.  We then examine the effect of drug efficacy on the recurrence time in the amplification model through simulations. Furthermore, we make comparisons between amplification and mutation-driven tumor recurrence timing and composition under biologically relevant parameters. Our results allow us to use the net growth rates, amplification and mutation rates, and other parameters defined in detail in Section~\ref{models} to characterize the ratio between the estimated recurrence times in the gene amplification and point mutation models.  %In the context of NSCLC, we find that it may not be possible to distinguish between mutation-driven resistance and amplification-driven resistance from the recurrence time alone. 
In Section~\ref{discussion}, we discuss our results and how they can potentially be used for parameter inference, as well as limitations and possible extensions of the models.  In Section~\ref{appendix}, we provide proofs of the main results. Because a typical solid tumor has between $10^7$ and $10^9$ cancer cells per cubic centimeter \cite{michaelson_breast_1999}, we focus on characterizing the tumor recurrence time in the large population limit. Throughout the paper, we use the following standard Landau notation for asymptotic behavior of non-negative functions: 
\begin{equation*}
f(t) \sim g(t) \text{ if } f(t)/g(t) \rightarrow 1 \text{ as } t \rightarrow \infty.
\end{equation*}

\section{Models and Preliminaries} \label{models}
In this section we introduce mathematical models of tumor populations, under continuous therapy, acquiring resistance via point mutation and gene amplification.   In both models, we start with an initial population of $n$ drug-sensitive cells and zero resistant cells at time $t=0$.  The population $X_s(t)$ of sensitive cells is modeled using a subcritical Markovian branching process that declines during treatment with birth rate $r_s$, death rate $d_s$, and net growth rate ${\lambda}_s = r_s - d_s < 0$. Note that a birth rate of $r_s$, for example, indicates that in an infinitesimal time interval $\Delta t$, the probability of a cell division is $\Delta r_s$. Since $\lambda_s < 0$, the sensitive cell population goes extinct with probability 1, unless resistant cells are produced that `save' the population from extinction. Throughout, we consider these processes on the approximate time scale of extinction of the sensitive cell process, $t_n = -\frac{1}{\lambda_s} \log n$. In the following sections we describe models and analyze recurrence timing for two mechanisms of resistance production: point mutation and genetic amplification. 

\subsection{Mutation Model} 
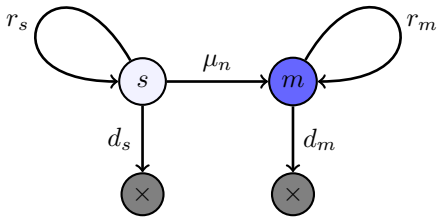
\begin{figure}[htb]
    \centering
    
    \begin{tikzpicture}[scale = 2]
        \node[vertex, inner sep = 4pt, fill = white!95!blue](s) at (2,0) {$s$};
        \node[vertex, inner sep = 3pt, fill = white!40!blue](m) at (3,0) {$m$};
        
        \node[vertex, inner sep = 2pt, fill = white!50!black](sd) at (2,-0.75) {$\times$};
        \node[vertex, inner sep = 2pt, fill = white!50!black](md) at (3,-0.75) {$\times$};
        
        \draw[line width = 1pt, ->] (s)--(m) node[midway,above] {$\mu_n$};

        \draw[line width = 1pt, ->] (s)--(sd) node[midway, left] {$d_s$};
        \draw[line width = 1pt, ->] (m)--(md) node[midway, right] {$d_m$};
        \tikzset{every loop/.style={min distance=10mm,in=180,out=120,looseness=10}}
        \path[->, line width = 1pt] (s) edge  [loop above ] node[above, left] {$r_s$} ();
        \tikzset{every loop/.style={min distance=10mm,in=0,out=60,looseness=10}}
        \path[->, line width = 1pt] (m) edge  [loop above] node[above, right] {$r_m$} ();
        
    \end{tikzpicture}
    \caption{Schematic of the gene mutation model. Sensitive type $s$ cells have birth rate $r_s$, death rate $d_s$, and can give rise to a mutated type $m$ cell with birth rate $r_m$ and death rate $d_m$ at rate $\mu_n$.}
    \label{fig:mutModel}
\end{figure}
% We first consider the scenario in which a sensitive cell may acquire a single point mutation, giving rise to a new resistant cell.
We first consider the scenario in which mutations occur at rate $\mu_n X_s(t)$, where $\mu_n =  n^{-\alpha}$ for $0 < \alpha < 1$. With this formulation, taking a limit as $n \to \infty$ simultaneously sends $\mu_n \to 0$, which achieves the  large population, small mutation rate asymptotic regime in which the model is biologically relevant.  Each mutant cell gives rise to a supercritical Markovian branching process with birth rate $r_m$, death rate $d_m$, and net growth rate $\lambda_m  = r_m - d_m > 0$.  The total population of mutant cells is denoted by $X_m(t)$.  In the following, all probability and expectation operators are conditioned on the initial conditions $X_s(0) = n$ and $X_m(0) = 0$. Lastly, we note that in this model formulation, mutations occur throughout the life cycle of the cell and are not specifically restricted to arise during cellular division; this assumption is made for notational convenience but the model can be easily restricted to this setting by multiplying the mutation rate by the cellular birth rate. A more detailed explanation of this adjustment is available in Appendix~\ref{note:restriction}.

Using the transition probabilities of these processes together with their moment generating functions, we derive a system of ODE's governing the first and second moments, which we then use to calculate the mean and variance of $X_s(t)$ and $X_m(t)$.

\begin{lemma}\label{lem:meanvarmut}
    For $i = s, m$, let $\phi_i(t) = \mathbb{E}[X_i(t)]$ and let $\psi_i(t) = \mathrm{Var}[X_i(t)]$.  Then for a fixed $z > 0$, we have:

\begin{align*}
\phi_s(z t_n) &= n^{1-z},\\
\phi_m(z t_n) &= \frac{ n^{1 - \alpha}}{\lambda_m  - \lambda_s} \left( n^{-\lambda_m  z/\lambda_s} - n^{-z} \right),\\
\psi_s(z t_n) &= n^{1-z} \left( \frac{r_s + d_s}{-\lambda_s} \right) \left(1 - n^{-z} \right),\\
\psi_m(z t_n) &\sim \frac{2r_m}{\lambda_m (2\lambda_m  - \lambda_s)} \cdot n^{1 - \alpha - 2\lambda_m  z/\lambda_s}.
\end{align*}
\end{lemma}
A proof of this lemma can be found in Appendix~\ref{proof:meanvarmut}.

% Note that the $n^{1-\alpha}$ term in $\phi_m(zt_n)$ necessitates our bounding of $\alpha$ above by 1. Otherwise, $\phi_m(zt_n) \to 0$ in the large population limit and thus, by the Markov Inequality, $\PP(X_m(t)$

\subsection{Amplification Model} 
\begin{figure}[htb]
    \centering
    \tikzset{every loop/.style={min distance=10mm,in=120,out=60,looseness=10}}
    \begin{tikzpicture}[scale = 1.5]
        \node[vertex, inner sep = 3pt, fill = white!95!blue](2) at (2,0) {2};
        \node[vertex, inner sep = 3pt, fill = white!80!blue](3) at (3,0) {3};
        \node[vertex, inner sep = 3pt, fill = white!60!blue](4) at (4,0) {4};
        \node[vertex, inner sep = 3pt, fill = white!40!blue](5) at (5,0) {5};
        \node[vertex, inner sep = 3pt, fill = white!20!blue](6) at (6,0) {6};
        \node[vertex, inner sep = 3pt, fill = white!20!blue](7) at (7,0) {7};

        % \node[vertex, inner sep = 3pt, fill = white!95!blue](2r1) at (2,1) {2};
        % \node[vertex, inner sep = 3pt, fill = white!95!blue](2r2) at (2.25,1.25) {2};
        % \node[vertex, inner sep = 3pt, fill = white!80!blue](3r1) at (3,1) {3};
        % \node[vertex, inner sep = 3pt, fill = white!80!blue](3r2) at (3.25,1.25) {3};
        % \node[vertex, inner sep = 3pt, fill = white!60!blue](4r1) at (4,1) {4};
        % \node[vertex, inner sep = 3pt, fill = white!60!blue](4r2) at (4.25,1.25) {4};
        % \node[vertex, inner sep = 3pt, fill = white!40!blue](5r1) at (5,1) {5};
        % \node[vertex, inner sep = 3pt, fill = white!40!blue](5r2) at (5.25,1.25) {5};
        % \node[vertex, inner sep = 3pt, fill = white!20!blue](6r1) at (6,1) {6};
        % \node[vertex, inner sep = 3pt, fill = white!20!blue](6r2) at (6.25,1.25) {6};
        % \node[vertex, inner sep = 3pt, fill = white!10!blue](7r1) at (7,1) {7};
        % \node[vertex, inner sep = 3pt, fill = white!10!blue](7r2) at (7.25,1.25) {7};

        \node[vertex, inner sep = 2pt, fill = white!50!black](2d) at (2,-1) {$\times$};
        \node[vertex, inner sep = 2pt, fill = white!50!black](3d) at (3,-1) {$\times$};
        \node[vertex, inner sep = 2pt, fill = white!50!black](4d) at (4,-1) {$\times$};
        \node[vertex, inner sep = 2pt, fill = white!50!black](5d) at (5,-1) {$\times$};
        \node[vertex, inner sep = 2pt, fill = white!50!black](6d) at (6,-1) {$\times$};
        \node[vertex, inner sep = 2pt, fill = white!50!black](7d)
        at (7,-1) {$\times$};
        
        \draw[line width = 1pt, ->] (2)--(3) node[midway,above] {$\nu_n$};
        \draw[line width = 1pt, ->] (3)--(4)node[midway,above] {$\nu_n$};
        \draw[line width = 1pt, ->] (4)--(5)node[midway,above] {$\nu_n$};
        \draw[line width = 1pt, ->] (5)--(6)node[midway,above] {$\nu_n$};
        \draw[line width = 1pt, ->] (6)--(7)node[midway,above] {$\nu_n$};

        \draw[line width = 1pt, ->] (2)--(2d) node[midway, left] {$d_2$};
        \draw[line width = 1pt, ->] (3)--(3d) node[midway, left] {$d_3$};
        \draw[line width = 1pt, ->] (4)--(4d) node[midway, left] {$d_4$};
        \draw[line width = 1pt, ->] (5)--(5d) node[midway, left] {$d_5$};
        \draw[line width = 1pt, ->] (6)--(6d) node[midway, left] {$d_6$};
        \draw[line width = 1pt, ->] (7)--(7d) node[midway, left] {$d_7$};

        \path[->, line width = 1pt] (2) edge  [loop above] node {$r_2$} ();
        \path[->, line width = 1pt] (3) edge  [loop above] node {$r_3$} ();
        \path[->, line width = 1pt] (4) edge  [loop above] node {$r_4$} ();
        \path[->, line width = 1pt] (5) edge  [loop above] node {$r_5$} ();
        \path[->, line width = 1pt] (6) edge  [loop above] node {$r_6$} ();
        \path[->, line width = 1pt] (7) edge  [loop above] node {$r_7$} ();
        % \draw[line width = 1pt, ->] (3)--(3r1) node[midway, left] {$r_3$};
        % \draw[line width = 1pt, ->] (4)--(4r1) node[midway, left] {$r_4$};
        % \draw[line width = 1pt, ->] (5)--(5r1) node[midway, left] {$r_5$};
        % \draw[line width = 1pt, ->] (6)--(6r1) node[midway, left] {$r_6$};
        % \draw[line width = 1pt, ->] (7)--(7r1) node[midway, left] {$r_7$};
    \end{tikzpicture}
    \caption{Schematic of the gene amplification model for $M = 7$. Type $k$ cells have birth rate $r_k$, death rate $d_k$ and can give rise to a type $k+1$ cell at rate $\nu_n$.}
    \label{fig:model}
\end{figure}
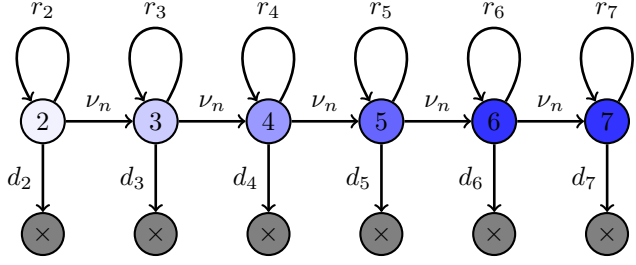
In the amplification model, we assume that sensitive cells each have two copies of the specific gene whose amplification is associated with drug resistance.  For this reason and for ease of notation, let a subscript of $s$ be equivalent to a subscript of $2$. Say type $k$ cells have $k$ gene copies and $M$ is the maximum number of gene copies a cell can have. Let $X_k(t)$, $k = 2, 3, 4, \ldots, M$, represent the population of type $k$ cells with birth rate $r_k$, death rate $d_k$, and net growth rate $\lambda_k = r_k - d_k$. We assume that the wild-type cells are drug sensitive, that is, $\lambda_2 < 0$. In this model, a type $k-1$ cell may gain a single gene copy, giving rise to a new type $k$ cell. With each additional gene copy, cells gain a fitness advantage, reaching a maximum of $\lambda_M$, which we will assume to be positive. More specifically we define 

\begin{equation*}
\lambda_k = \lambda_2 + (k-2)D, \quad D = \frac{\lambda_M - \lambda_2}{M-2}.
\end{equation*}
Let $k'$ be the smallest $k$ for which $\lambda_k> 0$. These amplification steps occur at rate $\nu_n X_{k-1}(t)$, where $\nu_n = n^{-\beta}$ for $0 < \beta < \frac{1}{k'-2}$. As in the mutation model, we allow amplification events to occur at any time for notational convenience.  Notice that as $\beta$ increases, the mutation rate $\nu_n =  n^{-\beta}$ decreases. In particular, when $\beta$ is large, the mutation rate may decrease to the point where very few mutations are actually generated, resulting in the entire population of cancer cells dying out before a resistant clone becomes established and hence preventing recurrence altogether. In Proposition~\ref{prop:extinction} below, we will establish that this bound on $\beta$ is necessary and sufficient to guarantee survival in the large population limit. However, for the sake of convenience, the results in \ref{subsection:recTimeGeneAmp} will focus on the case where $\beta <\frac{1}{M-2}$.

% This bound on $\beta$ is sufficient and necessary to guarantee survival in the large population limit, which we will show below with Proposition~\ref{prop:extinction}.

We have already seen the formula for the mean of the sensitive cells in the previous section.  Now we examine the means of the other populations.  

\begin{lemma}
\label{lem:Means}
For $k = 3, 4, \ldots, M$, 
\begin{equation*}
\mathbb{E}[X_k(t)] =  n^{1-(k-2)\beta} (-1)^k S_k(t), 
\end{equation*}
\noindent where 
\begin{equation*}
S_k(t) = \sum_{i=2}^k \frac{e^{\lambda_i t}}{P_{i,k}}, \quad P_{i,k} = \prod_{j=2, j \neq i}^k (\lambda_j - \lambda_i).
\end{equation*}
\end{lemma}

An inductive proof of this lemma can be found in Appendix~\ref{proof:Means}. The proof uses the observation that for $\ell = 3, 4, \ldots, M$, the means satisfy the following ordinary differential equation:
\[
    \frac{d}{dt} \EE[X_\ell (t)] = \lambda_k \EE[X_\ell(t)] +  n^{-\beta} \EE[X_{\ell-1}(t)]. 
\]

\noindent For $k = 2,3,\ldots, M$ let $\phi_k(t) = \mathbb{E}[X_k(t)]$.  We already have that $\phi_2(t) = \phi_s(t)$.  For $k = 3, 4, \ldots, M$  

\begin{equation*}
\phi_k(z t_n) = \frac{ (-1)^k}{D^{k-2}} n^{1-(k-2)\beta} \tilde{S}_k(z), 
\end{equation*}

\noindent where 

\begin{equation*}
\tilde{S}_k(z) = \sum_{i=2}^k \frac{n^{-\lambda_iz/\lambda_2}}{\tilde{P}_{i,k}}, \quad \tilde{P}_{i,k} = \prod_{j=2, j \neq i}^k (j-i).
\end{equation*}

\noindent Note that for fixed $z$ and $n$ sufficiently large, 

\begin{equation*}
\max_{2 \leq k \leq M} \phi_k(zt_n) \in \{ \phi_2(zt_n), \phi_M(zt_n) \}. 
\end{equation*}
This is because $ \phi_2(zt_n) = n^{1-z}$ and for large $n$, $ \phi_k(zt_n) $ has largest degree $n^{1-(k-2)\beta -\lambda_k z/\lambda_2}$. Since $\lambda_M > 0 > \lambda_2$ and $\lambda_M$ is the largest net growth rate, the maximum will be $\phi_M(zt_n)$ if $z > \frac{\beta(k-2)}{1-\lambda_M/\lambda_2}$ and $\phi_2(zt_n)$ otherwise.

\section{Results} \label{results} 
\subsection{Conditions for tumor escape via amplification-driven resistance} 

We now present sufficient and necessary conditions for a tumor in the gene amplification model to escape extinction and achieve recurrence. Recall that the amplification rate is $\nu_n = n^{-\beta}$. As such, a threshold on $\beta$ is necessary to control the fluctuations between the cell populations and their means and to guarantee survival in the large population limit. 

Fix the initial population size $n$. Define $s_{n,k}$ to be the extinction probability of the lineage generated by a single type $k$ cell with birth rate $r_k$, death rate $d_k$, and mutation rate $\nu_n = n^{-\beta}$, as defined above for our gene amplification model. That is, $s_{n,k}$ is the probability that the original cell and all of its descendants eventually become extinct. Then we can write the extinction probability of the entire process starting with $n$ type 2 cells as $q_n = (s_{n,2})^n$ because individual cells behave independently. Recall that $k'$ is the first $k$ for which the type $k$ cells are supercritical, that is $r_k > d_k$. The following proposition relates the probability of extinction in the large population limit to $\beta$ and $k'$.

% We will show below with Proposition~\ref{prop:EstimateAmp} that there always exists a unique positive solution $\tilde{v}_n$ to the equation $\sum_{k=2}^M \phi_k(\tilde{v}_n t_n) = n$.  This means that, at the estimate for the recurrence time, the sum of the means is always equal to $n$. 

% Recall, however, that, 

% In this case, the total number of cells at the recurrence time estimate would be zero, causing the fluctuations between the actual populations and their means to become infinitely large in the large population limit.  On the other hand, by placing a threshold on $\beta$, we guarantee that the mutation rate is large enough to generate enough mutations in order to drive recurrence.  Thus the fluctuations between the cell populations and their means are controlled in this parameter regime.  

\begin{proposition}\label{prop:extinction}
    Let $q = \lim_{n\to \infty} q_n$ be the extinction probability of the gene amplification process in the large population limit. Then 
    \[
    q = \begin{cases}
        0 & \text{ if } 0< \beta< \frac{1}{k'-2},\\
        \displaystyle{\exp\left(-\frac{1-d_{k'}/r_{k'}}{\prod_{i=2}^{k'-1} d_i-r_i} \right)} & \text{ if } \beta = \frac{1}{k'-2}, \\
        1 & \text{ if } \beta > \frac{1}{k'-2}.
    \end{cases}
    \]
\end{proposition}
A complete proof of this proposition is given in Appendix~\ref{proof:extinction}. However, a sketch of the proof is as follows: By Theorem 2.1 of \cite{hautphenne_extinction_2013}, we know that for $2 \le k \le M$, $s_{n,k}$ is the minimal non-negative solution to
\[
    s_{n,k} =
    \begin{cases}
        \displaystyle{\frac{r_k}{r_k + d_k + \nu_n}(s_{n,k})^2 + \frac{d_k}{r_k + d_k + \nu_n} + \frac{\nu_n}{r_k + d_k + \nu_n}s_{n,k}s_{n,k+1}}& \text{ if } 2 \le k < M,\\
        \displaystyle{\frac{r_M}{r_M + d_M}(s_{n,M})^2 + \frac{d_M}{r_M + d_M}}& \text{ if } k = M.
    \end{cases}
\]
% The intuition for why this relation holds is that the terms represent whether the next event is a birth, a death, or an amplification, respectively. If the next event is a birth event, then there are now two independent type $k$ cells each with extinction probability $s_{n,k}$. If the next event is a death, then the particle becomes extinct. If the next event is a gene amplification, then the type $k$ cell becomes a type $k+1$ cell, whose extinction probability is $s_{n,k+1}$. In the case that $k = M$, no further amplification events are possible so only the first two terms exist. 
Intuitively, these equations represent conditioning on whether the next event is a birth, death, or amplification. By induction, we can establish that if $\ell = k'-k$,
\[
    s_{n,k} = 
    \begin{cases}
        \displaystyle{\frac{d_k}{r_k}} + O(\nu_n) & \text{ if } k' \le k \le M,\\
        \displaystyle{1 - \frac{1-d_{k'}/r_{k'}}{\prod_{i=k}^{k'-1} d_i-r_i} \nu_n^\ell + O (\nu_n^{\ell+1})} & \text{ if } 2 < k < k'.
    \end{cases}
\]
Then taking $q = \lim_n q_n = \lim_n (s_{n,2})^n$ and using that $\nu_n = n^{-\beta}$ produces our desired result. In fact, we also prove that the same result holds for extinction under a variation of the model where amplification can happen at any point during the cell cycle.

This result guarantees the survival of the process in the large population limit when $0 < \beta < \frac{1}{k'-2}$ and guarantees its extinction when $\beta > \frac{1}{k'-2}$. Note that this bound relies on $k'$ and not $M$, since the key to survival is producing a population of supercritical cells before the entire tumor becomes extinct. This result can also be applied to the gene mutation model and justifies bounding $0 < \alpha < 1$ to guarantee survival of the process.

\subsection{Recurrence Time in Gene Amplification Model} \label{subsection:recTimeGeneAmp}
To compare the resistance mechanisms, we analyze the tumor recurrence time in both models.  Tumor recurrence time is defined intuitively as the amount of time it takes from the beginning of treatment until the total number of cancer cells surpasses the initial population size $n$, neglecting possible brief instances of this early on due to the stochasticity of the processes.  A more precise mathematical definition of recurrence time will be given below.  

Let 
\begin{equation}
\frac{-\lambda_2 \beta (M-2)}{\lambda_M - \lambda_2} < d < \frac{-\lambda_2 \beta (M-2)}{\lambda_M}.
\label{restrict3} 
\end{equation}
We will prove that, with high probability, for sufficiently large $n$ there are no permanent recurrence events in the time interval $[0, dt_n]$. In other words, either the number of cancer cells does not surpass $n$ during this time or it surpasses $n$ only briefly before subsequently falling to $n$ or lower once again.  To show this, we  prove that for $n$ large, with high probability the total number of cancer cells does not exceed $n$ at time $dt_n$.   
\begin{lemma}
\label{NoPermRecurAmp}
\begin{equation*}
\lim_{n \rightarrow \infty} \mathbb{P} \left( \sum_{k=2}^M X_k(dt_n) - n \leq 0 \right) = 1.
\end{equation*} 
\end{lemma}
As a sketch of the proof, note that 

\begin{align*}
\mathbb{P} \left( \sum_{k=2}^M X_k(dt_n) - n > 0 \right) 
&= \mathbb{P} \left( \sum_{k=2}^M \hat{A}_k(n) + \hat{A}_{\phi}(n) > 0 \right),
\end{align*}

\noindent where 

\begin{align*}
\hat{A}_k(n) &= n^{\beta(M-2) + \lambda_Md/\lambda_2 - 1} \left( X_k(dt_n) - \phi_k(dt_n) \right), \\
\hat{A}_{\phi}(n) &= n^{\beta(M-2) + \lambda_Md/\lambda_2 - 1} \left( \sum_{k=2}^M \phi_k(dt_n) - n \right).
\end{align*}
We can then show that $\left| \hat{A}_k(n) \right|$ converges to zero in probability using Proposition~\ref{prop:FlucAmp} below and that  $\hat{A}_{\phi}(n)$ is negative in the large population limit, which suffices to complete the proof. A full proof of this result can be found in Appendix~\ref{proof:NoPermRecurAmp}. Then for sufficiently large $n$, the total number of cancer cells is less than or equal to $n$ at time $dt_n$ with high probability, implying that permanent recurrence occurs at time $dt_n$ or later. 
\begin{definition}
Define the \textbf{recurrence time in the gene amplification model} as 
\begin{equation*}
\omega_n = \inf \left\{ t \geq dt_n : \sum_{k=2}^M X_k(t) > n \right\}, 
\end{equation*}
\end{definition}

\noindent We restrict to $t \ge dt_n$ in order to ignore any initial fluctuations that may temporarily bring the total population above $n$. 

To obtain an estimate of the stochastic recurrence time, we start with the following proposition. 

\begin{proposition}
\label{prop:EstimateAmp}
There exists a unique $\tilde{v}_n > 0$ satisfying $\sum_{k=2}^M \phi_k(\tilde{v}_n t_n) = n$.  Moreover, $b_n < \tilde{v}_n < B_n$ where: 

\begin{align*}
b_n &= \frac{1}{\lambda_M} \left[ -\lambda_2 \beta (M-2) - \frac{1}{t_n} \log \left[ \frac{ (-1)^M}{D^{M-2}\tilde{P}_{M,M}} \left( 1 - \frac{\lambda_M}{\lambda_2} \right) \right] \right],\\
B_n &= \frac{1}{\lambda_M} \left[ -\lambda_2 \beta (M-2) - \frac{1}{t_n} \log \left[ \frac{ (-1)^M}{D^{M-2}\tilde{P}_{M,M}} \right] \right]. 
\end{align*}

\noindent Hence $\tilde{v}_n \rightarrow -\frac{\lambda_2}{\lambda_M} \beta (M-2)$ as $n \rightarrow \infty$.  
\end{proposition}

The idea behind the proof of this proposition is to define $\hat{f}_n(z) = \sum_{k=2}^M \phi_k(zt_n) - n$. We can find its critical points and show that $\hat{f}_n(b_n) < 0 < \hat{f}_n(B_n)$. From this, we can show that there exists a unique $\tilde{v}_n > 0$ with $\hat{f}_n(\tilde{v}_n) = 0$ and that $\tilde{v}_n \in (b_n, B_n)$. Finally, since $b_n \to -\frac{\lambda_2}{\lambda_M} \beta (M-2)$ and $B_n \to -\frac{\lambda_2}{\lambda_M} \beta (M-2)$, we can conclude that $\tilde{v}_n \rightarrow -\frac{\lambda_2}{\lambda_M} \beta (M-2)$ as $n \rightarrow \infty$ as well.  A full proof of this result can be found in Appendix~\ref{proof:EstimateAmp}.

\begin{figure}[htb]
    \centering
    \begin{subfigure}{0.49\textwidth}
        \centering
        \includegraphics[width=\linewidth]{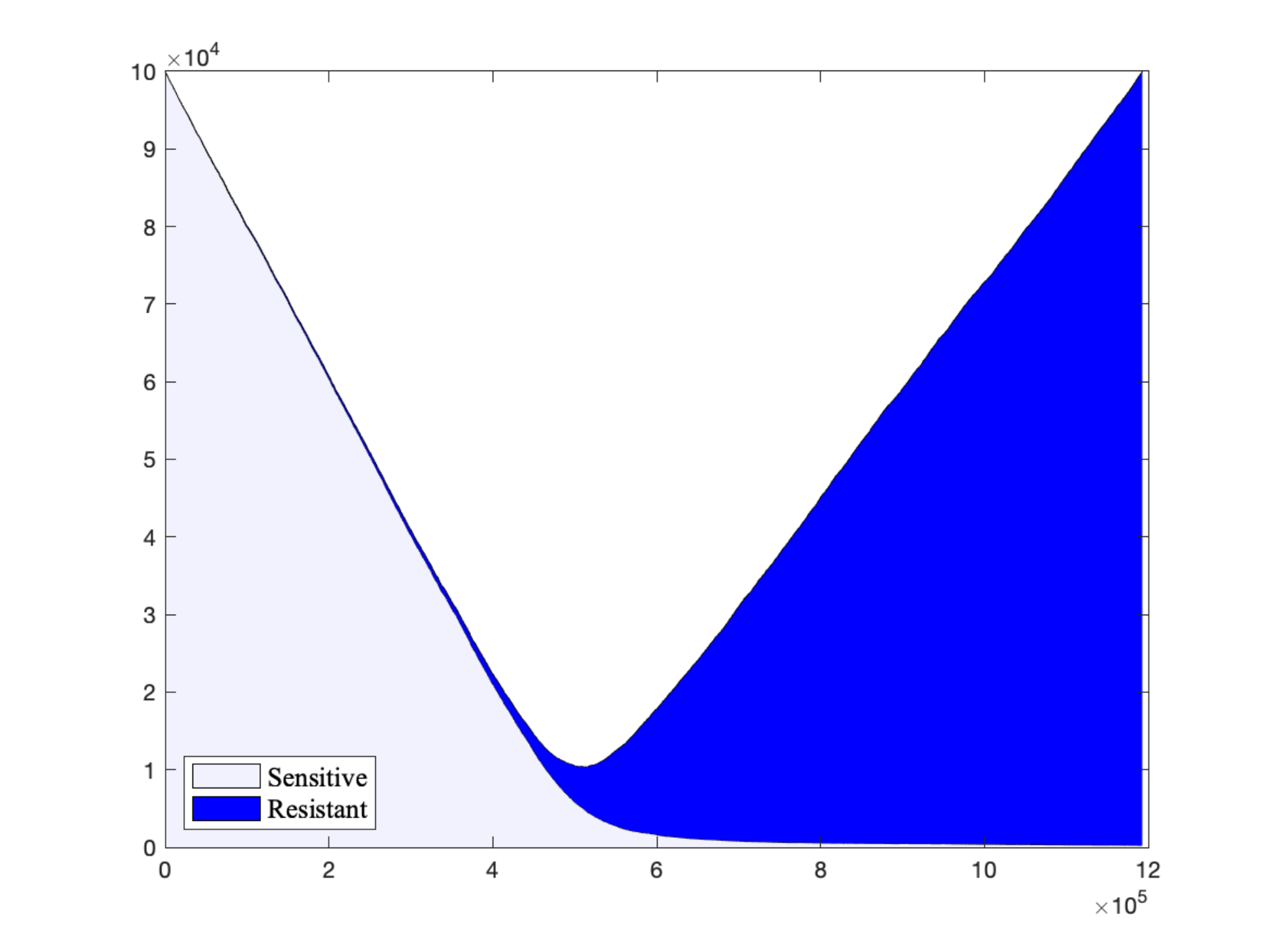}
        \caption{Gene mutation model.}
        \label{fig:GeneMutRecSim}
    \end{subfigure}%
    \begin{subfigure}{0.49\textwidth}
        \centering
        \includegraphics[width=\linewidth]{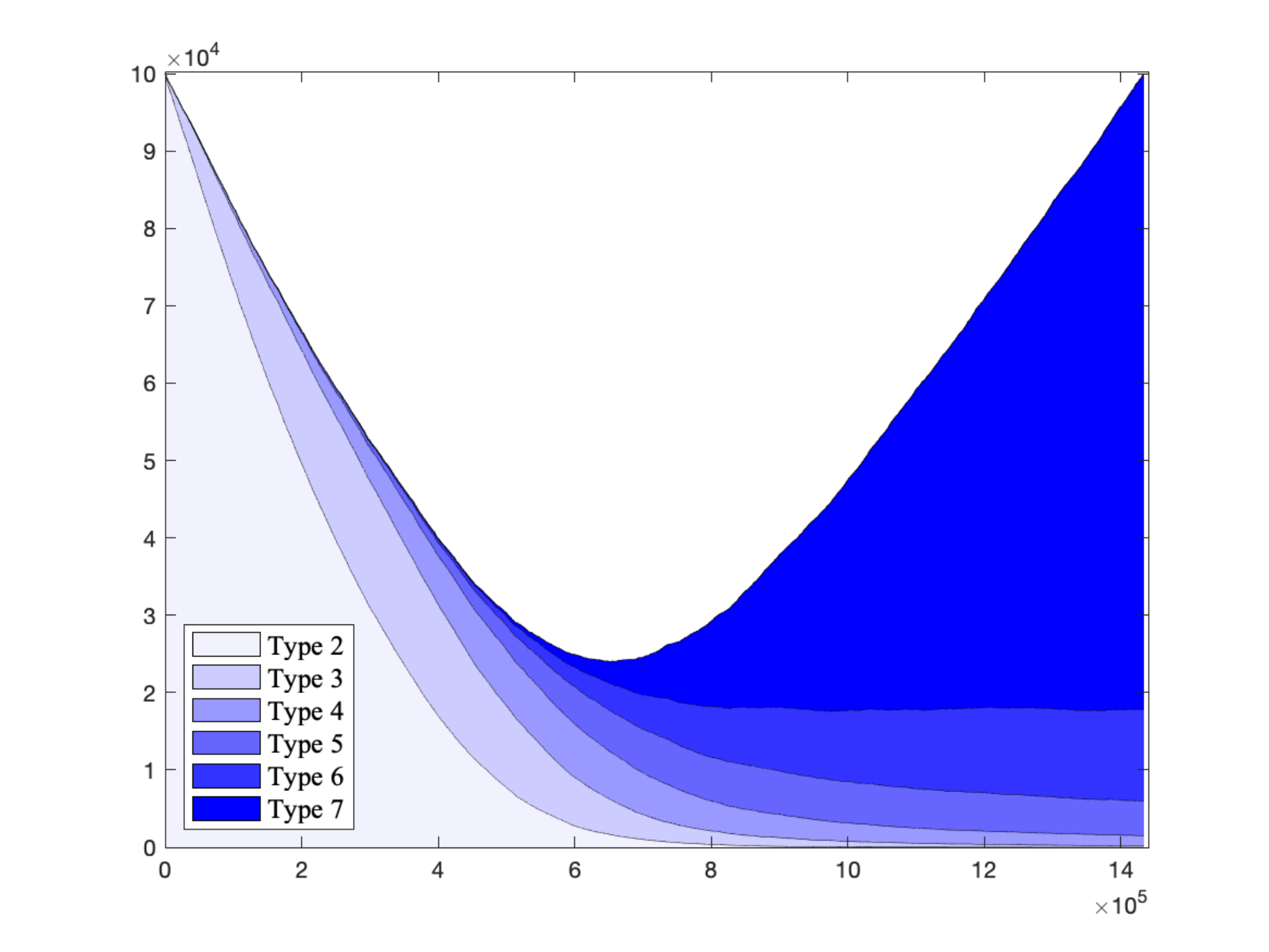}
        \caption{Gene amplification model.}
        \label{fig:GeneAmpRecSim}
    \end{subfigure}
    \caption{(a) An example simulation of a tumor under the gene mutation model with $n = 10^5$, $\alpha = 0.5$, $r_s = 1$, $r_m = 2$ and $d_s = d_m = 1.5$. (b) An example simulation of a tumor under the gene amplification model with $M = 7$, $n = 10^5$, $\beta = 0.1$, and birth rates ranging from $r_2 = 1.0$ to $r_7 = 2.0$ and uniform death rates of $d_k = 1.5$.}
    \label{fig:bothSim}
    
\end{figure}

For ease of notation, let $v_n = \tilde{v}_n t_n$ represent our estimate of the tumor recurrence time. Given $\varepsilon > 0$, define 
\begin{align*}
v_n^-(\varepsilon) &= \frac{v_n - \varepsilon}{t_n},\quad 
v_n^+(\varepsilon) = \frac{v_n + \varepsilon}{t_n}.
\end{align*}
Recall also that we assume $\beta < \frac{1}{M-2}.$

We will now state a proposition pertaining to the fluctuations of these populations about their means. 

\begin{proposition}
\label{prop:FlucAmp}
For $\delta > 0$ and $2 \leq k \leq M$, 
\begin{equation*}
\lim_{n \rightarrow \infty} \mathbb{P} \left( \sup_{z \in [d, v_n^+(\varepsilon)]} n^{\beta(M-2) + \lambda_M z/\lambda_2 - 1} \left| X_k(zt_n) - \phi_k(zt_n) \right| > \delta \right) = 0.
\end{equation*}
\end{proposition} 

\noindent The proof of this proposition can be found in Appendix~\ref{proof:FlucAmp}. 

We now state our main law of large numbers result regarding the convergence of the stochastic recurrence time in the gene amplification model: 

\begin{theorem}
\label{LLNAmp} 

\noindent For all $\varepsilon > 0$, 
\begin{equation*}
\lim_{n \rightarrow \infty} \mathbb{P} \left( \left| \omega_n - v_n \right| > \varepsilon \right) = 0. 
\end{equation*}
\end{theorem}

\noindent In other words, the difference between the stochastic recurrence time and its estimate converges to 0 in probability as $n \rightarrow \infty$. Recall that $v_n = \tilde{v}_n t_n$ and $\tilde{v}_n \to -\frac{\lambda_2}{\lambda_M}\beta(M-2)$ as $n \to \infty$ by Proposition~\ref{prop:EstimateAmp}. So in the large population limit, the recurrence time is not influenced by the birth and death rates directly. Instead, the estimate depends on just the net growth rates $\lambda_2$ and $\lambda_M$ (which also determine the net growth rates $\lambda_k$ for $2 < k < M$), $\beta$, and $M$. 

% This suggests that drugs that slow growth by reducing birth rates and drugs that do so by increasing death rates both would have an equivalent effect on the recurrence time in the large population limit. 

In order to show the desired result, we must show 

\begin{equation*}
\lim_{n \rightarrow \infty} \mathbb{P} (\omega_n > v_n + \varepsilon) + \lim_{n \rightarrow \infty} \mathbb{P} (\omega_n < v_n - \varepsilon) = 0.
\end{equation*}

\noindent We can observe that 

\begin{align*}
\mathbb{P} (\omega_n < v_n - \varepsilon)
&\leq \mathbb{P} \left( \sup_{z \in [d, v_n^-(\varepsilon)]} \left( \sum_{k=2}^M X_k(zt_n) - n \right) > 0 \right) \\
&= \mathbb{P} \left( \sup_{z \in [d, v_n^-(\varepsilon)]} n^{\beta(M-2) + \lambda_Mz/\lambda_2 - 1} \left( \sum_{k=2}^M X_k(zt_n) + \sum_{k=2}^M \phi_k(zt_n) - \sum_{k=2}^M \phi_k(zt_n) - n \right) > 0 \right) \\
&\leq \mathbb{P} \left( \sum_{k=2}^M \hat{B}_k(n, \varepsilon) + \hat{B}_{\phi}(n, \varepsilon) > 0 \right),
\end{align*}

\noindent where 

\begin{align*}
\hat{B}_k(n, \varepsilon) &= \sup_{z \in [d, v_n^-(\varepsilon)]} n^{\beta(M-2) + \lambda_Mz/\lambda_2 - 1} \left( X_k(zt_n) - \phi_k(zt_n) \right), \\
\hat{B}_{\phi}(n, \varepsilon) &= \sup_{z \in [d, v_n^-(\varepsilon)]} n^{\beta(M-2) + \lambda_Mz/\lambda_2 - 1} \left( \sum_{k=2}^M \phi_k(zt_n) - n \right). 
\end{align*}

\noindent Using Proposition~\ref{prop:FlucAmp}, we can show that the $\hat{B}_k(n, \varepsilon)$'s converge to zero in probability. Then it suffices to show that $\hat{B}_{\phi}(n, \varepsilon)$ is negative in the large population limit. This can be done by examining the critical points of $\hat{g}_n(z) = n^{\beta(M-2) + \lambda_M z/\lambda_2 - 1} \left( \sum_{k=2}^M \phi_k(z t_n) - n \right)$. A complete proof of this theorem can be found in Appendix~\ref{proof:LLNAmp}.

\subsection{Recurrence Time in Gene Mutation Model} 

Results analogous to those presented above for the gene amplification model also exist for the gene mutation model. In fact, some of the results are almost direct corollaries of the results from the gene amplification model in the case where $M = 3$. However, there is a slight difference in the definition of the recurrence time in this context, which we will give below. As such, full proofs of the following results are available in the appendix for the sake of both completeness and illumination. 

Let 
\begin{equation*}
c = \frac{\alpha \lambda_s(\lambda_s - 2\lambda_m )}{2\lambda_m (\lambda_m  - \lambda_s)}. 
\end{equation*}
The following lemma is analogous to Lemma~\ref{NoPermRecurAmp} for the gene amplification model and shows that with high probability there are no permanent recurrence events in the interval $[0,ct_n]$.

\begin{lemma}
\label{NoPermRecurMut}
\begin{equation*}
\lim_{n \rightarrow \infty} \mathbb{P} \left( X_s(c t_n) + X_m(c t_n) - n \leq 0 \right) = 1.
\end{equation*} 
\end{lemma}

\begin{definition}
Define the \textbf{recurrence time in the gene mutation model} as 
\begin{equation*}
\tau_n = \inf \{ t \geq ct_n : X_s(t) + X_m(t) > n \}.
\end{equation*}
\end{definition}
\noindent As in the gene amplification model, we restrict $t$ in order to ignore any initial fluctuations that may temporarily bring the total population above $n$. 

\begin{proposition}
\label{prop:EstimateMut}
There exists a unique $\tilde{u}_n > 0$ satisfying $\phi_s(\tilde{u}_n t_n) + \phi_m(\tilde{u}_n t_n) = n$.  Moreover, $a_n < \tilde{u}_n < A_n$ where: 

\begin{align*}
a_n &= \frac{1}{\lambda_m } \left( -\lambda_s \alpha + \frac{1}{t_n} \log \left( -\lambda_s \right) \right),\\
A_n &= \frac{1}{\lambda_m } \left( -\lambda_s \alpha + \frac{1}{t_n} \log \left( {\lambda_m  - \lambda_s} \right) \right).
\end{align*}

\noindent Hence $\tilde{u}_n \rightarrow -\frac{\lambda_s }{\lambda_m }\alpha$ as $n \rightarrow \infty$.  
\end{proposition}

% that will aid in the proof of Lemma~\ref{NoPermRecurMut} and Theorem~\ref{LLNMut} below.
For ease of notation, let $u_n = \tilde{u}_n t_n$ represent our estimate of the tumor recurrence time. Given $\varepsilon > 0$, define 

\begin{align*}
u_n^-(\varepsilon) = \frac{u_n - \varepsilon}{t_n}, \quad 
u_n^+(\varepsilon) = \frac{u_n + \varepsilon}{t_n}.
\end{align*}
\begin{proposition}
\label{prop:FlucMut}
For $\delta > 0$ and $i \in \{ s, m \}$,  

\begin{equation*}
\lim_{n \rightarrow \infty} \mathbb{P} \left( \sup_{z \in [c, u_n^+(\varepsilon)]} n^{\alpha + \lambda_m  z/\lambda_s - 1} \left| X_i(zt_n) - \phi_i(zt_n) \right| > \delta \right) = 0.
\end{equation*}
\end{proposition} 

We now state our main law of large numbers result regarding the convergence of the stochastic recurrence time in the gene mutation model:

\begin{theorem}
\label{LLNMut} 

\noindent For all $\varepsilon > 0$, 
\begin{equation*}
\lim_{n \rightarrow \infty} \mathbb{P} \left( \left| \tau_n - u_n \right| > \varepsilon \right) = 0. 
\end{equation*}
\end{theorem}

In other words, the difference between the stochastic recurrence time and its estimate converges to 0 in probability as $n \rightarrow \infty$ in the gene mutation model as well. Recall that $u_n = \tilde{u}_n t_n$ and $\tilde{u}_n \to -\frac{\lambda_s}{\lambda_m }\alpha$ as $n \to \infty$.

\subsection{Model Comparisons and Simulations}

{\bf Recurrence time comparison.} We first compare recurrence estimates under the amplification and mutation models using our analytic results. Consider the ratio between the estimated recurrence times $-\frac{\lambda_2}{\lambda_M}\beta(M-2)$ and $-\frac{\lambda_s}{\lambda_m}\alpha$, given by Proposition~\ref{prop:EstimateAmp} and Proposition~\ref{prop:EstimateMut} respectively. This ratio will be denoted $v/u$ and can be simplified to 
\[
v/u = \frac{-\frac{\lambda_2}{\lambda_M}\beta(M-2)}{-\frac{\lambda_s}{\lambda_m}\alpha} = \frac{\lambda_m}{\lambda_M} \cdot \frac{\beta(M-2)}{\alpha},
\]
because we assume $\lambda_2 = \lambda_s$.

We first consider the case where $\lambda_M = \lambda_m $, which represents the scenario where amplification and mutation can eventually confer the same degree of resistance. In the amplification model resistance is attained incrementally, whereas in the mutation model resistance is attained in a single step. For example, Figure~\ref{fig:bothSim} shows a simulation under the gene mutation model with $\lambda_s = -0.5$ and $\lambda_m = 0.5$ and a simulation under the gene amplification model with $M=7$ and $\lambda_k \in [-0.5,0.5]$. If $\lambda_M = \lambda_m $, then $v/u = \frac{\beta(M-2)}{\alpha}$, so the ratio between the amplification and mutation-driven recurrence times increases linearly with $M$ and the amplification parameter $\beta$.  This agrees with intuition that acquiring resistance more gradually (either via more steps or a lower amplification rate) leads to longer times to recurrence than acquiring resistance via mutation. However, amplification events typically occur more frequently than mutation events so the ratio of the rates $\beta/\alpha$ plays a key role in determining the mechanism under which recurrence is more rapid for any specific system.

 % Recall that the amplification rate is $\nu_n = n^{-\beta}$. As a rough heuristic, since $M-2$ amplification events are required for a wild type cell to attain $M$ copies via amplification, the rate at which a wild type cell will give rise to a fully amplified cell is $(\nu_n)^{M-2} = n^{-\beta(M-2)}$. Since the rate at which a wild type cell gives rise to a mutated cell in the mutation model is $\mu_n = n^{-\alpha}$. Comparing these two rates provides intuition for why $v/u = \frac{\beta(M-2)}{\alpha}$ in this scenario. 
        
        % Case $\lambda_2 = \lambda_s$ and $\lambda_M = \lambda_m $. We compare the case where the total potential fitness gained is the same, but it happens with one big step in the mutation model and several small steps in the amplification model. In this scenario $u_n/v_n = \frac{\alpha}{\beta (M-2)}$. 

        % Takeaway: The relative recurrence speeds in a ``one big step" vs "several small steps" scenario depends on the size of $\alpha$ relative to $\beta(M-2)$.

Next consider the case where $\lambda_k = \lambda_m $ for some fixed $k < M$. This would represent a scenario where cells have the possibility of increased fitness through continued amplification relative to mutation. That is, $k$ amplified copies is enough to have acquired the same fitness as a mutated cell, but increasing beyond $k$ copies allows the cell to achieve net growth rates greater than that of mutated cells. In this scenario,
$\lambda_M = \frac{\lambda_k -\lambda_2}{k-2} (M-2) + \lambda_2$, so 
\[
v/u = \frac{\beta(M-2)}{\alpha }\cdot \frac{\lambda_m }{\frac{\lambda_k -\lambda_2}{k-2} (M-2) + \lambda_2} = \frac{\beta}{\alpha }\cdot \frac{\lambda_m  (k-2)}{\lambda_m  -\lambda_2  + (k-2)\lambda_2/(M-2)}.
\]
Since $\lambda_2 < 0$, notice that $(k-2)\lambda_2/(M-2) < 0$ so $v/u$ is bounded below as $M$ increases. This means that even if cells can acquire higher net growth rates through continued amplification, the time it takes to accumulate the mutations is enough to prevent recurrence before the given bound. In other words, the potential for increased fitness through continued amplification has diminishing returns on the recurrence time. This is a useful observation because $M$ may be unknown in practice. It may be difficult or impossible to determine whether the largest empirically observed copy number is truly the maximum copy number possible. This observation means that even if there is a gap between the true value of $M$ and what is observed, the effect it will have on the estimated recurrence time is limited by this bound.

{\em Non-small cell lung cancer (NSCLC) example.} In NSCLC, both MET amplification and T790M mutation have been recognized as mechanisms of acquiring resistance to EFGR inhibitors such as gefitinib and erlotinib \cite{bean_met_2007}. We estimate $n = 10^{10}$ as typical solid tumors have between $10^7$ and $10^9$ cells per cubic centimeter \cite{michaelson_breast_1999}. For a given locus, amplification rates have been estimated to be to $10^{-3}$ or $10^{-4}$ events per division \cite{stark_gene_1984, tlsty_differences_1989} while mutation rates are estimated to be on the order of $10^{-10}$ events per division \cite{brown_poised_2014}. Solving for $\beta, \alpha $ from $\nu_n = n^{-\beta}$ and $\mu_n = n^{-\alpha}$, respectively, yields an estimate of $\beta = 0.3$ and $\alpha = 1$. In a comparison of gefitinib resistance in non-small cell lung cancer arising from T790M point mutation and from MET amplification, Engelman et al.\ \cite{engelman_met_2007} observed that MET was amplified by a factor of 5 to 10 in resistant cells , so we can estimate $5 \le M \le 10$. In a comparison of erlotinib resistance driven by T790M mutation and MET amplification, Mumenthaler et al.\ \cite{mumenthaler_impact_2015} observed similar levels of amplification and documented the net growth rates of amplified and mutated cells at various drug concentrations. From their data, we can estimate $0.8 \le \frac{\lambda_m }{\lambda_M} \le 0.90$ for concentrations of erlotinib up to 1 micromolar. Then all together, we can estimate $0.7 \le v_n/u_n \le 2.2$. This suggests that the recurrence times under these two mechanisms may not be distinguishable. 

%However, the specific mutation and amplification rates $\mu_n$ and $\nu_n$ for non-small cell lung cancer may differ from generic estimates, the observed maximum copy number may differ from the true maximum copy number $M$, and other mechanisms \emph{in vivo} may influence recurrence as well. 
 
\begin{figure}
\centering 
\begin{subfigure}{\linewidth}
\centering
\includegraphics[width=0.5\linewidth]{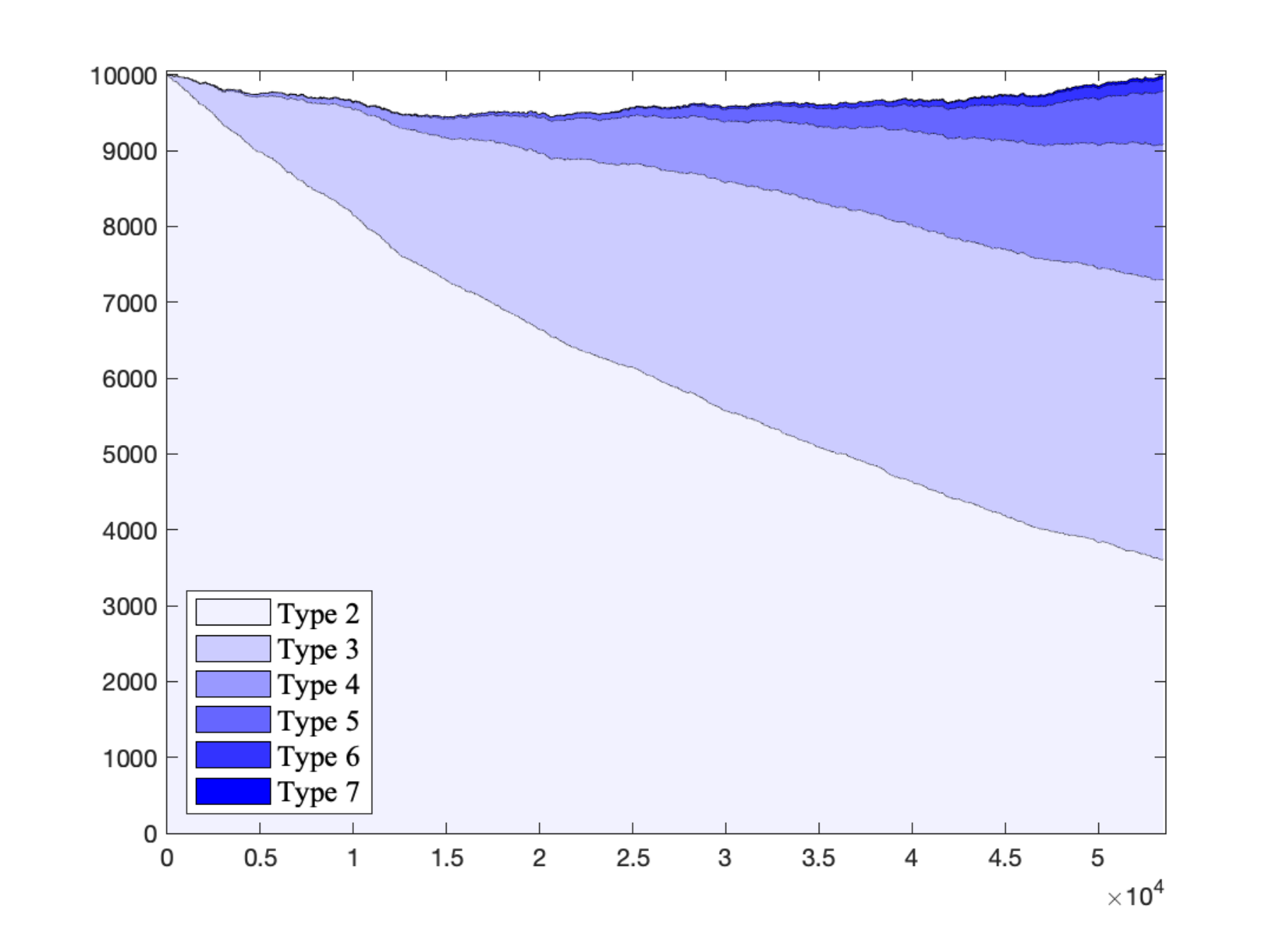}
\hspace{0.05\linewidth}
\includegraphics[width = 0.4\linewidth]{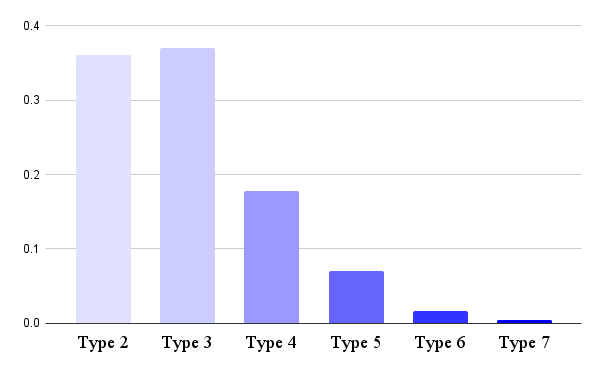}
\caption{$d_k = 1.1$}
\label{fig:TumorStackedKPrime3}
\end{subfigure}%

\begin{subfigure}{\linewidth}
\centering
\includegraphics[width=0.5\linewidth]{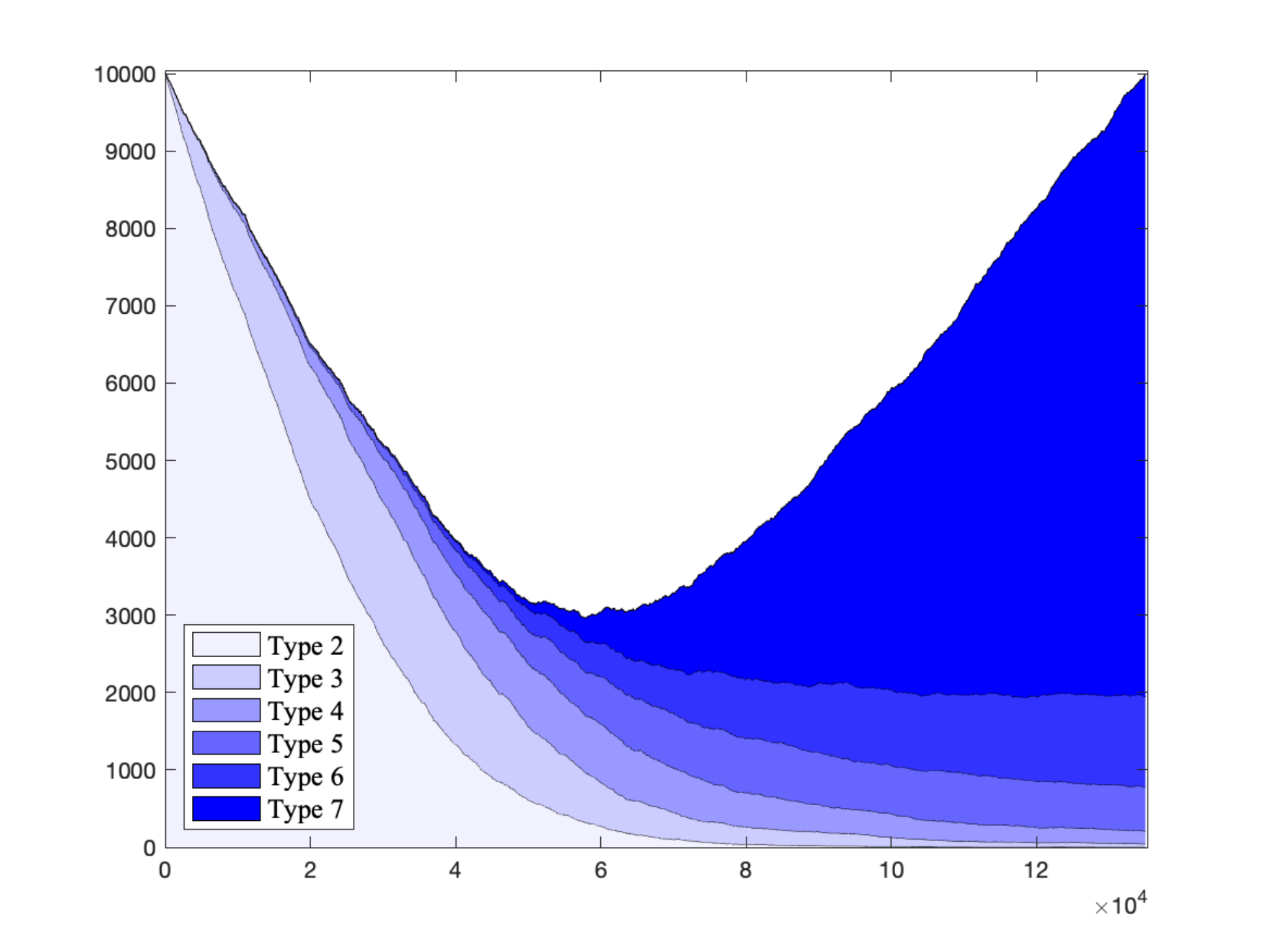}
\hspace{0.05\linewidth}
\includegraphics[width = 0.4\linewidth]{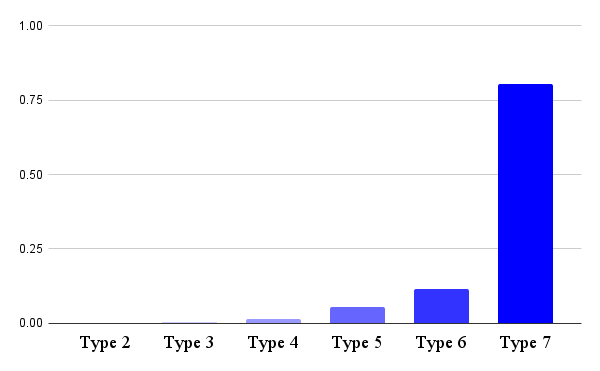}
\caption{$d_k = 1.5$}
\label{fig:TumorStackedKPrime5}
\end{subfigure}%

\begin{subfigure}{\linewidth}
\centering
\includegraphics[width=0.5\linewidth]{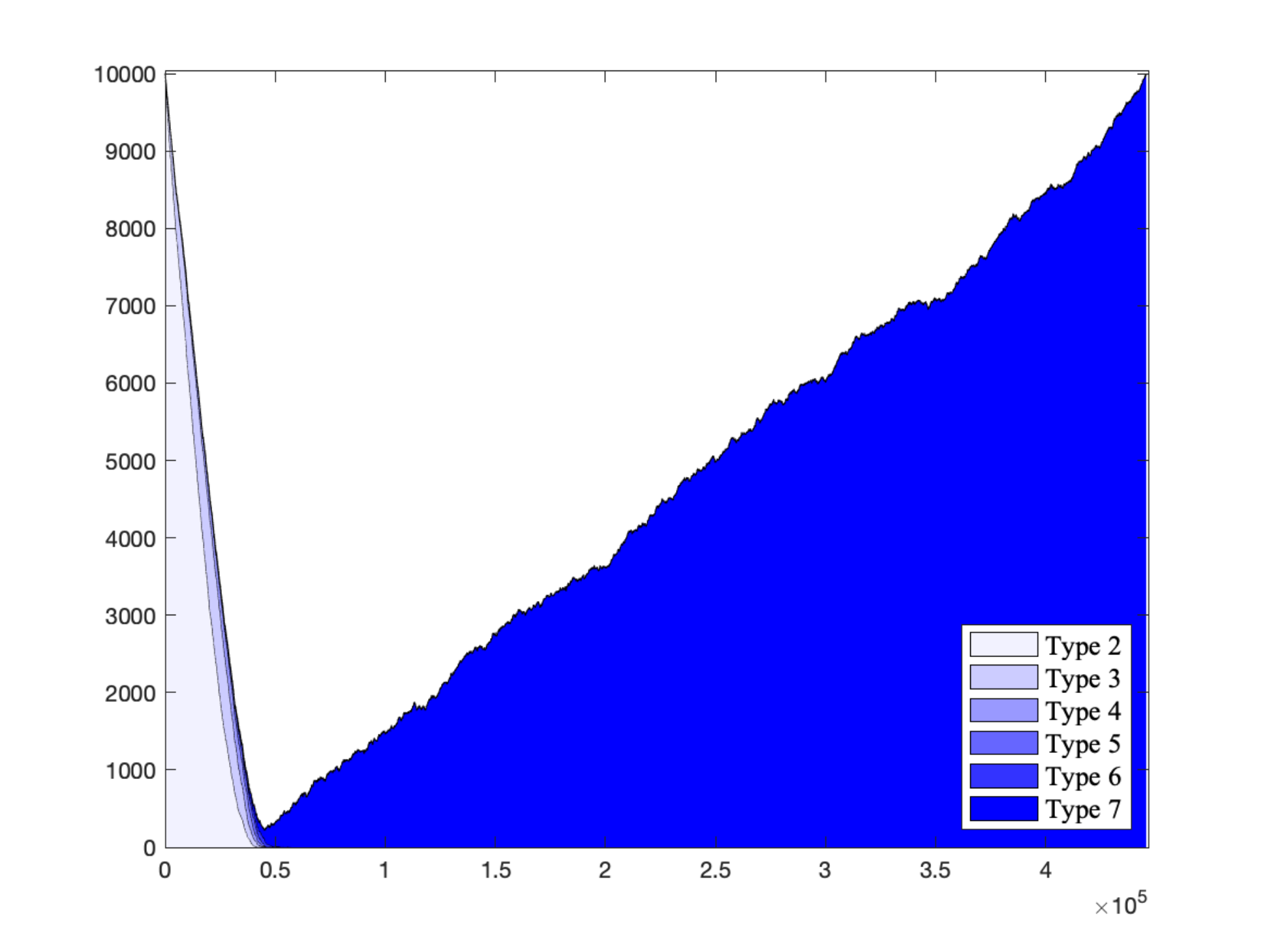}
\hspace{0.05\linewidth}
\includegraphics[width = 0.4\linewidth]{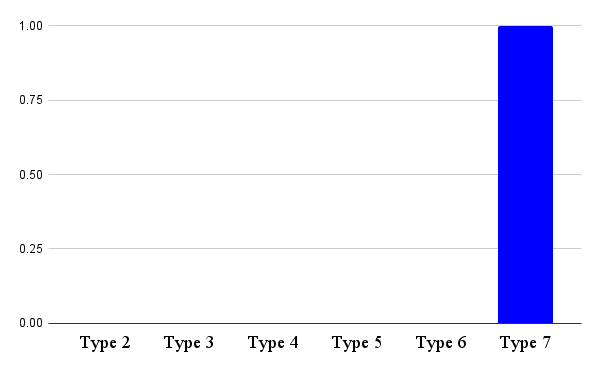}
\caption{$d_k = 1.9$}
\label{fig:TumorStackedKPrime7}
\end{subfigure}%
\caption{The composition of simulated tumors over time and at recurrence is shown for (a) $d_k = 1.1$, (b) $d_k = 1.5$, and (c) $d_k = 1.9$. The simulations were all conducted using the gene amplification model with $M = 7$, $n = 10^4$, $\beta = 0.1$, and birth rates ranging from $r_2 = 1.0$ to $r_7 = 2.0$. The figures on the right show the tumor composition at the time of recurrence. }
\label{fig:geneAmpStackedKPrime} 
\end{figure}

{\bf Composition of recurrent tumors driven by amplification resistance mechanisms.}
Using simulations, we first examine the effect of drug concentration and efficacy on the composition of the recurrent tumor in the amplification model. In particular, we study how recurrence timing and composition of the recurrent tumor depend on the net growth rates. Recall that $\lambda_k$ is the net growth rate of the type $k$ cells. 
In Figure~\ref{fig:TumorStackedKPrime3}, we consider the amplification model with $M = 7$ types with birth rates $r_k \in [1.0, 2.0]$ and all types having death rate $d_k = 1.1$. This results in net growth rates $\lambda_k \in [-0.1, 0.9]$ and $k' = 3$ being the first cell type to have a positive net growth rate in this scenario. We then increase the death rates uniformly to $d_k = 1.5$ in \ref{fig:TumorStackedKPrime5} and $d_k = 1.9$ in \ref{fig:TumorStackedKPrime7}, decreasing all the net growth rates by 0.2 each time, to simulate increased drug concentration or efficacy. We observe that the minimum tumor size decreases and the composition consists of more cells with a high number of amplified copies. Moreover, we observe that increasing the $d_k$'s results in a decrease in heterogeneity. In \ref{fig:TumorStackedKPrime3}, all cell types are present at recurrence. In \ref{fig:TumorStackedKPrime5}, the majority of the cells at recurrence are type 7 and less than 5\% of the cells are of type 2, 3, or 4. In \ref{fig:TumorStackedKPrime7}, all the cells are type 7 at recurrence.

% We can use our results for the amplification model and simulations to examine potential effects of drug concentration and efficacy on the recurrence time. One way to model this is by shifting each $\lambda_k$ by a uniform negative constant. This would represent decreasing the birth rate and/or increasing the death rate for all cells. Since the estimated recurrence time in the amplification model is $-\frac{\lambda_2}{\lambda_M} \beta(M-2)$ and $\lambda_2 < 0 < \lambda_M$, decreasing $\lambda_2$ and $\lambda_M$ will naturally increase the estimated recurrence time. In \citet{foo_dynamics_2013}, we define the ``turnaround time" as the time at which the tumor population as a whole switches from subcritical to supercritical. 

Overall, we observe that increasing drug concentration, in the amplification-driven resistance model, leads to 
a stronger initial reduction in tumor burden, but that the eventual recurrent tumor population is less heterogeneous, more aggressive and harbors high levels of drug-resistance.  To explain this, we note that decreasing the net growth rates $\lambda_k$ has a two-fold effect. First, the tumor size will decrease more quickly before it reaches its minimum size and increase more slowly afterwards because the negative growth rates will be amplified in magnitude while the positive growth rates will decrease in magnitude. Second, $k'$, the number of gene copies necessary to become supercritical, may increase depending on the size of the shift. That is, a sensitive tumor cell may need to undergo more amplification events to have a positive net growth rate. This would increase the time it takes for the overall tumor size to increase as the tumor has to ``wait" longer for more amplifications to occur. Together these two effects result in the tumor having a smaller minimum size and an increased number of gene copies at recurrence. Additionally, the heterogeneity of the tumor at recurrence will decrease as $k'$ increases as fewer cell types will have sufficient drug resistance to maintain growth. However, the extinction results given in Proposition~\ref{prop:extinction} mean that if it is still the case that $\lambda_M > 0$ and the amplification parameter $\beta < \frac{1}{k'-2}$, tumor survival and recurrence will not be prevented. 

% Our observations also suggest that the tumor may consist of an increased proportion of resistant cells at recurrence. 
        
        % How does drug efficacy affect tumor development and composition? One way to model this question is by examining the effect of shifting each $\lambda_k$ by a uniform negative constant.  Doing so may increase $k'$, the number of gene copies necessary to become supercritical. The result is a faster decrease in population size and a smaller size at turnaround time (introduce definition). This is because more amplifications are needed to become supercritical so the tumor is decreasing in size for longer and is doing so at a faster rate. The increased $k'$ and delayed recurrence results in an increased number of copies at recurrence. For example in Figure~\ref{fig:geneAmpStackedKPrime}, we decrease each $\lambda_k$ by $0.2$ (twice). We see that the size at turnaround time decreases and the composition consists of more cells with a high number of amplified copies.

\begin{figure}
\centering 
\begin{subfigure}{\linewidth}
\centering
\includegraphics[width=0.5\linewidth]{Figures/newstacked11d.pdf}
\hspace{0.05\linewidth}
\includegraphics[width = 0.4\linewidth]{Figures/newcomp11d.png}
\caption{$\beta = 0.1$}
\label{fig:geneAmpStacked02}
\end{subfigure}%

\begin{subfigure}{\linewidth}
\centering
\includegraphics[width=0.5\linewidth]{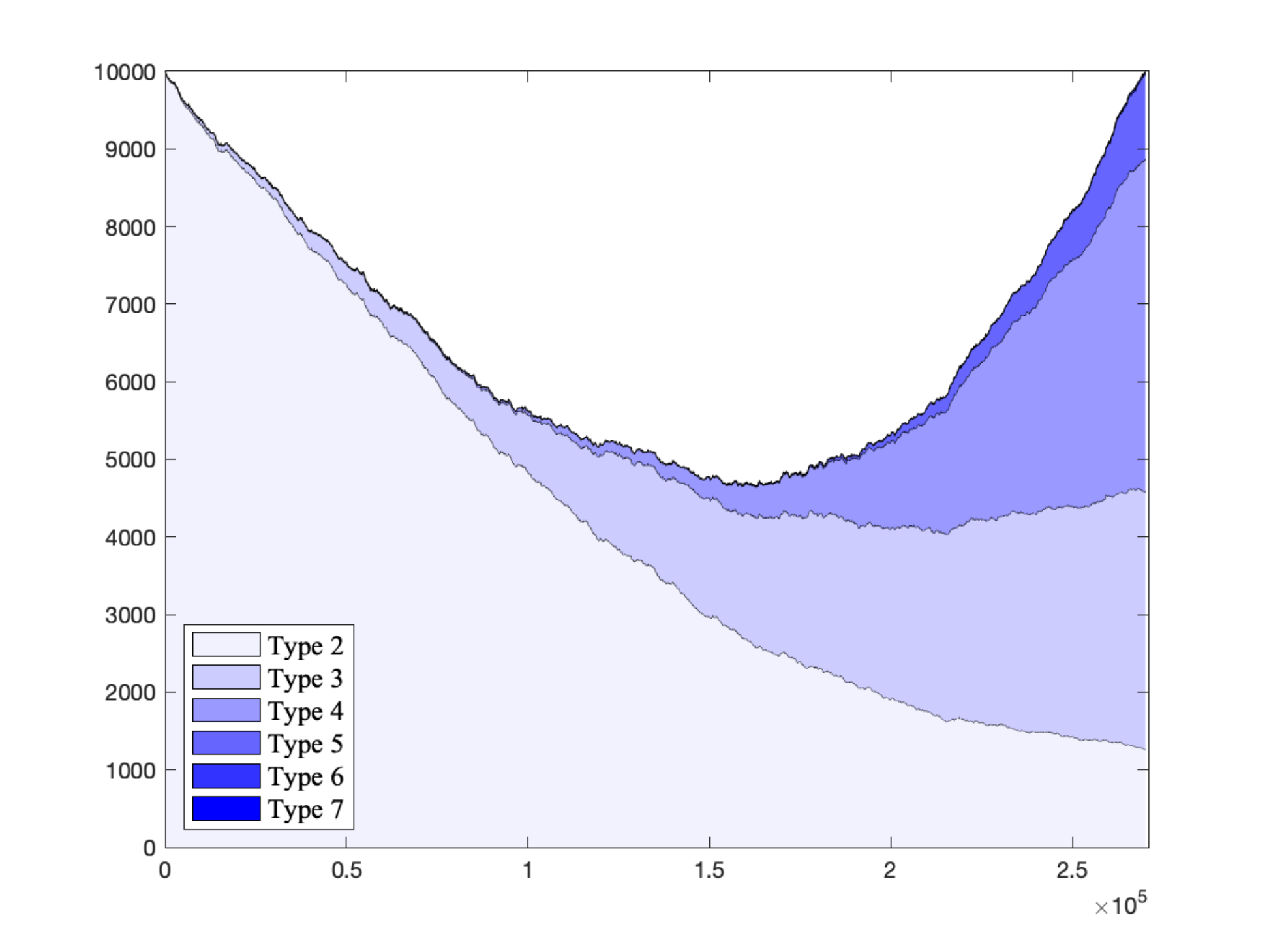}
\hspace{0.05\linewidth}
\includegraphics[width = 0.4\linewidth]{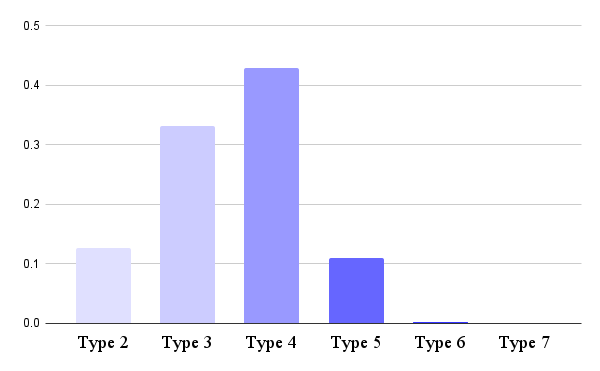}
\caption{$\beta = 0.5$}
\label{fig:geneAmpStacked05}
\end{subfigure}%

\begin{subfigure}{\linewidth}
\centering
\includegraphics[width=0.5\linewidth]{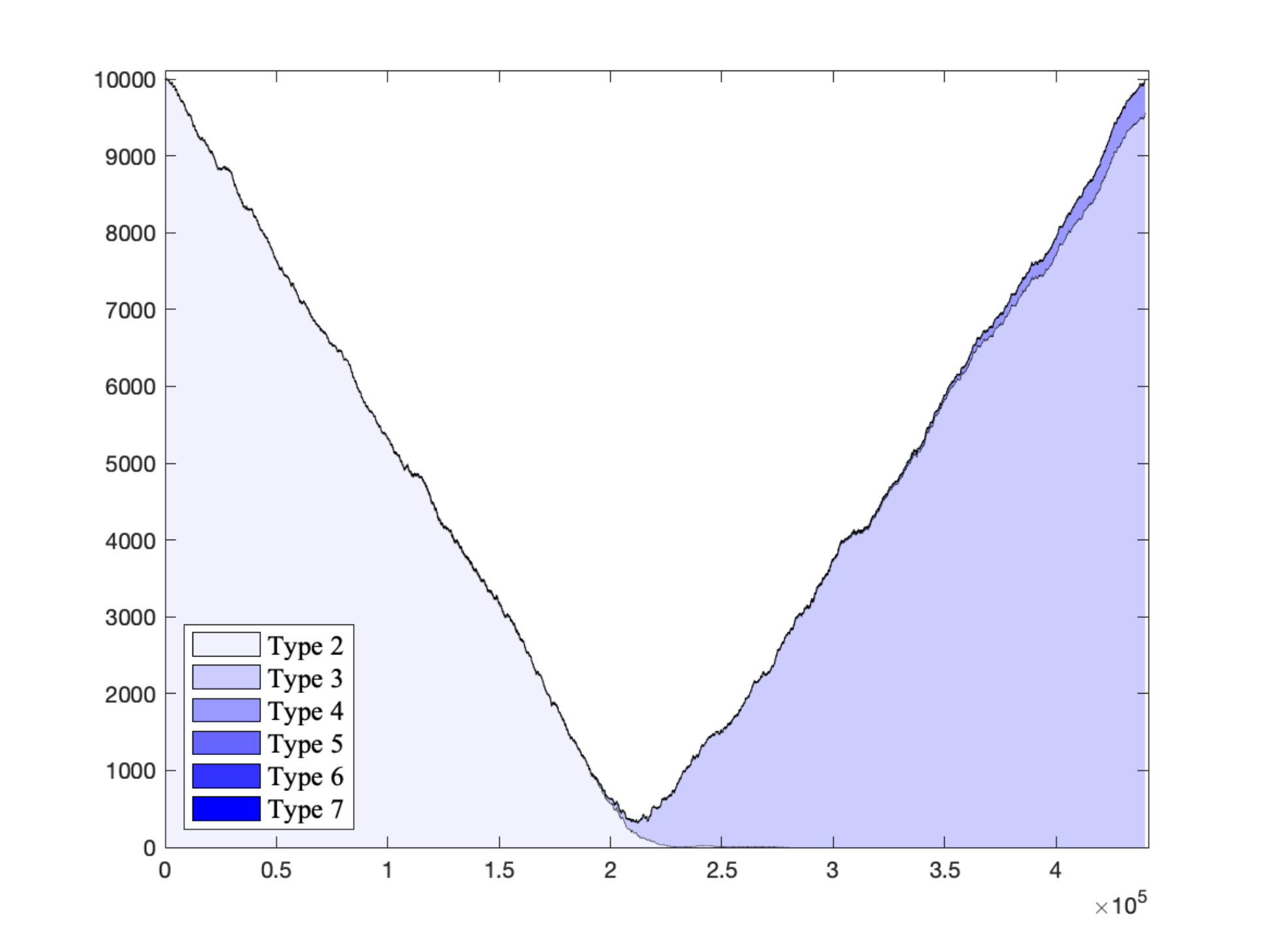}
\hspace{0.05\linewidth}
\includegraphics[width = 0.4\linewidth]{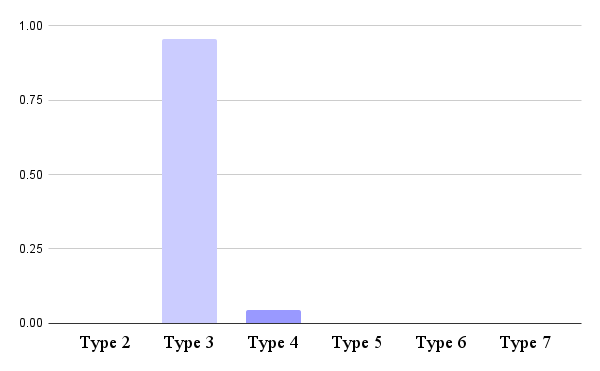}
\caption{$\beta = 0.9$}
\label{fig:geneAmpStacked08}
\end{subfigure}%
\caption{The composition of simulated tumors over time and at recurrence is shown for (a) $\beta = 0.1$, (b) $\beta = 0.5$, and (c) $\beta = 0.9$. The simulations were all conducted using the gene amplification model with $M = 7$, $n = 10^4$, $\beta = 0.1$, and birth rates ranging from $r_2 = 1.0$ to $r_7 = 2.0$. The figures on the right show the tumor composition at the time of recurrence. }
\label{fig:geneAmpStackedBeta} 
\end{figure}

        % Takeaway: In practice, increasing drug dosage may result in a smaller tumor size at turnaround, but as long as $\lambda_M > 0$ still, the tumor will eventually reach recurrence and may have an increased proportion of resistant cells at recurrence. 
The previous observations about the minimum tumor size and tumor composition may prompt the question: ``Can minimum tumor size be used to predict tumor composition at recurrence?" Unfortunately, it cannot, as shown by the example in Figure~\ref{fig:geneAmpStackedBeta}. We begin with the same initial scenario as in Figure~\ref{fig:TumorStackedKPrime3} and decrease the amplification rate $\nu_n = n^{-\beta}$ by increasing $\beta$. This also results in a smaller minimum tumor size and decreased heterogeneity. However, in contrast to the effect of increasing the death rates $d_k$, increasing $\beta$ results in an increased proportion of cells with low copy numbers. The increase in $\beta$ decreases the amplification rate so the time at which the tumor is able to achieve growth again is delayed, resulting in a smaller minimum tumor size. When the tumor develops drug resistant cells, the regrowth generated happens ``faster" than the amplification events so recurrence happens before the tumor is fully amplified.

\section{Discussion}\label{discussion}
In this work, we have defined two branching process models that represent drug resistance arising from gene mutation, a single event that allows subcritical cells to become supercritical, and gene amplification, which allows cells to gain resistance through multiple incremental amplification events. For each model, we derived the mean of each cell type and established necessary and sufficient conditions for non-extinction of the processes in the large population limit. The extinction result for the gene amplification process establishes the importance of the relationship between the amplification parameter $\beta$ and $k'$, the number of gene copies necessary for a cell to become supercritical. Furthermore, since the result is independent of the net growth rates beyond the assumption that $\lambda_M > 0$, non-extinction is still guaranteed in the large population limit in exceedingly unfavorable scenarios such as $\lambda_M$ being very small and $k'$ being very large, as long as $\beta < \frac{1}{k'-2}$. This suggests that even drugs that are very effective at killing sensitive cells may be unable to prevent recurrence when it is still possible for cells to eventually become drug insensitive through amplification.

We then proved law of large numbers results regarding the convergence of the stochastic recurrence times to their mean in both the mutation-driven resistance and the amplification-driven resistance models. In particular, the estimated recurrence time in the gene mutation model is $-\frac{\lambda_s}{\lambda_m }\alpha$ and the estimated recurrence time in the gene amplification model is $-\frac{\lambda_2}{\lambda_M}\beta(M-2)$. These results establish how different parameters of the model influence the stochastic recurrence time and could aid prediction of the recurrence time in practice, given sufficient knowledge of the relevant parameters. 

Finally, we examined the effects of various parameter regimes on tumor composition in various simulations as well as on the estimated recurrence time in each model. In simulations, we found that modeling increased drug efficacy by uniformly increasing the death rates across all types caused the tumor to reach a smaller minimum size and have less heterogeneity and an increased proportion of cells with high copy numbers at recurrence. Increasing $\beta$ also caused the tumor to achieve a smaller minimum size and have less heterogeneity at recurrence. However, this was associated with a different tumor composition at recurrence. In particular, a greater proportion of cells with low copy numbers. An area for further work is the exploration of estimates for the minimum tumor size and the relationship between the minimum size, recurrence time, tumor population, and tumor composition. Next, we compared the estimated recurrence times under different scenarios analytically. We examined the scenario where $\lambda_m  = \lambda_M$ to explore how taking incremental steps to achieve the same resistance affects the estimated recurrence time. In this scenario, we established that the ratio between the estimated recurrence time in the amplification model and the estimated recurrence time in the point mutation model is $\frac{\beta(M-2)}{\alpha}$. We also examined the scenario where continued amplification allowed cells to potentially attain more resistance than through mutation. We found that in this scenario increasing the maximum number of gene copies, $M$, had a limited effect on reducing the estimated recurrence time. Thus, our results are somewhat robust to discrepancies between the true and observed values for $M$. 

A limitation of this work is that many of the relevant parameters needed to estimate the recurrence time may be difficult to attain in practice. As such, the results may have more clinical relevance in the context of inference of the parameters of the evolutionary processes during tumor progression. For example, consider the scenario where it is known that a tumor has developed drug resistance through gene amplification and the initial net growth rate of the drug sensitive cells $\lambda_2$ and the gene amplification parameter $\beta$ are known. Suppose it is observed that a population of resistant cells has emerged with net growth rate $\lambda_k > 0$, but that the number of gene copies $k$ in the cells and the maximum possible number of gene copies $M$ are both unknown. Then it may be possible to estimate $k$ and $M$ by comparing an observed recurrence time with our estimated recurrence time. Intuitively, if $k$ is increased, the estimated recurrence time increases, and if $M$ increases relative to $k$, the estimated recurrence time decreases to an extent. 

Another context in which inference may be possible is in the estimation of $k'$, the number of amplified copies of a particular gene necessary for a tumor cell to become supercritical in the presence of drug. For NSCLC, the number of MET amplifications necessary to achieve resistance to tyrosine kinase inhibitors is still unknown \cite{kolesar_integration_2022}. Since Proposition~\ref{prop:extinction} guarantees survival of the tumor when $\beta < \frac{1}{k'-2}$ and guarantees extinction when $\beta > \frac{1}{k'-2}$ in the large population limit, knowledge of $\beta$ in conjunction with observations of recurrence or extinction of tumors could provide estimates of $k'$. Another possibility is through analysis of tumor composition at recurrence. Changing $d_k$, as in Figure \ref{fig:geneAmpStackedKPrime}, has the effect of shifting $k'$. It may be possible to exploit the relationship between $k'$ and tumor composition to perform inference of $k'$ in practice.

As we saw in our discussion of Figure~\ref{fig:geneAmpStackedKPrime} and Figure~\ref{fig:geneAmpStackedBeta}, inference of tumor composition from just the minimum tumor size is not possible in general as smaller minimum size leads to opposite trends in the composition of the tumor in the two examples. However, it is possible to conclude that the growth of the tumor from its minimum size until the recurrence time consists of only treatment-insensitive cells. This could have important clinical implications. An area for future work is further exploration of the effect of different parameters on the tumor composition at recurrence. 

Other areas for future work include generalizations of the models to better capture other nuances in these biological systems. For example, in the geometric amplification model from Kimmel and Axelrod \cite{kimmel_mathematical_1990}, the number of gene copies doubles with each amplification event. We could account for this by allowing the values of the step sizes $\lambda_k - \lambda_{k-1}$ to vary rather than stay constant. Our extinction result would still hold as it does not rely on the values of $r_k$ and $d_k$, but other results may need to be altered. Another extension would be to explore the behavior of the model when allowing for  multi-drug combinations, as in Iwasa et al.\ \cite{tomasetti_elementary_2010}, Komarova and Wodarz \cite{komarova_drug_2005}, and Komarova \cite{komarova_stochastic_2006}, as well as periods off drug to represent drug holidays, which are commonly used to manage toxicity. This would involve new parameters for the growth rates off drug and could also include the potential for cells to revert their typing and lose their resistance. We also assume that the models are mutually exclusive in this paper. However, work by Bean et al.\ \cite{bean_met_2007} shows that MET amplification can occur alongside T790M mutation in lung tumors with resistance to gefitinib or erlotinib. Thus, it may be more biologically relevant to extend and combine the models to accommodate both point mutations and amplification events in a single model. 

\section{Acknowledgements}
J.F. was partially supported by NSF DMS 2052465, NSF CMMI 2228034, NIH R01CA241134, and Research Council of Norway Grant 309273. A.L. was partially supported by NSF DMS 2052465. We thank Einar Bjarki Gunnarsson (University of Iceland), Anna Kraut (University of Minnesota), and Kevin Leder (University of Minnesota) for their thoughtful comments and suggestions.

\section{Data availability statement.} Data sharing not applicable to this article as no datasets were generated or analysed during the current study.
\section{Appendix} \label{appendix}
\subsection{Restriction of mutation and amplification to cell division}\label{note:restriction}
To restrict mutations to arise only during cell division, let $\tilde{\mu}_n$ be the probability that a cell undergoing division produces a mutated cell. Then the rate at which mutated cells arise in the population is $r_s \tilde{\mu}_n X_s(t)$. The bounds in Proposition~\ref{prop:EstimateMut} then become 
\begin{align*}
a_n &= \frac{1}{\lambda_m } \left( -\lambda_s \alpha + \frac{1}{t_n} \log \left(\frac{ -\lambda_s}{r_s} \right) \right),\\
A_n &= \frac{1}{\lambda_m } \left( -\lambda_s \alpha + \frac{1}{t_n} \log \left( \frac{{\lambda_m  - \lambda_s}}{r_s} \right) \right).
\end{align*}
However, it is still the case that $\tilde{u}_n \rightarrow -\frac{\lambda_s }{\lambda_m }\alpha$ as $n \rightarrow \infty$ so our estimated recurrence time in the large population limit does not change and the convergence in Theorem~\ref{LLNMut} will still hold. 

Similarly, to restrict amplification events to only occur during cell division, let $\tilde{\nu}_n$ be the probability that a cell undergoing division produces an amplified cell. Then the rate at which amplified cells with $k$ copies of the gene arise is $r_{k-1} \tilde{\nu}_n X_{k-1}(t)$. For the proof of Proposition~\ref{prop:extinction}, $s_{n,k}$ will instead be of the form
\[
    s_{n,k} = 
    \begin{cases}
        \displaystyle{\frac{d_k}{r_k}} + O(\tilde{\nu}_n) & \text{ if } k' \le k \le M,\\
        \displaystyle{1 - \frac{1-d_{k'}/r_{k'}}{\prod_{i=k}^{k'-1} d_i-r_i} \left(\prod_{i=k}^{k'-1} r_i\right) \tilde{\nu}_n^\ell + O (\tilde{\nu}_n^{\ell+1})} & \text{ if } 2 < k < k'.
    \end{cases}
\]
Thus, we will have an extinction probability of 
\[
    q = \lim_{n \to \infty} \left(s_{n,2}\right)^n =  \begin{cases}
        0 & \text{ if } 0< \beta< \frac{1}{k'-2},\\
        \displaystyle{\exp\left(-\frac{1-d_{k'}/r_{k'}}{\prod_{i=2}^{k'-1} d_i-r_i} \left(\prod_{i=k}^{k'-1} r_i\right)\right)} & \text{ if } \beta = \frac{1}{k'-2}, \\
        1 & \text{ if } \beta > \frac{1}{k'-2}.
    \end{cases}
\]
That is, the only difference occurs when $\beta = \frac{1}{k'-2}$. 

For Proposition~\ref{prop:EstimateAmp}, the bounds become 
\begin{align*}
b_n &= \frac{1}{\lambda_M} \left[ -\lambda_2 \beta (M-2) - \frac{1}{t_n} \log \left[ \frac{ \prod_{i=2}^M -r_i }{D^{M-2}\tilde{P}_{M,M}} \left( 1 - \frac{\lambda_M}{\lambda_2} \right) \right] \right],\\
B_n &= \frac{1}{\lambda_M} \left[ -\lambda_2 \beta (M-2) - \frac{1}{t_n} \log \left[ \frac{ \prod_{i=2}^M -r_i}{D^{M-2}\tilde{P}_{M,M}} \right] \right]. 
\end{align*}

\noindent Hence we still have that $\tilde{v}_n \rightarrow -\frac{\lambda_2}{\lambda_M} \beta (M-2)$ as $n \rightarrow \infty$. As in the mutation model, our estimated recurrence time is unchanged and our convergence in Theorem~\ref{LLNAmp} will still hold as well.

\subsection{Proof of Lemma~\ref{lem:meanvarmut}}\label{proof:meanvarmut}
We know that the first moment of the process generated by a single sensitive cell is $m_1^{s}(t) = e^{\lambda_s t}$. Notice then that 
\[
\phi_s(zt_n) = n m^s_1(zt_n) = ne^{\lambda_s z t_n} = ne^{-z \log n } = n^{1-z}.
\]
By Equation 5 in Chapter III Part 4 of \cite{athreya_one_1972}, we know that the second moment of the process generated by a single sensitive cell is given by 
\[
m_2^s(t) = u''(1) \frac{e^{2\lambda_s t} - e^{\lambda_s t}}{\lambda_s} + e^{\lambda_s t},
\]
where $u(x) = d_s + r_s x^2 -x(r_s+d_s)$. Then
\begin{align*}
    \var[X_s(t)] &= n\left(m_2^s(t) - \left(m_1^s(t)\right)^2\right) = 2 n  r_s \frac{e^{2\lambda_s t} - e^{\lambda_s t}}{\lambda_s} + n e^{\lambda_s t} - ne^{2 \lambda_s t}.
\end{align*}
Thus,
\begin{align*}
    \psi_s(zt_n) &= \mathrm{Var}[X_s(zt_n)]\\
    &= 2 n  r_s \frac{e^{2\lambda_s zt_n} - e^{\lambda_s zt_n}}{\lambda_s} + n e^{\lambda_s zt_n} - n e^{2 \lambda_s zt_n}\\
    &=2   r_s \frac{n^{1-2z} - n^{1-z}}{\lambda_s} + n^{1-z} - n^{1-2z}\\
    &= -\left( n^{1-z} - n^{1-2z}\right) \left(\frac{2r_s}{\lambda_s} - 1\right) \\
    &= n^{1-z} (1-n^{-z}) \frac{2r_s - r_s +d_s}{-\lambda_s}\\
    &= n^{1-z} \frac{r_s+d_s}{-\lambda_s} (1-n^{-z}) ,
\end{align*}
as desired.

By definition, $\EE[X_m(t)]$ satisfies
\[
\frac{d}{dt} \EE[X_m(t)] = \lambda_m \EE[X_m(t)] +  n^{-\alpha} \EE[X_s(t)],
\]
with initial condition $\EE[X_m(t)] = 0$. Notice then that $\frac{ n^{1-\alpha}}{\lambda_m - \lambda_s} \left({e^{\lambda_m t}}-{e^{\lambda_s t}}\right) $ satisfies the differential equation and initial condition:
\begin{align*}
    \frac{d}{dt}\left[ \frac{ n^{1-\alpha}}{\lambda_m - \lambda_s} \left({e^{\lambda_m t}}-{e^{\lambda_s t}}\right) \right]
    &= \frac{ n^{1-\alpha}}{\lambda_m - \lambda_s}\left({\lambda_m e^{\lambda_m t}}-{\lambda_s e^{\lambda_s t}}\right)\\
    &= \frac{n^{1-\alpha}}{\lambda_m - \lambda_s}\left({\lambda_m e^{\lambda_m t}}-{\lambda_m e^{\lambda_s t}}\right) +\frac{ n^{1-\alpha}}{\lambda_m - \lambda_s} \left({(\lambda_m-\lambda_s) e^{\lambda_s t}}\right)\\
    &= \lambda_m \left( \frac{n^{1-\alpha}}{\lambda_m - \lambda_s} \left({e^{\lambda_m t}}-{e^{\lambda_s t}}\right)  \right) + n^{-\alpha} \EE[X_s(t)].
\end{align*}
So indeed, $\EE[X_m(t)] = \frac{ n^{1-\alpha}}{\lambda_m - \lambda_s} \left({e^{\lambda_m t}}-{e^{\lambda_s t}}\right) .$ Thus,
\[
\phi_m(zt_n) = \EE[X_m(zt_n)] = \frac{n^{1-\alpha}}{\lambda_m - \lambda_s} \left({e^{\lambda_m zt_n}}-{e^{\lambda_s zt_n}}\right) = \frac{ n^{1-\alpha}}{\lambda_m - \lambda_s} \left({n^{-\frac{\lambda_m}{\lambda_s} z }}-n^{-z}\right).
\]

To calculate $\var[X_m(t)]$, let $B(t)$ represent the population size at $t$ of a process generated by a single resistant cell, and let $N(t)$ be the number of resistant cells generated by the sensitive cells before time $t$. Then we have that
\begin{align*}
    \EE[X_m(t)^2\mid (X_s(s))_{s \le t}] &= \sum_{k=0}^\infty \PP(N(t) = k\mid (X_s(s))_{s \le t})\cdot \EE[X_m(t)^2 \mid N(t) = k,  (X_s(s))_{s \le t}]\\
    &= \sum_{k=0}^\infty \PP(N(t) = k\mid (X_s(s))_{s \le t})\cdot  \EE\left[\left(\sum_{i = 1}^{N(t)} B(t-\tau_i)\right)^{\!\!2} \,\middle| \,N(t) = k, (X_s(s))_{s \le t}\right],
\end{align*}
where $\tau_i$ represents the time that the $i$-th resistant cell is generated. Notice that
\[
P(N(t) = k \mid (X_s(s))_{s \le t}) =\frac{\rho^k e^{-\rho}}{k!}, 
\]
where $\rho = \int_0^t X_s(s)\mu \, ds$. Let us consider the conditional expectation on its own. Conditioned on $N(t) = k$, the $\tau_i$'s are distributed as order statistics of $\{T_i\mid 1 \le i \le k\}$, where the $T_i$'s are i.i.d. random variables with PDF
\[
f(s) = \frac{X_s(s)\mu_n}{\int_0^t X_s(s)\mu_n \,ds} = \frac{1}{\rho } X_s(s)\mu_n,
\]
supported on $s \in (0,t]$.
We have that 
\begin{align*}
    \EE\left[\left(\sum_{i = 1}^{N(t)} B(t-\tau_i)\right)^{\!\!2} \,\middle| \,N(t) = k, (X_s(s))_{s\le t}\right]  &\overset{(d)}{=} \EE\left[\left(\sum_{i = 1}^{k} B(t-T_{(i)})\right)^{\!\!2}\, \middle| (X_s(s))_{s \le t} \right], 
\end{align*}
and we can reorder terms as necessary and expand the square to write
%f(s) = \frac{\EE[X_s(s)]\mu }{\int_0^t \EE[X_s(s)] \mu ds } = \frac{ne^{\lambda_s s}}{\int_0^t ne^{\lambda_s s} ds} = \frac{\lambda_s e^{\lambda_s s}}{e^{\lambda_s t}-1} \textbf{ or } 
%&= \EE\left[\sum_{i = 1}^k B(t-T_i)^2 + 2\sum_{1 \le i < j \le k} B(t-T_i) B(t-T_j)\,\middle| (X_s(s))_{s \le t}\right]\\
\begin{align*}
    \EE\left[\left(\sum_{i = 1}^{k} B(t-T_i)\right)^{\!\!2} \,\middle| (X_s(s))_{s \le t}\right] 
    & =\sum_{i = 1}^k \EE\left[B(t-T_i)^2 \,\middle| (X_s(s))_{s \le t}\right] \\
    &\phantom{==}+ 2\sum_{1 \le i < j \le k} \EE\left[B(t-T_i) B(t-T_j) \,\middle|(X_s(s))_{s \le t}\right]. 
\end{align*}
In particular, we have that 
\begin{multline}\label{eq:2mom}
    \EE[X_m(t)^2\mid (X_s(s))_{s \le t}] \overset{(d)}{=}\sum_{k=0}^\infty \PP(N(t) = k\mid (X_s(s))_{s \le t}) \left(\sum_{i = 1}^k \EE\left[B(t-T_i)^2 \,\middle| (X_s(s))_{s \le t}\right]\right)  \\
    +\sum_{k=0}^\infty \PP(N(t) = k\mid (X_s(s))_{s \le t}) \left( 2\sum_{1 \le i < j \le k} \EE\left[B(t-T_i) B(t-T_j) \,\middle|(X_s(s))_{s \le t}\right]\right).
\end{multline}
Notice that
\(
    \EE\left[B(t-T_i)^2 \,\middle|(X_s(s))_{s \le t}\right] = \int_0^t f(s) \EE\left[B(t-s)^2 \right] ds.
\)
Following the process above for the sensitive process, we can get that for $s \le t$,
\begin{align*}
    \EE[ B(t-s)^2] &= m_2^{B}(t-s)=\frac{2r_m}{\lambda_m} e^{2\lambda_m(t-s)} + \left( \frac{-d_m-r_m}{\lambda_m}\right) e^{\lambda_m(t-s)}.
\end{align*}
We have then that
\begin{align*}
    \EE\left[B(t-T_i)^2 \,\middle|(X_s(s))_{s \le t}\right] &=  \int_0^t\frac{1}{\rho } X_s(s)\mu_n \left(\frac{2r_m}{\lambda_m} e^{2\lambda_m(t-s)} + \left( \frac{-d_m-r_m}{\lambda_m}\right) e^{\lambda_m(t-s)} \right)ds.
\end{align*}
Then 
\begin{equation}\label{eq:square}
    \sum_{i = 1}^k \EE\left[B(t-T_i)^2 \,\middle| (X_s(s))_{s \le t}\right] = k \cdot \frac{1}{\rho } \int_0^t X_s(s)\mu_n \left(\frac{2r_m}{\lambda_m} e^{2\lambda_m(t-s)} + \left( \frac{-d_m-r_m}{\lambda_m}\right) e^{\lambda_m(t-s)} \right)ds.
\end{equation}
Similarly, by independence
\begin{align*}
     \EE\left[B(t-T_i) B(t-T_j) \,\middle| (X_s(s))_{s \le t}\right] &= \EE\left[B(t-T_i) \,\middle| (X_s(s))_{s \le t}\right] \EE\left[B(t-T_j) \,\middle| (X_s(s))_{s \le t}\right] \\
     &= \EE\left[B(t-T_i)\,\middle| (X_s(s))_{s \le t}\right]^2\\
     &= \left(\int_{-\infty}^\infty f(s) \EE\left[B(t-s) \right] ds\right)^2\\
     &=\frac{1}{\rho^2 } \left(\int_0^t  X_s(s)\mu_n
     e^{\lambda_m(t-s)} \,ds\right)^2.
\end{align*}
So 
\begin{equation}\label{eq:cross}
    2\sum_{1 \le i < j \le k} \EE\left[B(t-T_i) B(t-T_j) \,\middle|(X_s(s))_{s \le t}\right] = k(k-1) \cdot \frac{1}{\rho^2 } \left(\int_0^t  X_s(s)\mu_n
     e^{\lambda_m(t-s)} \,ds\right)^2.
\end{equation}
Then inserting (\ref{eq:square}) and (\ref{eq:cross}) into (\ref{eq:2mom}) gets us that 
\begin{align*}
    \EE[X_m(t)^2 &\mid (X_s(s))_{s \le t}]\\ &\overset{(d)}{=} \sum_{k=0}^\infty \frac{\rho^k e^{-\rho}}{k!} \left( k \cdot \frac{1}{\rho } \int_0^t X_s(s)\mu_n \left(\frac{2r_m}{\lambda_m} e^{2\lambda_m(t-s)} + \left( \frac{-d_m-r_m}{\lambda_m}\right) e^{\lambda_m(t-s)} \right)ds\right)  \\
    &\phantom{==}+\sum_{k=0}^\infty \frac{\rho^k e^{-\rho}}{k!} \left(k(k-1) \cdot \frac{1}{\rho^2 } \left(\int_0^t  X_s(s)\mu_n
     e^{\lambda_m(t-s)} \,ds\right)^2\right)\\
    % &= \left(\int_0^t X_s(s)\mu_n \left(\frac{2r_m}{\lambda_m} e^{2\lambda_m(t-s)} + \left( \frac{-d_m-r_m}{\lambda_m}\right) e^{\lambda_m(t-s)} \right)ds\right) e^{-\rho}\sum_{k=1}^\infty \frac{\rho^{k-1} }{(k-1)!}   \\
    % &\phantom{==}+ \left(\left(\int_0^t  X_s(s)\mu_n
    %  e^{\lambda_m(t-s)} \,ds\right)^2\right) e^{-\rho}\sum_{k=2}^\infty \frac{\rho^{k-2} }{(k-2)!}\\
     &= \int_0^t X_s(s)\mu_n \left(\frac{2r_m}{\lambda_m} e^{2\lambda_m(t-s)} + \left( \frac{-d_m-r_m}{\lambda_m}\right) e^{\lambda_m(t-s)} \right)ds + \left(\int_0^t  X_s(s)\mu_n
     e^{\lambda_m(t-s)} \,ds\right)^2.
\end{align*}
Repeating the ideas from the above calculation, we can calculate 
\begin{align*}
    \EE[X_m(t) \mid (X_s(s))_{s\le t}]^2 &= \left(\sum_{k=0}^\infty \PP(N(t) = k\mid (X_s(s))_{s \le t})\cdot  \EE\left[\sum_{i = 1}^{N(t)} B(t-\tau_i)\,\middle| \,N(t) = k, (X_s(s))_{s \le t}\right]\right)^2\\
    &\overset{(d)}{=}\left(\sum_{k=0}^\infty \frac{\rho^k e^{-\rho}}{k!} \cdot  \EE\left[\sum_{i = 1}^{k} B(t-T_i)\,\middle| \, (X_s(s))_{s \le t}\right]\right)^2\\
     &= \left(\int_0^t  X_s(s)\mu_n
     e^{\lambda_m(t-s)} \,ds\right)^2.
\end{align*}
Then we have that 
\begin{align*}
    \var\left(X_m(t) \middle|(X_s(s))_{s\le t}\right)
    &=\EE[X_m(t)^2\mid (X_s(s))_{s \le t}] - \EE[X_m(t) \mid (X_s(s))_{s\le t}]^2\\
    &= \int_0^t X_s(s)\mu_n \left(\frac{2r_m}{\lambda_m} e^{2\lambda_m(t-s)} + \left( \frac{-d_m-r_m}{\lambda_m}\right) e^{\lambda_m(t-s)} \right)ds.
\end{align*}
Taking the expectation and then interchanging the expectation of the integral, we have that 
\begin{align*}
    \EE\left[\var\left(X_m(t) \middle|(X_s(s))_{s\le t}\right)\right] &= \EE\left[\int_0^t X_s(s)\mu_n \left(\frac{2r_m}{\lambda_m} e^{2\lambda_m(t-s)} + \left( \frac{-d_m-r_m}{\lambda_m}\right) e^{\lambda_m(t-s)} \right)ds\right]\\
    % &= \int_0^t \EE[X_s(s)]\mu_n \left(\frac{2r_m}{\lambda_m} e^{2\lambda_m(t-s)} + \left( \frac{-d_m-r_m}{\lambda_m}\right) e^{\lambda_m(t-s)} \right)ds\\
    &= \mu_n \int_0^t ne^{\lambda_s(s)}\left(\frac{2r_m}{\lambda_m} e^{2\lambda_m(t-s)} + \left( \frac{-d_m-r_m}{\lambda_m}\right) e^{\lambda_m(t-s)} \right)ds\\
    % &= n\mu_n \int_0^t \left(\frac{2r_m}{\lambda_m} e^{2\lambda_m t + (\lambda_s -2\lambda_m)s} + \left( \frac{-d_m-r_m}{\lambda_m}\right) e^{\lambda_m t + (\lambda_s-\lambda_m)s} \right)ds\\
    &= n\mu_n \left[  \frac{2r_m e^{2\lambda_m t + (\lambda_s -2\lambda_m)s}}{\lambda_m(\lambda_s -2\lambda_m)}  + \frac{(-d_m-r_m)e^{\lambda_m t + (\lambda_s-\lambda_m)s}}{\lambda_m (\lambda_s-\lambda_m)} \right]_{s=0}^t\\
    &= \frac{n\mu_n}{\lambda_m} \left(  \frac{2r_m e^{\lambda_s t}}{\lambda_s -2\lambda_m}  +  \frac{(-d_m-r_m)e^{\lambda_s t}}{ \lambda_s-\lambda_m} -    \frac{2r_m e^{2\lambda_m t}}{\lambda_s -2\lambda_m}  -  \frac{(-d_m-r_m)e^{\lambda_m t}}{\lambda_s-\lambda_m}  \right).
\end{align*}
Let $g(t) = \EE\left[\var\left(X_m(t) \middle|(X_s(s))_{s\le t}\right)\right]$. Notice that since $\lambda_s < 0 < \lambda_m$, we have that 
\begin{align*}
    g(zt_n) &\sim n \mu_n \frac{2r_m e^{2\lambda_m zt_n}}{\lambda_m(2\lambda_m -\lambda_s)}
    % &=n^{1-\alpha}\frac{2r_m e^{-2\lambda_m z \log n /\lambda_s}}{\lambda_m(2\lambda_m -\lambda_s)}\\
    = \frac{2r_m}{\lambda_m(2\lambda_m -\lambda_s)} n^{1-\alpha -2 \lambda_mz/\lambda_s}.
\end{align*}
The Law of Total Variance gives that 
\[
\var[X_m(t)] = g(t) + h(t),
\]
where $h(t) = \var[\EE[X_m(t) \mid (X_s(s))_{s\le t}]]$. Define $h_1(t) = \EE[\EE[X_m(t) \mid (X_s(s))_{s\le t}]^2] $ and define $h_2(t) = \EE[\EE[X_m(t) \mid (X_s(s))_{s\le t}]]^2$. Then $h(t) = h_1(t) - h_2(t)$. Notice that 
\begin{align*}
    h_2(t) &= \EE[\EE[X_m(t) \mid (X_s(s))_{s\le t}]]^2= \EE[X_m(t)]^2= \left( \frac{ n^{1-\alpha}}{\lambda_m - \lambda_s} \left({e^{\lambda_m t}}-{e^{\lambda_s t}}\right) \right)^2.
\end{align*}
We also have that
% \begin{align*}
%     \var[\EE[X_m(t) \mid (X_s(s))_{s\le t}]] &= \EE[\EE[X_m(t) \mid (X_s(s))_{s\le t}]^2] - \EE[\EE[X_m(t) \mid (X_s(s))_{s\le t}]]^2\\
%     &= \EE\left[\left(\int_0^t  X_s(s)\mu_n
%     e^{\lambda_m(t-s)} \,ds\right)^2\right] - \EE[X_m(t)]^2\\
%     &= \EE\left[\left(\int_0^t  X_s(s)\mu_n
%     e^{\lambda_m(t-s)} \,ds\right)^2\right] - \left( \frac{ n^{1-\alpha}}{\lambda_m - \lambda_s} \left({e^{\lambda_m t}}-{e^{\lambda_s t}}\right) \right)^2.
% \end{align*}
% Let $h(t) = \EE\left[\left(\int_0^t  X_s(s)\mu_n
%     e^{\lambda_m(t-s)} \,ds\right)^2\right]$. Then 
\begin{align*}
    h_1(t)&=\EE\left[\left(\int_0^t  X_s(s)\mu_n
    e^{\lambda_m(t-s)} \,ds\right)^2\right]\\
    % &= \mu_n^2 \,\EE\left[\int_0^t\int_0^t X_s(s)X_s(y) e^{\lambda_m(t-s)} e^{\lambda_m(t-y)}\, dsdy\right]\\
    &=\mu_n^2 \int_0^t\int_0^t \EE[X_s(s)X_s(y)] e^{\lambda_m(t-s)} e^{\lambda_m(t-y)}\, dsdy\\
    &= \mu_n^2 \int_0^t\int_0^y \EE[X_s(s)\EE[X_s(y)|X_s(s)]] e^{\lambda_m(t-s)} e^{\lambda_m(t-y)}\, dsdy\\
    &\phantom{==}+ \mu_n^2 \int_0^t\int_y^t \EE[\EE[X_s(s)|X_s(y)]X_s(y)] e^{\lambda_m(t-s)} e^{\lambda_m(t-y)}\, dsdy\\
    % &= \mu_n^2 \int_0^t\int_0^y \EE[X_s(s)^2]e^{\lambda_s(y-s)} e^{\lambda_m(t-s)} e^{\lambda_m(t-y)}\, dsdy,
    &= 2\mu_n^2 \int_0^t\int_y^t \EE[X_s(y)^2]e^{\lambda_s(s-y)} e^{\lambda_m(t-s)} e^{\lambda_m(t-y)}\, dsdy.
\end{align*}
by symmetry. Note that 
\(
    \EE[X_s(t)^2] = \var[X_s(t)] + \EE[X_s(t)]^2 
    % &=2 n  r_s \frac{e^{2\lambda_s t} - e^{\lambda_s t}}{\lambda_s} + n e^{\lambda_s t} - ne^{2 \lambda_s t} + n^2 e^{2 \lambda_s t} \\
    % &= \left(n^2-n + \frac{2nr_s}{\lambda_s}\right)e^{2\lambda_s t} + \left(n - \frac{2nr_s}{\lambda_s}\right)e^{\lambda_s t}\\
    =\left(n^2+\frac{r_s+d_s}{\lambda_s}n\right)e^{2\lambda_s t} + \frac{-r_s-d_s}{\lambda_s}ne^{\lambda_s t}.
\)
Then 
\begin{align*}
    \int_0^t\int_y^t &\EE[X_s(y)^2]e^{\lambda_s(s-y)} e^{\lambda_m(t-s)} e^{\lambda_m(t-y)}\, dsdy 
    \\
    &= \int_0^t\EE[X_s(y)^2]e^{\lambda_m(t-y)}\int_y^t e^{\lambda_s(s-y)} e^{\lambda_m(t-s)} \, dsdy\\
    % &=\int_0^t\EE[X_s(y)^2]\frac{e^{\lambda_m(t-y)}}{\lambda_s-\lambda_m}\left[ e^{\lambda_s(s-y)} e^{\lambda_m(t-s)} \right]_{s=y}^t \, dy\\
    &=\int_0^t\left(\left(n^2+\frac{r_s+d_s}{\lambda_s}n\right)e^{2\lambda_s y} + \frac{-r_s-d_s}{\lambda_s}ne^{\lambda_s y}\right)\frac{e^{\lambda_m(t-y)}}{\lambda_s-\lambda_m}\left( e^{\lambda_s(t-y)} -e^{\lambda_m(t-y)} \right) \, dy\\
    % &= \int_0^t\left(\left(n^2+\frac{r_s+d_s}{\lambda_s}n\right)e^{2\lambda_s y} + \frac{-r_s-d_s}{\lambda_s}ne^{\lambda_s y}\right)\left( \frac{e^{(\lambda_s+\lambda_m)(t-y)}}{\lambda_s-\lambda_m}-\frac{e^{2\lambda_m(t-y)}}{\lambda_s-\lambda_m}\right) \, dy\\
    &= \int_0^t  \left(\frac{n^2}{\lambda_s-\lambda_m}+\frac{(r_s+d_s)n}{\lambda_s(\lambda_s-\lambda_m)}\right)e^{(\lambda_s+\lambda_m)t + (\lambda_s-\lambda_m)y} \\
    &\phantom{==}+ 
    \left(\frac{n^2}{\lambda_m-\lambda_s}+\frac{(r_s+d_s)n}{\lambda_s(\lambda_m-\lambda_s)}\right)e^{2\lambda_mt + (2\lambda_s-2\lambda_m)y} \\
    &\phantom{==} + \frac{(r_s+d_s)n}{\lambda_s(\lambda_m-\lambda_s)} e^{(\lambda_s+\lambda_m)t - \lambda_m y} + \frac{(r_s+d_s)n}{\lambda_s(\lambda_s-\lambda_m)} e^{2\lambda_mt + (\lambda_s-2\lambda_m)y} \, dy\\
    % &=\frac{1}{(\lambda_m-\lambda_s)^2}\left[\left(n^2+\frac{(r_s+d_s)n}{\lambda_s}\right)e^{(\lambda_s+\lambda_m)t + (\lambda_s-\lambda_m)y}     + \left(\frac{n^2}{-2}+\frac{(r_s+d_s)n}{-2\lambda_s}\right)e^{2\lambda_mt + (2\lambda_s-2\lambda_m)y}  
    % \right]_{y=0}^t\\
    % &\phantom{==}  +\left[ \frac{(r_s+d_s)n}{\lambda_s(-\lambda_m)(\lambda_m-\lambda_s)} e^{(\lambda_s+\lambda_m)t - \lambda_m y} + \frac{(r_s+d_s)n}{\lambda_s(\lambda_s-\lambda_m)(\lambda_s-2\lambda_m)} e^{2\lambda_mt + (\lambda_s-2\lambda_m)y}  \right]_{y=0}^t\\
    &=\frac{1}{(\lambda_m-\lambda_s)^2}\left(\left(n^2+\frac{(r_s+d_s)n}{\lambda_s}\right)e^{2 \lambda_s t}     + \left(\frac{n^2}{-2}+\frac{(r_s+d_s)n}{-2\lambda_s}\right)e^{2\lambda_s t}\right.\\ 
    &\phantom{==}\left.   -\left(n^2+\frac{(r_s+d_s)n}{\lambda_s}\right)e^{(\lambda_s+\lambda_m)t}   
     + \left(\frac{n^2}{2}+\frac{(r_s+d_s)n}{2\lambda_s}\right)e^{2\lambda_mt }  
    \right) + \frac{(r_s+d_s)n}{\lambda_s(-\lambda_m)(\lambda_m-\lambda_s)} e^{\lambda_s t} \\
    &\phantom{==}  +   \frac{(r_s+d_s)n}{\lambda_s(\lambda_s-\lambda_m)(\lambda_s-2\lambda_m)} e^{\lambda_s t}  + \frac{(r_s+d_s)n}{\lambda_s\lambda_m(\lambda_m-\lambda_s)} e^{(\lambda_s+\lambda_m)t} - \frac{(r_s+d_s)n}{\lambda_s(\lambda_s-\lambda_m)(\lambda_s-2\lambda_m)} e^{2\lambda_mt }.
\end{align*}
% Similarly,
% \[
%     \int_0^t\int_0^y \EE[X_s(s)^2]e^{\lambda_s(y-s)} e^{\lambda_m(t-s)} e^{\lambda_m(t-y)}\, dsdy = \int_0^t\int_s^t \EE[X_s(s)^2]e^{\lambda_s(y-s)} e^{\lambda_m(t-s)} e^{\lambda_m(t-y)}\, dyds,
% \]
% which is equivalent to the previous integral by change of variables, so 
Then 
\begin{align*}
    h_1(t) 
    % &= 
    % 2\mu_n^2\left[\frac{1}{(\lambda_m-\lambda_s)^2}\left(\left(n^2+\frac{(r_s+d_s)n}{\lambda_s}\right)e^{2 \lambda_s t}     + \left(\frac{n^2}{-2}+\frac{(r_s+d_s)n}{-2\lambda_s}\right)e^{2\lambda_s t} \right.\right.\\ 
    % &\phantom{==}\left.  -\left(n^2+\frac{(r_s+d_s)n}{\lambda_s}\right)e^{(\lambda_s+\lambda_m)t}   
    % + \left(\frac{n^2}{2}+\frac{(r_s+d_s)n}{2\lambda_s}\right)e^{2\lambda_mt }  
    % \right) + \frac{(r_s+d_s)n}{\lambda_s(-\lambda_m)(\lambda_m-\lambda_s)} e^{\lambda_s t}  \\
    % &\phantom{==}\left.
    % +   \frac{(r_s+d_s)n}{\lambda_s(\lambda_s-\lambda_m)(\lambda_s-2\lambda_m)} e^{\lambda_s t}  + \frac{(r_s+d_s)n}{\lambda_s\lambda_m(\lambda_m-\lambda_s)} e^{(\lambda_s+\lambda_m)t} - \frac{(r_s+d_s)n}{\lambda_s(\lambda_s-\lambda_m)(\lambda_s-2\lambda_m)} e^{2\lambda_mt }\right]\\
    &= \left({n^2}+\frac{(r_s+d_s)n}{\lambda_s}\right)\left(\frac{\mu_n (e^{\lambda_m t}- e^{\lambda_s t})}{\lambda_m-\lambda_s}\right)^2 + 2\mu_n^2\left[\frac{(r_s+d_s)n}{\lambda_s(-\lambda_m)(\lambda_m-\lambda_s)} e^{\lambda_s t}  \right.\\
    &\phantom{==}\left.
    +   \frac{(r_s+d_s)n}{\lambda_s(\lambda_s-\lambda_m)(\lambda_s-2\lambda_m)} e^{\lambda_s t}  + \frac{(r_s+d_s)n}{\lambda_s\lambda_m(\lambda_m-\lambda_s)} e^{(\lambda_s+\lambda_m)t} - \frac{(r_s+d_s)n}{\lambda_s(\lambda_s-\lambda_m)(\lambda_s-2\lambda_m)} e^{2\lambda_mt }\right]\\
    &= h_2(t) +\left(\frac{(r_s+d_s)n}{\lambda_s}\right)\left(\frac{\mu_n (e^{\lambda_m t}- e^{\lambda_s t})}{\lambda_m-\lambda_s}\right)^2 + 2\mu_n^2\left[\frac{(r_s+d_s)n}{\lambda_s(-\lambda_m)(\lambda_m-\lambda_s)} e^{\lambda_s t}  \right.\\
    &\phantom{==}\left.
    +   \frac{(r_s+d_s)n}{\lambda_s(\lambda_s-\lambda_m)(\lambda_s-2\lambda_m)} e^{\lambda_s t}  + \frac{(r_s+d_s)n}{\lambda_s\lambda_m(\lambda_m-\lambda_s)} e^{(\lambda_s+\lambda_m)t} - \frac{(r_s+d_s)n}{\lambda_s(\lambda_s-\lambda_m)(\lambda_s-2\lambda_m)} e^{2\lambda_mt }\right].
\end{align*}
Then 
% \begin{align*}
%     h(t) &= h_1(t) - h_2(t) \\
%     &= \left(\frac{(r_s+d_s)n}{\lambda_s}\right)\left(\frac{\mu_n (e^{\lambda_m t}- e^{\lambda_s t})}{\lambda_m-\lambda_s}\right)^2 + 2\mu_n^2\left[\frac{(r_s+d_s)n}{\lambda_s(-\lambda_m)(\lambda_m-\lambda_s)} e^{\lambda_s t}  \right.\\
%     &\phantom{==}\left.
%     + \frac{(r_s+d_s)n}{\lambda_s\lambda_m(\lambda_m-\lambda_s)} e^{(\lambda_s+\lambda_m)t} - \frac{(r_s+d_s)n}{\lambda_s(\lambda_s-\lambda_m)(\lambda_s-2\lambda_m)} e^{2\lambda_mt }\right].
% \end{align*}
% So 
\begin{align*}
    h(zt_n) &= \left(\frac{(r_s+d_s)n}{\lambda_s}\right)\left(\frac{n^{-\alpha} (e^{\lambda_m zt_n}- e^{\lambda_s zt_n})}{\lambda_m-\lambda_s}\right)^2 + 2n^{-2\alpha}\left[\frac{(r_s+d_s)n}{\lambda_s(-\lambda_m)(\lambda_m-\lambda_s)} e^{\lambda_s zt_n}  \right.\\
    &\phantom{==}\left.
    + \frac{(r_s+d_s)n}{\lambda_s\lambda_m(\lambda_m-\lambda_s)} e^{(\lambda_s+\lambda_m)zt_n} - \frac{(r_s+d_s)n}{\lambda_s(\lambda_s-\lambda_m)(\lambda_s-2\lambda_m)} e^{2\lambda_m zt_n }\right]\\
    &= n^{1-2\alpha}\left(\frac{(r_s+d_s)}{\lambda_s}\right)\left(\frac{ (n^{-\lambda_m z/\lambda_2}- n^{-z})}{\lambda_m-\lambda_s}\right)^2 + 2n^{1-2\alpha}\left[\frac{(r_s+d_s)}{\lambda_s(-\lambda_m)(\lambda_m-\lambda_s)} n^{-z}  \right.\\
    &\phantom{==}\left.
    + \frac{(r_s+d_s)}{\lambda_s\lambda_m(\lambda_m-\lambda_s)} n^{-(1 +\lambda_m/\lambda_2)z} - \frac{(r_s+d_s)}{\lambda_s(\lambda_s-\lambda_m)(\lambda_s-2\lambda_m)} n^{-2\lambda_m z/\lambda_s}\right]\\
    &= O(n^{1-2\alpha -2\lambda_mz/\lambda_s}),
\end{align*}
because $-\lambda_mz/\lambda_s > 0$. Then since $g(zt_n) \sim \frac{2r_m}{\lambda_m(2\lambda_m-\lambda_s)}n^{1-\alpha-2\lambda_mz/\lambda_s}$, we have that 
\begin{align*}
    \psi_m(zt_n) &= \var[X_m(zt_n)]\\
    &= g(zt_n) + h(zt_n)\\
    &\sim \frac{2r_m}{\lambda_m(2\lambda_m-\lambda_s)}n^{1-\alpha-2\lambda_mz/\lambda_s},
\end{align*}
as desired.

\subsection{Proof of Lemma~\ref{lem:Means}} \label{proof:Means}

\begin{proof}

We will prove this by induction.  Let's start by proving the base case: $k=3$.  We need to show that 

\begin{equation*}
\mathbb{E}[X_3(t)] = n^{1-\beta} \left( \frac{e^{\lambda_2 t}}{\lambda_2 - \lambda_3} - \frac{e^{\lambda_3 t}}{\lambda_2 - \lambda_3} \right). \label{E3}
\end{equation*}

\noindent Because $X_3(t)$ is a birth-death process with mutation, its mean is governed by the following ODE: 

\begin{equation*}
\frac{d}{dt} \mathbb{E}[X_3(t)] = \lambda_3 \mathbb{E}[X_3(t)] + n^{-\beta} \mathbb{E}[X_2(t)].
\end{equation*}

\noindent Sure enough, plugging $\mathbb{E}[X_2(t)]$ and $\mathbb{E}[X_3(t)]$ into the above ODE yields 

\begin{align*}
\frac{d}{dt} \left[ n^{1-\beta} \left( \frac{e^{\lambda_2 t}}{\lambda_2 - \lambda_3} - \frac{e^{\lambda_3 t}}{\lambda_2 - \lambda_3} \right) \right] &= \lambda_3 n^{1-\beta} \left( \frac{e^{\lambda_2 t}}{\lambda_2 - \lambda_3} - \frac{e^{\lambda_3 t}}{\lambda_2 - \lambda_3} \right) + n^{-\beta} n e^{\lambda_2 t}\\
n^{1-\beta} \left( \frac{\lambda_2 e^{\lambda_2 t}}{\lambda_2 - \lambda_3} - \frac{\lambda_3 e^{\lambda_3 t}}{\lambda_2 - \lambda_3} \right) &= n^{1-\beta} \left( \frac{\lambda_3 e^{\lambda_2 t}}{\lambda_2 - \lambda_3} - \frac{\lambda_3 e^{\lambda_3 t}}{\lambda_2 - \lambda_3} + \frac{(\lambda_2 - \lambda_3)e^{\lambda_2 t}}{\lambda_2 - \lambda_3} \right),
\end{align*} 

\noindent which is clearly true.  So we do indeed have $\mathbb{E}[X_3(t)] = n^{1-\beta} \left( \frac{e^{\lambda_2 t}}{\lambda_2 - \lambda_3} - \frac{e^{\lambda_3 t}}{\lambda_2 - \lambda_3} \right)$, as desired.  Now assume

\begin{equation*}
\mathbb{E}[X_\ell(t)] =  n^{1-(\ell-2)\beta} (-1)^l S_\ell(t).  \label{Ek}
\end{equation*}

\noindent We want to show that 

\begin{equation*}
\mathbb{E}[X_{\ell+1}(t)] =  n^{1-(\ell-1)\beta} (-1)^{\ell+1} S_{\ell+1}(t). \label{Ek+1}
\end{equation*}

\noindent Because $X_{\ell+1}(t)$ is a birth-death process with mutation, its mean is governed by the following ODE: 

\begin{equation*}
\frac{d}{dt} \mathbb{E}[X_{\ell+1}(t)] = \lambda_{\ell+1} \mathbb{E}[X_{\ell+1}(t)] + n^{-\beta} \mathbb{E}[X_\ell(t)].
\end{equation*}

\noindent Plugging $\mathbb{E}[X_\ell(t)]$ and $\mathbb{E}[X_{\ell+1}(t)]$ into the above ODE yields 

\begin{align*}
\frac{d}{dt} \left[  n^{1-(\ell-1)\beta} (-1)^{\ell+1} S_{\ell+1}(t) \right] &= \lambda_{\ell+1}  n^{1-(\ell-1)\beta} (-1)^{\ell+1} S_{\ell+1}(t) \\
&\hspace{4mm}+ n^{-\beta}  n^{1-(\ell-2)\beta} (-1)^l S_\ell(t) \\
 n^{1-(\ell-1)\beta} (-1)^{\ell+1} \sum_{i=2}^{\ell+1} \frac{\lambda_i e^{\lambda_i t}}{P_{i, \ell+1}} &=  n^{1-(\ell-1)\beta} (-1)^{\ell+1} \left[ \lambda_{\ell+1} S_{\ell+1}(t) - S_\ell(t) \right].
\end{align*}

\noindent Dividing both sides by $ n^{1-(\ell-1)\beta} (-1)^{\ell+1}$, we get 

\begin{align*}
\sum_{i=2}^{\ell+1} \frac{\lambda_i e^{\lambda_i t}}{P_{i, \ell+1}} &= \lambda_{\ell+1} S_{\ell+1}(t) - S_\ell(t) \\
&= \sum_{i=2}^l \frac{\lambda_{\ell+1} e^{\lambda_i t}}{P_{i,l+1}} + \frac{\lambda_{\ell+1} e^{\lambda_{\ell+1} t}}{P_{\ell+1,l+1}} - \sum_{i=2}^l \frac{(\lambda_{\ell+1} - \lambda_i) e^{\lambda_i t}}{P_{i,l+1}},
\end{align*}

\noindent which is clearly true.  So we have shown the desired result.  \end{proof}

\subsection{Proof of Proposition~\ref{prop:extinction}}\label{proof:extinction}
\begin{proof}
    We first will solve for $\hat s_{n,k}$, the extinction probability of a type $k$ cell under a slightly different model. Consider a model where amplification events involve replacing a type $k$ cell with a type $k+1$ cell. That is, when a new amplified cell arises, a type $k$ cell is removed. Then the following relation will hold
\begin{equation}\label{eq:ParticleExtinction}
    \hat s_{n,k} =
    \begin{cases}
        \displaystyle{{\frac{r_k}{r_k + d_k + \nu_n}(\hat s_{n,k})^2 + \frac{d_k}{r_k + d_k + \nu_n} + \frac{\nu_n}{r_k + d_k + \nu_n}\hat s_{n,k+1}}}& \text{ if } 2 \le k < M,\\
        \displaystyle{{\frac{r_M}{r_M + d_M}(\hat s_{n,M})^2 + \frac{d_M}{r_M + d_M}}}& \text{ if } k = M.
    \end{cases}
\end{equation}
The intuition for why this relation holds is that the terms represent whether the next event is a birth, a death, or an amplification, respectively. If the next event is a birth event, then there are now two independent type $k$ cells each with extinction probability $\hat s_{n,k}$. If the next event is a death, then the particle becomes extinct. If the next event is a gene amplification, then the type $k$ cell becomes a type $k+1$ cell, whose extinction probability is $\hat s_{n,k+1}$. In the case that $k = M$, no further amplification events are possible so only the first two terms exist. By Theorem 2.1 of \cite{hautphenne_extinction_2013}, for $2 \le k < M$, $\hat s_{n,k}$ is the minimal non-negative solution to (\ref{eq:ParticleExtinction}).

Applying the quadratic formula to \ref{eq:ParticleExtinction}, the minimal non-negative solution is $\hat s_{n,M} = \frac{d_M}{r_M}$. 
% \begin{align*}
%     \hat s_{n,M} &= \frac{r_M + d_M}{2r_M} \pm \frac{1}{2}\sqrt{\frac{(r_M+d_M)^2}{{r_M}^2} - 4 \frac{d_M}{r_M}}\\
%     &=\frac{r_M + d_M}{2r_M} \pm \frac{1}{2r_M}\sqrt{{(r_M+d_M)^2} - 4r_Md_M}\\
%     &= \frac{r_M + d_M}{2r_M} \pm \frac{1}{2r_M}\sqrt{{(r_M-d_M)^2} }\\
%     &= \frac{r_M + d_M}{2r_M} \pm \frac{1}{2r_M}\sqrt{{(r_M-d_M)^2} }\\
%     &= \frac{r_M + d_M}{2r_M} \pm \frac{r_M-d_M}{2r_M},
% \end{align*}
% which yields either $1$ or $\frac{d_M}{r_M}$. Since the type $M$ process is assumed to be supercritical, $r_M > d_M$ so 
Similarly, for $2 \le k < M$,
\begin{align*}
    \hat s_{n,k} &= \frac{r_k+d_k + \nu_n}{2r_k} \pm \frac{1}{2} \sqrt{\frac{(r_k+d_k + \nu_n)^2}{{r_k}^2} - 4 \frac{d_k + \nu_n \hat s_{n, k+1}}{r_k}}\\
    &= \frac{r_k+d_k + \nu_n}{2r_k} \pm \frac{1}{2r_k} \sqrt{(r_k + d_k)^2 + 2(r_k+d_k)\nu_n + {\nu_n}^2 - 4 r_k d_k - 4 r_k\nu_n \hat s_{n,k+1}}\\
    &= \frac{r_k+d_k + \nu_n}{2r_k} \pm \frac{|r_k-d_k|}{2r_k} \sqrt{1 + \frac{2(r_k+d_k)}{(r_k-d_k)^2}\nu_n + \frac{1}{{(r_k-d_k)^2}}{\nu_n}^2 - \frac{4 r_k}{{(r_k-d_k)^2}}\nu_n \hat s_{n,k+1}}\\
    &= \frac{(r_k+d_k)+\nu_n}{2r_k} \pm \frac{|r_k-d_k|}{2r_k} \sqrt{1+ x_k}
\end{align*}
where 
\[
    x_k = \frac{2(r_k+d_k)}{(r_k-d_k)^2}\nu_n + \frac{1}{(r_k-d_k)^2} \nu_n^2 - \frac{4r_k}{(r_k-d_k)^2}  \hat s_{n,k+1}\nu_n.
\]
Since $\nu_n = n^{-\beta} \to 0$ in the large population limit, the Taylor expansion
\(
\sqrt{1+x} = \sum_{i=0}^\infty \binom{1/2}{i} x^i,
\)
gives
\[
\hat s_{n,k} =  \frac{(r_k+d_k)+\nu_n}{2r_k} \pm \frac{|r_k-d_k|}{2r_k} \left( \sum_{i=0}^\infty \binom{1/2}{i} {x_k}^i\right).
\]

Now consider $k$ such that $k' \le k < M$. In particular, $r_k > d_k$ because the cells with $k'$ or more copies are supercritical by definition. Then, 
\begin{align*}
    \hat s_{n,k} &= \frac{(r_k+d_k)+\nu_n}{2r_k} \pm \frac{r_k-d_k}{2r_k} \left( \sum_{i=0}^\infty \binom{1/2}{i} {x_k}^i\right)\\
    &= \frac{(r_k+d_k)+\nu_n}{2r_k} \pm \frac{r_k-d_k}{2r_k} \left(1 + \sum_{i=1}^\infty \binom{1/2}{i} {x_k}^i\right)\\
    &= 1 + O(\nu_n) \text{ or } \frac{d_k}{r_k} + O(\nu_n),
\end{align*}
because $x_k$ is a multiple of $\nu_n$. Since $r_k > d_k$, the minimal non-negative root is thus of the form $\hat s_{n,k} =  \frac{d_k}{r_k} + O(\nu_n)$. Intuitively, this aligns with the knowledge that without mutation, the extinction probability would be $\frac{d_k}{r_k}$.

Now consider $k$ such that $2 \le k < k'$. We will show with induction that 
\[
    \hat s_{n, k} =  1- \frac{1-d_{k'}/r_{k'}}{\prod_{i=k}^{k'-1} d_i-r_i} {\nu_n}^\ell + O\left(\nu_n^{\ell+1}\right),
\]
where $\ell = k'-k$. Note that in these cases $r_k < d_k$.

For our base case, consider $\hat s_{n, k'-1}$. Then
% We know from above that $\hat s_{n, k'} = \frac{d_{k'}}{r_{k'}} + O(\nu_n)$. Then
\begin{align*}
    \hat s_{n,k'-1} &= \frac{(r_{k'-1}+d_{k'-1})+\nu_n}{2r_{k'-1}} \pm \frac{d_{k'-1}-r_{k'-1}}{2r_{k'-1}} \left( 1+\sum_{i=1}^\infty \binom{1/2}{i} {x_{k'-1}}^i\right) \\
    &=1 + \frac{\nu_n}{2r_{k'-1}} - \frac{d_{k'-1}-r_{k'-1}}{2r_{k'-1}} \left(\sum_{i=1}^\infty \binom{1/2}{i} {x_{k'-1}}^i\right), \\
    &\phantom{=}\text{ or } \frac{d_{k'-1}}{r_{k'-1}} + \frac{\nu_n}{2r_{k'-1}} + \frac{d_{k'-1}-r_{k'-1}}{2r_{k'-1}} \left(\sum_{i=1}^\infty \binom{1/2}{i} {x_{k'-1}}^i\right).
\end{align*}
Since $r_{k'-1} < d_{k'-1}$, we know that the former must be the correct root so
\[
    \hat s_{n,k'-1} =1 + \frac{\nu_n}{2r_{k'-1}} - \frac{d_{k'-1}-r_{k'-1}}{2r_{k'-1}} \left(\sum_{i=1}^\infty \binom{1/2}{i} {x_{k'-1}}^i\right),
\]
where
\begin{align*}
    x_{k'-1} &= \frac{2(r_{k'-1}+d_{k'-1})}{(r_{k'-1}-d_{k'-1})^2}\nu_n + \frac{1}{(r_{k'-1}-d_{k'-1})^2} \nu_n^2 - \frac{4r_{k'-1}}{(r_{k'-1}-d_{k'-1})^2} \left(\frac{d_{k'}}{r_{k'}} + O(\nu_n)\right)\nu_n
\end{align*}
Intuitively, this makes sense because the lineage of a subcritical cell without mutation is guaranteed to go extinct. Notice that the coefficient of $\nu_n$ in $\hat s_{n, k'-1}$ is 
\begin{align*}
    &\frac{1}{2r_{k'-1}} - \frac{d_{k'-1}-r_{k'-1}}{2r_{k'-1}} \binom{1/2}{1}\left(\frac{2(r_{k'-1}+d_{k'-1})-4r_{k'-1}\cdot {d_{k'}}/{r_{k'}}}{2(r_{k'-1}-d_{k'-1})^2} \right) \\
    &= \frac{1}{2r_{k'-1}} \left( 1+ \frac{(r_{k'-1}+d_{k'-1})-2r_{k'-1} \cdot {d_{k'}}/{r_{k'}}}{(r_{k'-1}-d_{k'-1})} \right)\\
    &=\frac{1}{r_{k'-1}} \left( \frac{2r_{k'-1}-2r_{k'-1} \cdot {d_{k'}}/{r_{k'}}}{(r_{k'-1}-d_{k'-1})} \right)\\
    &= \frac{1-d_{k'}/r_{k'}}{r_{k'-1}-d_{k'-1}}\\
    &= -\frac{1-d_{k'}/r_{k'}}{d_{k'-1}-r_{k'-1}}.
\end{align*}
Thus, we indeed have that
\[
\hat s_{n, k'-1} = 1 - \frac{1-d_{k'}/r_{k'}}{d_{k'-1}-r_{k'-1}} \nu_n + O\left(\nu_n^2\right),
\]
as desired. Now assume that for some $2 < k+1 < k'$ that 
\[
    \hat s_{n, k+1} =  1- \frac{1-d_{k'}/r_{k'}}{\prod_{i=k+1}^{k'-1} d_i-r_i} {\nu_n}^{\ell-1} + O\left(\nu_n^{\ell}\right),
\]
where $\ell = k'-k$. Then we know

\begin{align*}
    \hat s_{n,k} &= \frac{(r_{k}+d_{k})+\nu_n}{2r_{k}} \pm \frac{d_{k}-r_{k}}{2r_{k}} \left( 1+\sum_{i=1}^\infty \binom{1/2}{i} {x_{k}}^i\right) \\
    &=1 + \frac{\nu_n}{2r_{k}} - \frac{d_{k}-r_{k}}{2r_{k}} \left(\sum_{i=1}^\infty \binom{1/2}{i} {x_{k}}^i\right) \text{ or } \frac{d_{k}}{r_{k}} + \frac{\nu_n}{2r_{k}} + \frac{d_{k}-r_{k}}{2r_{k}} \left(\sum_{i=1}^\infty \binom{1/2}{i} {x_{k}}^i\right),
\end{align*}
and that the former root is the correct one because $r_k < d_k$. So 
\[
    \hat s_{n,k} =1 + \frac{\nu_n}{2r_{k}} - \frac{d_{k}-r_{k}}{2r_{k}} \left(\sum_{i=1}^\infty \binom{1/2}{i} {x_{k}}^i\right),
\]
where 
\begin{align*}
    x_k &= \frac{2(r_k+d_k)}{(r_k-d_k)^2}\nu_n + \frac{1}{(r_k-d_k)^2} \nu_n^2 - \frac{4r_k}{(r_k-d_k)^2}  \left(1- \frac{1-d_{k'}/r_{k'}}{\prod_{i=k+1}^{k'-1} d_i-r_i} {\nu_n}^{\ell-1} + O\left(\nu_n^{\ell}\right)\right)\nu_n\\
    &=\frac{2(r_k+d_k)-4r_k}{(r_k-d_k)^2}\nu_n + \frac{1}{(r_k-d_k)^2} \nu_n^2 + \frac{4r_k}{(r_k-d_k)^2}  \frac{1-d_{k'}/r_{k'}}{\prod_{i=k+1}^{k'-1} d_i-r_i} {\nu_n}^{\ell} + O\left(\nu_n^{\ell+1}\right)\\
    &= \frac{-2}{(r_k-d_k)}\nu_n + \frac{1}{(r_k-d_k)^2} \nu_n^2 + \frac{4r_k}{(r_k-d_k)^2} \cdot \frac{1-d_{k'}/r_{k'}}{\prod_{i=k+1}^{k'-1} d_i-r_i} {\nu_n}^{\ell} + O\left(\nu_n^{\ell+1}\right).
\end{align*}
Notice that if we collect the terms in the sum that arise from the first two terms of $x_k$, we can write
\begin{align*}
    \sum_{i=1}^\infty \binom{1/2}{i} {x_{k}}^i &= \sum_{i=1}^\infty \binom{1/2}{i} \left(\frac{-2}{(r_k-d_k)}\nu_n + \frac{1}{(r_k-d_k)^2} \nu_n^2 \right)^i\\
    &\phantom{=}+\binom{1/2}{1} \frac{4r_k}{(r_k-d_k)^2} \cdot \frac{1-d_{k'}/r_{k'}}{\prod_{i=k+1}^{k'-1} d_i-r_i} {\nu_n}^{\ell} + O\left(\nu_n^{\ell+1}\right)\\
    &= \sqrt{1+\frac{-2}{(r_k-d_k)}\nu_n + \frac{1}{(r_k-d_k)^2} \nu_n^2 } -1 + \frac{2r_k}{(r_k-d_k)^2} \cdot \frac{1-d_{k'}/r_{k'}}{\prod_{i=k+1}^{k'-1} d_i-r_i} {\nu_n}^{\ell} + O\left(\nu_n^{\ell+1}\right)\\
    &= \left|1-\frac{\nu_n}{r_k-d_k}\right| - 1 + \frac{2r_k}{(r_k-d_k)^2} \cdot \frac{1-d_{k'}/r_{k'}}{\prod_{i=k+1}^{k'-1} d_i-r_i} {\nu_n}^{\ell} + O\left(\nu_n^{\ell+1}\right)\\
    &= \frac{\nu_n}{d_k-r_k}+ \frac{2r_k}{(r_k-d_k)^2} \cdot \frac{1-d_{k'}/r_{k'}}{\prod_{i=k+1}^{k'-1} d_i-r_i} {\nu_n}^{\ell} + O\left(\nu_n^{\ell+1}\right),
\end{align*}
because $r_k < d_k$. Then, we have that 
\begin{align*}
    \hat s_{n,k} &=1 + \frac{\nu_n}{2r_{k}} - \frac{d_{k}-r_{k}}{2r_{k}} \left(\frac{\nu_n}{d_k-r_k}+ \frac{2r_k}{(r_k-d_k)^2} \cdot \frac{1-d_{k'}/r_{k'}}{\prod_{i=k+1}^{k'-1} d_i-r_i} {\nu_n}^{\ell} + O\left(\nu_n^{\ell+1}\right)\right)\\
    &= 1 + \frac{\nu_n}{2r_k} - \frac{\nu_n}{2r_k} - \frac{1}{d_k-r_k} \cdot \frac{1-d_{k'}/r_{k'}}{\prod_{i=k+1}^{k'-1} d_i-r_i} {\nu_n}^{\ell} + O\left(\nu_n^{\ell+1}\right)\\
    &= 1- \frac{1-d_{k'}/r_{k'}}{\prod_{i=k}^{k'-1} d_i-r_i} {\nu_n}^{\ell} + O\left(\nu_n^{\ell+1}\right),
\end{align*}
as desired. This completes our inductive step so indeed we have that for $2 \le k < k'$,
\[
    \hat s_{n, k} =  1- \frac{1-d_{k'}/r_{k'}}{\prod_{i=k}^{k'-1} d_i-r_i} {\nu_n}^\ell + O\left(\nu_n^{\ell+1}\right),
\]
where $\ell = k'-k$.

To complete the proof, we will derive an answer for $s_{n,k}$ in terms of $\hat s_{n,k}$. Notice that instead of (\ref{eq:ParticleExtinction}), $s_{n,k}$ satisfies
\begin{equation}\label{eq:ParticleExtinction2}
   s_{n,k} =
\begin{cases}
    \displaystyle{\frac{r_k}{r_k+d_k+\nu_n} (s_{n,k})^2 + \frac{d_k}{r_k+d_k+\nu_n} + \frac{\nu_n}{r_k+d_k+\nu_n}s_{n,k} s_{n,k+1}} & \text{ if }2 \le k < M,\\
    \displaystyle{\frac{r_M}{r_M+d_M}( s_{n,M})^2 + \frac{d_M}{r_M+d_M} }& \text{ if } k = M.
\end{cases} 
\end{equation}
So $s_{n,M} = \frac{d_k}{r_k}$ as above.
Let $k' \le k < M$. Then $r_k > d_k$. Notice that 
\[
    0 = r_k (s_{n,k})^2 - (r_k+d_k) s_{n,k} + d_k + O(\nu_n).
\]
So the minimal non-negative root is of the form $s_{n,k} = \frac{d_k}{r_k} + O(\nu_n)$.

Now let $2 \le k < k'$. Notice that by (\ref{eq:ParticleExtinction}) and (\ref{eq:ParticleExtinction2}),
\begin{align*}
    (r_k + d_k + \nu_n)(\hat s_{n,k} - s_{n,k}) &= r_k((\hat s_{n,k})^2 - (s_{n,k})^2) + \nu_n(\hat s_{n,k+1} - s_{n,k} s_{n,k+1})\\
    &=r_k(\hat s_{n,k} + s_{n,k})(\hat s_{n,k} - s_{n,k}) + \nu_n \left(\hat s_{n,k+1} - \hat s_{n,k}\hat s_{n,k+1} + \hat s_{n,k}\hat s_{n,k+1}\right.\\
    &\phantom{==} \left.-s_{n,k}\hat s_{n,k+1} + s_{n,k}\hat s_{n,k+1}  - s_{n,k} s_{n,k+1}\right)\\
    &= r_k(\hat s_{n,k} + s_{n,k})(\hat s_{n,k} - s_{n,k}) \\
    &\phantom{==}+ \nu_n \left(\hat s_{n,k+1}(1 - \hat s_{n,k}) +\hat s_{n,k+1}( \hat s_{n,k} -s_{n,k}) + s_{n,k}(\hat s_{n,k+1}  - s_{n,k+1})\right).
\end{align*}
Then 
\begin{align*}
    (r_k - r_k(\hat s_{n,k} + s_{n,k}) + d_k + \nu_n(1-s_{n,k+1}))(\hat s_{n,k} - s_{n,k}) &= \nu_n \left(\hat s_{n,k+1}(1 - \hat s_{n,k}) + s_{n,k}(\hat s_{n,k+1}  - s_{n,k+1})\right).
\end{align*}
Then 
\begin{align*}
    |r_k - r_k(\hat s_{n,k} + s_{n,k}) + d_k + \nu_n(1-s_{n,k+1})|\cdot |\hat s_{n,k} - s_{n,k}| &= \nu_n \left|\hat s_{n,k+1}(1 - \hat s_{n,k}) + s_{n,k}(\hat s_{n,k+1}  - s_{n,k+1})\right|\\
    &\le \nu_n|1-\hat s_{n,k}| + \nu_n|\hat s_{n,k+1}  - s_{n,k+1}|,
\end{align*}
because $|\hat s_{n,k+1}|, |s_{n,k}| \le 1$. Since $\hat s_{n,k} + s_{n.k} \le 2$ and $1-s_{n,k+1} \ge 0$, 
\[
    r_k - r_k(\hat s_{n,k} + s_{n,k}) + d_k + \nu_n(1-s_{n,k+1}) \ge d_k - r_k.
\]
Since $k < k'$, we know that $d_k > r_k$. So 
\begin{equation}\label{eq:ExtInduction}
     |\hat s_{n,k} - s_{n,k}| \le \frac{\nu_n}{d_k - r_k}\left(|1-\hat s_{n,k}| + |\hat s_{n,k+1}  - s_{n,k+1}|\right).
\end{equation}
Our goal is to show that for $2 \le k < k'$ that $|\hat s_{n,k} - s_{n,k}| = O(\nu_n^{\ell+1})$, where $\ell = k'-k$. Proceed by induction on $\ell$, recall  that
\[
    \hat s_{n, k} =
    \begin{cases}
        \displaystyle 1- \frac{1-d_{k'}/r_{k'}}{\prod_{i=k}^{k'-1} d_i-r_i} {\nu_n}^\ell + O\left(\nu_n^{\ell+1}\right) & \text{ if $2 \le k < k'$,}\\
        \displaystyle\frac{d_k}{r_k} + O(\nu_n)& \text{ if $k' \le k \le M$.}
    \end{cases}
\]
For our base case, consider $\ell = 1$, that is $k = k'-1$. Then $\hat s_{n,k'-1} = 1+ O(\nu_n)$ so $|1-\hat s_{n,k'-1}| = O(\nu_n)$. We also know from above that $\hat s_{n,k'} = \frac{d_{k'}}{r_{k'}} + O(\nu_n)$ and $ s_{n,k'} = \frac{d_{k'}}{r_{k'}} + O(\nu_n)$, so $|\hat s_{n,k'} - s_{n,k'}|= O(\nu_n)$ as well. Then by (\ref{eq:ExtInduction}), 
\[
|\hat s_{n,k'-1} - s_{n,k'-1}| \le \frac{\nu_n}{d_k - r_k}\left(|1-\hat s_{n,k'-1}| + |\hat s_{n,k'}  - s_{n,k'}|\right) = \frac{\nu_n}{d_k - r_k}O(\nu_n) = O(\nu_n^2).
\]
Now consider $ \ell > 1$, i.e.\ $k < k'-1$. Then by induction, $|\hat s_{n,k+1} - s_{n,k+1}| = O\left(\nu_n^{(\ell-1) + 1}\right) = O(\nu_n^\ell)$.  We know $\hat s_{n,k} = 1 + O(\nu_n^\ell)$ so $|1- \hat s_{n,k}| = O(\nu_n^\ell)$ as well. Then by (\ref{eq:ExtInduction}), 
\[
|\hat s_{n,k} - s_{n,k}| \le \frac{\nu_n}{d_k - r_k}\left(|1-\hat s_{n,k}| + |\hat s_{n,k+1}  - s_{n,k+1}|\right) = \frac{\nu_n}{d_k - r_k}O(\nu_n^\ell) = O\left(\nu_n^{\ell+1}\right),
\]
as desired. So indeed  $|\hat s_{n,k} - s_{n,k}| = O(\nu_n^{\ell+1})$ for $2 \le k < k'$. Then since 
\[
\hat s_{n,k} = 1- \frac{1-d_{k'}/r_{k'}}{\prod_{i=k}^{k'-1} d_i-r_i} {\nu_n}^\ell + O\left(\nu_n^{\ell+1}\right),
\]
we know that 
\[
s_{n,k} = \hat s_{n,k} - (\hat s_{n,k} - s_{n,k}) = 1- \frac{1-d_{k'}/r_{k'}}{\prod_{i=k}^{k'-1} d_i-r_i} {\nu_n}^\ell + O\left(\nu_n^{\ell+1}\right),
\]
as well.

In particular, we have that 
\begin{align*}
    s_{n, 2} &= 1- \frac{1-d_{k'}/r_{k'}}{\prod_{i=2}^{k'-1} d_i-r_i} {\nu_n}^{k'-2} + O\left(\nu_n^{k'-1}\right)\\
    &= 1- \frac{1-d_{k'}/r_{k'}}{\prod_{i=2}^{k'-1} d_i-r_i} { n^{-\beta(k'-2)}}+ O\left(n^{-\beta(k'-2+1)}\right).
\end{align*}
We know that our extinction probability starting with $n$ particles of of type 2 is $q_n = (s_{n,2})^n$. Notice that since $r_{k'} > d_{k'}$ and $r_i < d_i$ for $2 \le i < k'$, we have $\frac{1-d_{k'}/r_{k'}}{\prod_{i=2}^{k'-1} d_i-r_i} > 0$. Then 
\begin{align*}
    q &= \lim_{n \to \infty} {q_n} \\
    &=\lim_{n \to \infty} \left(1- \frac{1-d_{k'}/r_{k'}}{\prod_{i=2}^{k'-1} d_i-r_i} { n^{-\beta(k'-2)}}+ O\left( n^{-\beta(k'-2+1)}\right)\right)^n\\
    &= \begin{cases}
        0 & \text{ if } 0< \beta< \frac{1}{k'-2},\\
        \displaystyle\exp\left(-\frac{1-d_{k'}/r_{k'}}{\prod_{i=2}^{k'-1} d_i-r_i} \right) & \text{ if } \beta = \frac{1}{k'-2}, \\
        1 & \text{ if } \beta > \frac{1}{k'-2}.
    \end{cases}
\end{align*}\qedhere
\end{proof}

\subsection{Proof of Proposition~\ref{prop:EstimateMut}}\label{proof:EstimateMut} 

\begin{proof}

Let $\bar{f}_n(z) = n^{z-1}(\phi_s(z t_n) + \phi_m(z t_n) - n)$.  Using the definitions of $\phi_s$ and $\phi_m$, we see that 

\begin{align*}
\bar{f}_n(z) &= n^{z-1} \left( n^{1-z} + \frac{ n^{1-\alpha}}{\lambda_m  - \lambda_s} \left( n^{-\lambda_m  z/\lambda_s} - n^{-z} \right) - n \right) \\
&= 1 + \frac{1}{n^{\alpha}(\lambda_m  - \lambda_s)} n^{z(1 - \lambda_m /\lambda_s)} - \frac{1}{n^{\alpha}(\lambda_m  - \lambda_s)} - n^z.
\end{align*}

\noindent Taking the derivative with respect to $z$ yields 

\begin{align*}
\bar{f}^{\prime}_n(z) &= \frac{1}{n^{\alpha}(\lambda_m  - \lambda_s)} \frac{\lambda_s - \lambda_m }{\lambda_s} n^{z(1 - \lambda_m /\lambda_s)} \log n - n^z \log n \\
&= n^z \log n \left( \frac{-1}{n^{\alpha} \lambda_s} n^{-\lambda_m  z/\lambda_s} - 1 \right).
\end{align*}

\noindent Next we set $\bar{f}'_n(z) = 0$ and solve for $z$ to find any local maxima or minima of $\bar{f}_n(z)$.  Since $n^z > 0$ and $\log n > 0$ for sufficiently large $n$, we have that

\begin{align*}
\bar{f}'_n(z) = 0 &\implies \frac{-1}{n^{\alpha} \lambda_s} n^{-\lambda_m  z/\lambda_s} = 1 \\
&\implies -\frac{\lambda_m }{\lambda_s} z \log n = \log \left( -{n^{\alpha} \lambda_s} \right).
\end{align*}

\noindent Solving the above equation for $z$ yields 

\begin{align*}
z &= \frac{-\lambda_s}{\lambda_m } \cdot \frac{ \log (-n^{\alpha} \lambda_s)}{ \log n} \\
&= \frac{-\lambda_s}{\lambda_m } \left( \alpha + \frac{\log(-\lambda_s)}{\log n} \right) \\
&= \frac{1}{\lambda_m } \left( -\lambda_s \alpha + \frac{1}{t_n} \log \left( {-\lambda_s} \right) \right).
\end{align*}

\noindent Hence $a_n = \frac{1}{\lambda_m } \left( -\lambda_s \alpha + \frac{1}{t_n} \log(-\lambda_s) \right)$ is the only maximum or minimum of $\bar{f}_n(z)$.  Moreover, we have that $a_n > 0$ for sufficiently large $n$.  

Next note that $\bar{f}'_n(a_n-1) = n^{a_n-1} \log n \left( \frac{-1}{n^{\alpha} \lambda_s} n^{-\lambda_m (a_n-1)/\lambda_s} - 1 \right)$.  Since $n^{a_n-1} > 0$ and $\log n > 0$ for sufficiently large $n$ and 

\begin{align*}
\frac{-1}{n^{\alpha} \lambda_s} n^{-\lambda_m (a_n-1)/\lambda_s} - 1 &= \frac{-1}{n^{\alpha} \lambda_s} n^{\frac{\lambda_m }{-\lambda_s} \cdot \frac{1}{\lambda_m } (-\lambda_s \alpha + \frac{1}{t_n} \log ({-\lambda_s}) - \lambda_m )} - 1 \\
&= \frac{-1}{n^{\alpha} \lambda_s} n^{\alpha + \frac{1}{\log n} \log ({-\lambda_s}) + \lambda_m /\lambda_s} - 1 \\
&= n^{\lambda_m /\lambda_s} - 1 \\
&< 0 \text{ for sufficiently large $n$},
\end{align*}

\noindent we have that $\bar{f}'_n(a_n-1) < 0$ for sufficiently large $n$.  Hence $\bar{f}_n(z)$ is monotonically decreasing for $z < a_n$.  Similarly, $\bar{f}'_n(a_n+1) = n^{a_n+1} \log n \left( \frac{-1}{n^{\alpha} \lambda_s} n^{-\lambda_m (a_n+1)/\lambda_s} - 1 \right)$.  Since $n^{a_n+1} > 0$ and $\log n > 0$ for sufficiently large $n$ and 

\begin{align*}
\frac{-1}{n^{\alpha} \lambda_s} n^{-\lambda_m (a_n+1)/\lambda_s} - 1 &= \frac{-1}{n^{\alpha} \lambda_s} n^{\frac{\lambda_m }{-\lambda_s} \cdot \frac{1}{\lambda_m } (-\lambda_s \alpha + \frac{1}{t_n} \log ({-\lambda_s}) + \lambda_m )} - 1 \\
&= \frac{-1}{n^{\alpha} \lambda_s} n^{\alpha + \frac{1}{\log n} \log ({-\lambda_s}) - \lambda_m /\lambda_s} - 1 \\
&= n^{-\lambda_m /\lambda_s} - 1 \\
&> 0 \text{ for sufficiently large $n$},
\end{align*}

\noindent we have that $\bar{f}'_n(a_n+1) > 0$ for sufficiently large $n$.  Hence $\bar{f}_n(z)$ is monotonically increasing for $z > a_n$.

Now let $A_n = \frac{1}{\lambda_m } \left(-\lambda_s \alpha + \frac{1}{t_n} \log (\lambda_m  - \lambda_s) \right)$.  Then 

\begin{align*}
\bar{f}_n(A_n) &= 1 + \frac{ n^{-\alpha}}{\lambda_m  - \lambda_s} \left( n^{A_n(1 - \lambda_m /\lambda_s)} - 1 \right) - n^{A_n} \\
&= 1 + \frac{ n^{-\alpha}}{\lambda_m  - \lambda_s} \left( n^{\frac{\lambda_m  - \lambda_s}{-\lambda_s\lambda_m } \left( -\lambda_s \alpha + \frac{1}{t_n}\log \left( {\lambda_m  - \lambda_s} \right) \right)} - 1 \right) - n^{\frac{1}{\lambda_m } \left( -\lambda_s \alpha + \frac{1}{t_n} \log \left( {\lambda_m  - \lambda_s}\right) \right)} \\
&= 1 + \frac{ n^{-\alpha}}{\lambda_m  - \lambda_s} \left( n^{\alpha(1 - \lambda_s/\lambda_m )} n^{\frac{\lambda_m  - \lambda_s}{\lambda_m  \log n} \log \left({\lambda_m  - \lambda_s} \right)} - 1 \right) - n^{-\lambda_s\alpha/\lambda_m } n^{\frac{-\lambda_s}{\lambda_m  \log n} \log \left( {\lambda_m  - \lambda_s}\right)} \\
&= 1 + \frac{ n^{-\alpha}}{\lambda_m  - \lambda_s} \left( n^{\alpha(1 - \lambda_s/\lambda_m )} \left( {\lambda_m  - \lambda_s} \right)^{1 - \lambda_s/\lambda_m } - 1 \right) - n^{-\lambda_s\alpha/\lambda_m } \left( {\lambda_m  - \lambda_s}\right)^{-\lambda_s/\lambda_m } \\
&= 1 + n^{-\lambda_s\alpha/\lambda_m } \left( {\lambda_m  - \lambda_s}\right)^{-\lambda_s/\lambda_m } - \frac{ n^{-\alpha}}{\lambda_m  - \lambda_s} - n^{-\lambda_s\alpha/\lambda_m } \left( {\lambda_m  - \lambda_s} \right)^{-\lambda_s/\lambda_m } \\
&= 1 - \frac{ n^{-\alpha}}{\lambda_m  - \lambda_s} \\
&> 0 \text{ for sufficiently large $n$}.
\end{align*}

Note that since $\bar{f}_n(0) = 0$ and $\bar{f}_n(z)$ is decreasing for all $z \in (0, a_n)$, we have that $\bar{f}_n(a_n) < 0$.  But $\bar{f}_n(z)$ is increasing for $z > a_n$ and we have that $\bar{f}_n(A_n) > 0$.  Therefore, there exists a unique $\tilde{u}_n \in (a_n, A_n)$ such that $\bar{f}_n(\tilde{u}_n) = 0$ by monotonicity.  Using the definition of $\bar{f}_n(z)$ and the fact that $n^{z-1} > 0$, this implies that there is a unique $\tilde{u}_n > 0$ such that $a_n < \tilde{u}_n < A_n$ and $\phi_s(\tilde{u}_n t_n) + \phi_m(\tilde{u}_n t_n) = n$.  Furthermore, $\lim_{n \rightarrow \infty} a_n = \lim_{n \rightarrow \infty} A_n = \frac{-\lambda_s \alpha}{\lambda_m }$.  Hence $\tilde{u}_n \rightarrow \frac{-\lambda_s \alpha}{\lambda_m }$ as $n \rightarrow \infty$.    
\end{proof}

\subsection{Proof of Theorem~\ref{LLNMut}}\label{proof:LLNMut} 

\begin{proof} 

In order to show the desired result, we must show 

\begin{equation*}
\lim_{n \rightarrow \infty} \mathbb{P} (\tau_n > u_n + \varepsilon) + \lim_{n \rightarrow \infty} \mathbb{P} (\tau_n < u_n - \varepsilon) = 0.
\end{equation*}

\noindent Let's start by proving that $\mathbb{P} (\tau_n < u_n - \varepsilon) \rightarrow 0$ as $n \rightarrow \infty$.  Note that 

\begin{align*}
\mathbb{P} \left(\tau_n < u_n - \varepsilon \right) &= \mathbb{P} \left( \frac{\tau_n}{t_n} < u_n^-(\varepsilon) \right)\\
&\leq \mathbb{P} \left( \sup_{z \in [c, u_n^-(\varepsilon)]} \left( X_s(z t_n) + X_m(z t_n) - n \right) > 0 \right)\\
&\begin{multlined}= \mathbb{P} \Biggl( \sup_{z \in [c, u_n^-(\varepsilon)]} n^{\alpha + \lambda_m  z/\lambda_s - 1} (X_s(z t_n) - \phi_s(z t_n) + \phi_s(z t_n) - n \\
+ \phi_m(z t_n) - \phi_m(z t_n) + X_m(z t_n)) > 0 \Biggr) \end{multlined}\\
&\leq \mathbb{P} \left( \bar{B}_1(n, \varepsilon) + \bar{B}_2(n, \varepsilon) + \bar{B}_3(n, \varepsilon) > 0 \right),
\end{align*}

\noindent where 

\begin{align*}
\bar{B}_1(n, \varepsilon) &= \sup_{z \in [c, u_n^-(\varepsilon)]} n^{\alpha + \lambda_m  z/\lambda_s - 1} (X_s(z t_n) - \phi_s(z t_n)),\\
\bar{B}_2(n, \varepsilon) &= \sup_{z \in [c, u_n^-(\varepsilon)]} n^{\alpha + \lambda_m  z/\lambda_s - 1} (\phi_s(z t_n) + \phi_m(z t_n) - n),\\
\bar{B}_3(n, \varepsilon) &= \sup_{z \in [c, u_n^-(\varepsilon)]} n^{\alpha + \lambda_m  z/\lambda_s - 1} (X_m(z t_n) - \phi_m(z t_n)).
\end{align*}

\noindent Note that, for $i \in \{s, m\}$, 

\begin{align*}
\MoveEqLeft[3] \sup_{z \in [c, u_n^-(\varepsilon)]} \left| n^{\alpha + \lambda_m z/\lambda_s - 1} \left( X_i(zt_n) - \phi_i(zt_n) \right) \right| \\
&\leq \sup_{z \in [c, u_n^+(\varepsilon)]} n^{\alpha + \lambda_m z/\lambda_s - 1} \left| X_i(zt_n) - \phi_i(zt_n) \right|,
\end{align*}

\noindent which converges to zero in probability by Proposition~\ref{prop:FlucMut}.  Now we just need to show that $\bar{B}_2(n, \varepsilon)$ is negative in the large population limit.  Let $\bar{g}_n(z) = n^{\alpha + \lambda_m  z/\lambda_s - 1} \left( \phi_s(z t_n) + \phi_m(z t_n) - n \right)$.  Using the definitions of $\phi_s$ and $\phi_m$, we see that 

\begin{align*}
\bar{g}_n(z) &= n^{\alpha + \lambda_m  z/\lambda_s -1} \left( n^{1-z} + \frac{ n^{1-\alpha}}{\lambda_m  - \lambda_s} \left( n^{-\lambda_m  z/\lambda_s} - n^{-z} \right) - n \right) \\
&= n^{\alpha + z(\lambda_m /\lambda_s - 1)} + \frac{1}{\lambda_m  - \lambda_s} - \frac{1}{\lambda_m  - \lambda_s} n^{z(\lambda_m /\lambda_s - 1)} - n^{\alpha + \lambda_m z/\lambda_s}. \\
\end{align*} 

\noindent Taking the derivative with respect to $z$ yields 

\begin{align*}
\bar{g}'_n(z) &= \frac{\lambda_m  - \lambda_s}{\lambda_s} n^{\alpha+z(\lambda_m /\lambda_s-1)} \log n - \frac{1}{\lambda_s} n^{z(\lambda_m /\lambda_s-1)} \log n - \frac{\lambda_m }{\lambda_s} n^{\alpha + \lambda_m  z/\lambda_s} \log n \\
&= \frac{-1}{\lambda_s} n^{\alpha+z(\lambda_m /\lambda_s-1)} \log n \left[ -\lambda_m  + \lambda_s +  n^{-\alpha} + \lambda_m  n^z \right].
\end{align*}

\noindent Next we set $\bar{g}'_n(z) = 0$ and solve for $z$ to find any local maxima or minima of $\bar{g}_n(z)$.  Since $-1/\lambda_s > 0$ and $n^{\alpha-z(1 - \lambda_m /\lambda_s)} > 0$ and $\log n > 0$ for sufficiently large $n$, we have that

\begin{align*}
\bar{g}'_n(z) = 0 &\implies \lambda_m  n^z = \lambda_m  - \lambda_s -  n^{-\alpha} \\
&\implies z \log n = \log \left( \frac{\lambda_m  - \lambda_s -  n^{-\alpha}}{\lambda_m } \right) \\
&\implies z = \frac{1}{\log n} \log \left( \frac{\lambda_m  - \lambda_s -  n^{-\alpha}}{\lambda_m } \right).
\end{align*}

\noindent Hence $\bar{q}_n = \frac{1}{\log n}  \log \left( \frac{\lambda_m  - \lambda_s -  n^{-\alpha}}{\lambda_m } \right)$ is the only maximum or minimum of $\bar{g}_n(z)$.  Next note that $\bar{g}'_n(\bar{q}_n+1) = \frac{-1}{\lambda_s} n^{\alpha-(\bar{q}_n+1)(1 - \lambda_m /\lambda_s)} \log n \left[ -\lambda_m  + \lambda_s +  n^{-\alpha} + \lambda_m  n^{\bar{q}_n+1} \right] $.  Since $-1/\lambda_s > 0$ and $n^{\alpha - (\bar{q}_n+1)(1 - \lambda_m /\lambda_s)} > 0$ and $\log n > 0$ for sufficiently large $n$ and 

\begin{align*}
-\lambda_m  + \lambda_s +  n^{-\alpha} + \lambda_m  n^{\bar{q}_n+1} &= \lambda_m  n \cdot n^{\frac{1}{\log n} \log \left( \frac{\lambda_m  - \lambda_s -  n^{-\alpha}}{\lambda_m } \right)} - \lambda_m  + \lambda_s +  n^{-\alpha} \\
&= \lambda_m  n \left( \frac{\lambda_m  - \lambda_s -  n^{-\alpha}}{\lambda_m } \right) - \lambda_m  + \lambda_s +  n^{-\alpha} \\
&= (n-1) \left( \lambda_m  - \lambda_s -  n^{-\alpha} \right) \\ 
&> 0 \text{ for sufficiently large $n$}, 
\end{align*}

\noindent we have that $\bar{g}'_n(\bar{q}_n+1) > 0$ for sufficiently large $n$.  Hence $\bar{g}_n(z)$ is monotonically increasing for $z > \bar{q}_n$.  Note that the only positive solution to $\bar{g}_n(z) = 0$ occurs at $z = \tilde{u}_n$ by Proposition~\ref{prop:EstimateMut}.  Also note that $\bar{q}_n < c < u_n^-(\varepsilon) < \tilde{u}_n$ for sufficiently large $n$.  Therefore, we have that $\bar{B}_2(n, \varepsilon) < 0$.  Moreover, we may rewrite $\bar{B}_2(n, \varepsilon)$ as 

\begin{equation*}
\bar{B}_2(n, \varepsilon) = n^{\alpha + \lambda_m  u_n^-(\varepsilon)/\lambda_s - 1} \left( \phi_s \left( u_n^-(\varepsilon) t_n \right) + \phi_m \left( u_n^-(\varepsilon) t_n \right) - n \right).
\end{equation*}

\noindent Then by the definitions of $\phi_s$ and $\phi_m$, we have 

\begin{align*}
\bar{B}_2(n, \varepsilon) &= n^{\alpha + \lambda_m  u_n^-(\varepsilon)/\lambda_s - 1} \left( n^{1 - u_n^-(\varepsilon)} + \frac{ n^{1-\alpha}}{\lambda_m  - \lambda_s} \left( n^{-\lambda_m  u_n^-(\varepsilon)/\lambda_s} - n^{-u_n^-(\varepsilon)} \right) - n \right) \\
&= n^{\alpha-u_n^-(\varepsilon)(1 - \lambda_m /\lambda_s)} + \frac{1}{\lambda_m  - \lambda_s} - \frac{1}{\lambda_m  - \lambda_s} n^{-u_n^-(\varepsilon)(1 - \lambda_m /\lambda_s)} - n^{\alpha + \lambda_m  u_n^-(\varepsilon)/\lambda_s}. 
\end{align*} 

\noindent Note that 

\begin{align*}
n^{\alpha + \lambda_m  u_n^-(\varepsilon)/\lambda_s} &= n^{\alpha} n^{\frac{\varepsilon \lambda_m }{\log n}} n^{\tilde{u}_n(\lambda_m /\lambda_s)} \\
&= e^{\varepsilon \lambda_m } n^{\alpha + \tilde{u}_n(\lambda_m /\lambda_s)} \\
&\geq e^{\varepsilon \lambda_m } n^{\alpha + A_n(\lambda_m /\lambda_s)} \\
&= e^{\varepsilon \lambda_m } n^{\alpha + \frac{\lambda_m }{\lambda_s} \cdot \frac{1}{\lambda_m } \left( -\lambda_s \alpha + \frac{1}{t_n} \log \left( {\lambda_m  - \lambda_s}\right) \right)} \\
&= e^{\varepsilon \lambda_m } n^{\alpha - \alpha - \frac{1}{\log n} \log \left( {\lambda_m  - \lambda_s} \right)} \\
&= e^{\varepsilon \lambda_m } \left( \frac{1}{\lambda_m  - \lambda_s} \right).
\end{align*}

\noindent Therefore, we have that 

\begin{align}
\bar{B}_2(n, \varepsilon) &\leq n^{\alpha-u_n^-(\varepsilon)(1 - \lambda_m /\lambda_s)} + \frac{1}{\lambda_m  - \lambda_s} - \frac{1}{\lambda_m  - \lambda_s} n^{-u_n^-(\varepsilon)(1 - \lambda_m /\lambda_s)} - e^{\varepsilon \lambda_m } \left( \frac{1}{\lambda_m  - \lambda_s} \right) \nonumber \\
&= \frac{1}{\lambda_m  - \lambda_s} \left(1 - e^{\varepsilon \lambda_m } \right) + n^{\alpha - u_n^-(\varepsilon)(1 - \lambda_m /\lambda_s)} \left( 1 - \frac{ n^{-\alpha}}{\lambda_m  - \lambda_s} \right). \label{B2barRHS}
\end{align}

\noindent Clearly, $1 -  n^{-\alpha}/(\lambda_m  - \lambda_s) \rightarrow 1$ as $n \rightarrow \infty$ since $\alpha > 0$.  We also have that 

\begin{align*}
\lim_{n \rightarrow \infty} n^{\alpha - u_n^-(\varepsilon)(1 - \lambda_m /\lambda_s)} &= \lim_{n \rightarrow \infty} n^{\alpha} n^{\frac{\varepsilon}{t_n}(1 - \lambda_m /\lambda_s)} n^{-\tilde{u}_n(1 - \lambda_m /\lambda_s)} \\
&= e^{\varepsilon (\lambda_m  - \lambda_s)} \lim_{n \rightarrow \infty} n^{\alpha - \tilde{u}_n(1 - \lambda_m /\lambda_s)} \\
&= e^{\varepsilon(\lambda_m  - \lambda_s)} \lim_{n \rightarrow \infty} e^{[\alpha - \tilde{u}_n(1 - \lambda_m /\lambda_s)] \log n} \\
&= 0 
\end{align*}

\noindent since $\tilde{u}_n \rightarrow -\alpha \lambda_s / \lambda_m $ as $n \rightarrow \infty$ by Proposition~\ref{prop:EstimateMut}.  Therefore, the right-hand side of equation (\ref{B2barRHS}) converges to $\frac{1}{\lambda_m  - \lambda_s}(1 - e^{\varepsilon \lambda_m }) < 0$.  So $\bar{B}_2(n, \varepsilon)$ is negative in the large population limit.  Putting this all together, we have our desired result: $\mathbb{P}(\bar{B}_1(n, \varepsilon) + \bar{B}_2(n, \varepsilon) + \bar{B}_3(n, \varepsilon) > 0) \rightarrow 0$ as $n \rightarrow \infty$.  Therefore, we have shown $\mathbb{P}(\tau_n < u_n - \varepsilon) \rightarrow 0$ as $n \rightarrow \infty$.  The proof that $\mathbb{P}(\tau_n > u_n + \varepsilon) \rightarrow 0$ as $n \rightarrow \infty$ follows using a similar argument.  
\end{proof} 

\subsection{Proof of Proposition~\ref{prop:FlucMut}} \label{proof:FlucMut}

\begin{proof}

Let's start by proving the $i=s$ case.  That is, we want to show that 

\begin{equation*}
\lim_{n \rightarrow \infty} \mathbb{P} \left( \sup_{z \in [c, u_n^+(\varepsilon)]} n^{\alpha + \lambda_m  z/\lambda_s - 1} \left| X_s(zt_n) - \phi_s(zt_n) \right| > \delta \right) = 0.
\end{equation*}

\noindent Because a branching process normalized by its mean is a martingale, we know that 
\[Z_s(z) = n^{\alpha + z - 1} \left( X_s(zt_n) - \phi_s(zt_n) \right)\]
is a martingale in $z$.  To prove our desired result for $i=s$, we need to show 

\begin{equation*}
\lim_{n \rightarrow \infty} \mathbb{P} \left( \sup_{z \in [c, u_n^+(\varepsilon)]} n^{\lambda_m  z/\lambda_s - z} |Z_s(z)| > \delta \right) = 0.
\end{equation*}

\noindent Note that 

\begin{align*}
\sup_{z \in [c, u_n^+(\varepsilon)]} n^{\lambda_m  z/\lambda_s - z} |Z_s(z)| &\leq \sup_{z \in [c, u_n^+(\varepsilon)]} n^{\lambda_m  z/\lambda_s - z} \cdot \sup_{z \in [c, u_n^+(\varepsilon)]} |Z_s(z)| \\
&= n^{\lambda_m  c/\lambda_s - c} \cdot \sup_{z \in [c, u_n^+(\varepsilon)]} |Z_s(z)|.
\end{align*}

\noindent Therefore, we have that 

\begin{align*}
\mathbb{P} \left( \sup_{z \in [c, u_n^+(\varepsilon)]} n^{\lambda_m  z/\lambda_s - z} |Z_s(z)| > \delta \right) &\leq \mathbb{P} \left( n^{\lambda_m  c/\lambda_s - c} \cdot \sup_{z \in [c, u_n^+(\varepsilon)]} |Z_s(z)| > \delta \right) \\
&= \mathbb{P} \left( \sup_{z \in [c, u_n^+(\varepsilon)]} |Z_s(z)| > \delta \cdot n^{c - \lambda_m  c/\lambda_s} \right) \\
&\leq \frac{1}{\delta^2} n^{2\lambda_m  c/\lambda_s - 2c} \cdot \mathbb{E} \left[ \left( Z_s(u_n^+(\varepsilon)) \right)^2 \right]  
\end{align*}

\noindent by Doob's Martingale Inequality.  Note that 

\begin{align*}
\mathbb{E} \left[ \left( Z_s(u_n^+(\varepsilon)) \right)^2 \right] &= \mathbb{E} \left[ \left( n^{\alpha + u_n^+(\varepsilon) - 1} X_s(u_n^+(\varepsilon) t_n) - n^{\alpha + u_n^+(\varepsilon) - 1} \phi_s(u_n^+(\varepsilon) t_n) \right)^2 \right] \\
&= \mathrm{Var} \left[ n^{\alpha + u_n^+(\varepsilon) - 1} X_s(u_n^+(\varepsilon) t_n) \right]. 
\end{align*}

\noindent Therefore, now we have 

\begin{align*}
\MoveEqLeft[2] \mathbb{P} \left( \sup_{z \in [c, u_n^+(\varepsilon)]} n^{\lambda_m  z/\lambda_s - z} |Z_s(z)| > \delta \right) \\
&\leq \frac{1}{\delta^2} n^{2\lambda_m  c/\lambda_s - 2c} \cdot \mathrm{Var} \left[ n^{\alpha + u_n^+(\varepsilon) - 1} X_s(u_n^+(\varepsilon) t_n) \right] \\
&= \frac{1}{\delta^2} n^{2\lambda_m  c/\lambda_s - 2c} n^{2\alpha + 2u_n^+(\varepsilon) - 2} n^{1 - u_n^+(\varepsilon)} \left( \frac{r_s + d_s}{-\lambda_s} \right) \left( 1 - n^{-u_n^+(\varepsilon)} \right) \\
&= \frac{1}{\delta^2} \left( \frac{r_s + d_s}{-\lambda_s} \right) \left( 1 - n^{-u_n^+(\varepsilon)} \right) n^{2\alpha + 2c(\lambda_m /\lambda_s - 1) + u_n^+(\varepsilon) - 1}. 
\end{align*}

\noindent Clearly, $\frac{1}{\delta^2} \left( \frac{r_s + d_s}{-\lambda_s} \right)$ is just a finite constant.  Note that 

\begin{align*}
\lim_{n \rightarrow \infty} \left( 1 - n^{-u_n^+(\varepsilon)} \right) &= 1 - \lim_{n \rightarrow \infty} n^{-\varepsilon/t_n} n^{-\tilde{u}_n} \\
&= 1 - e^{\varepsilon \lambda_s} \lim_{n \rightarrow \infty} e^{-\tilde{u}_n \log n} \\
&= 1
\end{align*}

\noindent since $\tilde{u}_n \rightarrow -\lambda_s \alpha/\lambda_m $ as $n \rightarrow \infty$ by Proposition~\ref{prop:EstimateMut}.  We also have that 

\begin{align*}
\lim_{n \rightarrow \infty} n^{2\alpha + 2c(\lambda_m /\lambda_s-1) + u_n^+(\varepsilon) - 1} &= \lim_{n \rightarrow \infty} n^{2\alpha + \frac{\alpha \lambda_s(\lambda_s - 2\lambda_m )}{\lambda_m (\lambda_m  - \lambda_s)} \cdot \frac{\lambda_m  - \lambda_s}{\lambda_s} - 1} n^{\tilde{u}_n} n^{\varepsilon/t_n} \\
&= e^{-\varepsilon \lambda_s} \lim_{n \rightarrow \infty} n^{2\alpha + \alpha(\lambda_s - 2\lambda_m )/\lambda_m  - 1 + \tilde{u}_n} \\ 
&= e^{-\varepsilon \lambda_s} \lim_{n \rightarrow \infty} e^{[\tilde{u}_n + \alpha \lambda_s/\lambda_m  - 1] \log n} \\
&= 0
\end{align*}

\noindent since $\tilde{u}_n \rightarrow -\lambda_s \alpha/\lambda_m $ as $n \rightarrow \infty$ by Proposition~\ref{prop:EstimateMut}.  So we are done with the $i=s$ case.  

Next, let's prove the desired result for $i=m$.  As a reminder, we want to show that 

\begin{equation*}
\lim_{n \rightarrow \infty} \mathbb{P} \left( \sup_{z \in [c, u_n^+(\varepsilon)]} n^{\alpha + \lambda_m  z/\lambda_s - 1} \left| X_m(zt_n) - \phi_m(zt_n) \right| > \delta \right) = 0.
\end{equation*}

Note that 

\begin{align*}
\phi_m(z t_n) &= \int_0^{zt_n}  n^{-\alpha} n e^{\lambda_s s} e^{\lambda_m (zt_n - s)} \, ds\\
&=  n^{1 -\lambda_m z/\lambda_s - \alpha} \int_0^{zt_n} e^{(\lambda_s - \lambda_m )s} \, ds.
\end{align*}

\noindent Therefore, 

\begin{align*}
\MoveEqLeft[3] n^{\alpha + \lambda_m z/\lambda_s - 1} \left( X_m(zt_n) - \phi_m(zt_n) \right) \\
&= n^{\alpha + \lambda_m z/\lambda_s - 1} X_m(zt_n) -  \int_0^{zt_n} e^{(\lambda_s - \lambda_m )s} \, ds\\
&= n^{\alpha + \lambda_m z/\lambda_s - 1} X_m(zt_n) - \frac{1}{n} \int_0^{zt_n} X_s(s) e^{-\lambda_m s} \, ds + \frac{1}{n} \int_0^{zt_n} \left( X_s(s) - ne^{\lambda_ss} \right) e^{-\lambda_m s} \, ds.
\end{align*}

\noindent Taking the absolute value of both sides and using the triangle inequality, we get 

\begin{align*}
\MoveEqLeft[3] n^{\alpha + \lambda_m z/\lambda_s - 1} \left| X_m(zt_n) - \phi_m(zt_n) \right| \\
&\leq \left| n^{\alpha + \lambda_m z/\lambda_s - 1} X_m(zt_n) - \frac{1}{n} \int_0^{zt_n} X_s(s) e^{-\lambda_m s} \, ds \right| + \frac{1}{n} \int_0^{zt_n} \left| X_s(s) - ne^{\lambda_ss} \right| e^{-\lambda_m s} \, ds.
\end{align*}

\noindent This implies that 

\begin{align*}
\MoveEqLeft[3] \sup_{z \in [c, u_n^+(\varepsilon)]} n^{\alpha + \lambda_m z/\lambda_s - 1} \left| X_m(zt_n) - \phi_m(zt_n) \right| \\
&\begin{multlined}\leq \sup_{z \in [c, u_n^+(\varepsilon)]} \left| n^{\alpha + \lambda_m z/\lambda_s - 1} X_m(zt_n) - \frac{1}{n} \int_0^{zt_n} X_s(s) e^{-\lambda_m s} \, ds \right| \\
+ \sup_{z \in [c, u_n^+(\varepsilon)]} \frac{1}{n} \int_0^{zt_n} \left| X_s(s) - ne^{\lambda_ss} \right| e^{-\lambda_m s} \, ds. \end{multlined} 
\end{align*}

\noindent Hence 

\begin{align*}
\MoveEqLeft[3] \mathbb{P} \left( \sup_{z \in [c, u_n^+(\varepsilon)]} n^{\alpha + \lambda_m z/\lambda_s - 1} \left| X_m(zt_n) - \phi_m(zt_n) \right| > \delta \right) \\
&\begin{multlined}\leq \mathbb{P} \left( \sup_{z \in [c, u_n^+(\varepsilon)]} \left| n^{\alpha + \lambda_m  z/\lambda_s - 1} X_m(zt_n) - \frac{1}{n} \int_0^{zt_n} X_s(s) e^{-\lambda_m s} \, ds \right| > \delta/2 \right) \\
+ \mathbb{P} \left(  \sup_{z \in [c, u_n^+(\varepsilon)]} \frac{1}{n} \int_0^{zt_n} \left| X_s(s) - ne^{\lambda_ss} \right| e^{-\lambda_m s} \, ds > \delta/2 \right). \end{multlined} 
\end{align*}

\noindent The process in the second term of the above sum is monotonically increasing in $z$.  So we may simplify the expression above to 

\begin{align*}
\MoveEqLeft[3] \mathbb{P} \left( \sup_{z \in [c, u_n^+(\varepsilon)]} n^{\alpha + \lambda_m z/\lambda_s - 1} \left| X_m(zt_n) - \phi_m(zt_n) \right| > \delta \right) \\
&\begin{multlined}\leq \mathbb{P} \left( \sup_{z \in [c, u_n^+(\varepsilon)]} \left| n^{\alpha + \lambda_m z/\lambda_s - 1} X_m(zt_n) - \frac{1}{n} \int_0^{zt_n} X_s(s) e^{-\lambda_m s} \, ds \right| > \delta/2 \right) \\
+ \mathbb{P} \left( \frac{1}{n} \int_0^{u_n^+(\varepsilon) t_n} \left| X_s(s) - ne^{\lambda_ss} \right| e^{-\lambda_m s} \, ds > \delta/2 \right). \end{multlined} 
\end{align*}

\noindent By Lemma 1 in \cite{durrett_evolution_2010}, we know that 

\begin{equation*}
e^{-\lambda_m t} X_m(t) - \int_0^t  n^{-\alpha} e^{-\lambda_m s} X_s(s) \, ds 
\end{equation*}

\noindent is a martingale.  Setting $t = zt_n = \frac{z}{-\lambda_s}\log n$, we get that 

\begin{equation*}
n^{\lambda_m z/\lambda_s}X_m(zt_n) - \frac{1}{n^{\alpha}} \int_0^{zt_n} e^{-\lambda_m s} X_s(s) \, ds 
\end{equation*}

\noindent is a martingale in $z$.  Since linear combinations of martingales are also martingales, 

\begin{equation*}
n^{\alpha + \lambda_m z/\lambda_s - 1} X_m(zt_n) - \frac{1}{n} \int_0^{zt_n} e^{-\lambda_m s} X_s(s) \, ds 
\end{equation*}

\noindent is also a martingale in $z$.  Therefore, 

\begin{equation*}
\left| n^{\alpha + \lambda_m z/\lambda_s - 1} X_m(zt_n) - \frac{1}{n} \int_0^{zt_n} X_s(s) e^{-\lambda_m s} \, ds \right|
\end{equation*}

\noindent is a non-negative submartingale in $z$, so we can apply Doob's Martingale Inequality to get 

\begin{align}
\MoveEqLeft[3] \mathbb{P} \left( \sup_{z \in [c, u_n^+(\varepsilon)]} n^{\alpha + \lambda_m z/\lambda_s - 1} \left| X_m(zt_n) - \phi_m(zt_n) \right| > \delta \right) \nonumber \\
&\begin{multlined}\leq \frac{4}{\delta^2} \cdot \mathbb{E} \left[ \left( n^{\alpha + \lambda_m  u_n^+(\varepsilon)/\lambda_s - 1} X_m (u_n + \varepsilon) - \frac{1}{n} \int_0^{u_n + \varepsilon} X_s(s) e^{-\lambda_m  s} \, ds \right)^2 \right] \\
+ \mathbb{P} \left( \frac{1}{n} \int_0^{u_n + \varepsilon} \left| X_s(s) - ne^{\lambda_ss} \right| e^{-\lambda_m s} \, ds > \delta/2 \right).\end{multlined} \label{flucprob} 
\end{align}

\noindent Let us start by showing that the second term in (\ref{flucprob}) converges to zero.  Since convergence in mean implies convergence in probability (by Markov's Inequality), it suffices to show that 

\begin{equation*}
\lim_{n \rightarrow \infty} \mathbb{E} \left[ \frac{1}{n} \int_0^{u_n+\varepsilon} \left| X_s(s) - ne^{\lambda_ss} \right| e^{-\lambda_m s} \, ds \right] = 0,
\end{equation*}

\noindent or, equivalently, 

\begin{equation*}
\lim_{n \rightarrow \infty} \frac{1}{n} \int_0^{u_n+\varepsilon} \mathbb{E} \left[ \left| X_s(s) - ne^{\lambda_ss} \right| \right] e^{-\lambda_m s} \, ds = 0. 
\end{equation*}

\noindent By the Cauchy-Schwarz Inequality, we know 

\begin{align*}
\mathbb{E} \left[ \left| X_s(s) - ne^{\lambda_ss} \right| \right] &\leq \sqrt{\mathbb{E} \left[ \left| X_s(s) - ne^{\lambda_ss} \right|^2 \right]}\\
&\leq \sqrt{ \text{Var} \left[ X_s(s) \right]}\\
&= n^{1/2} \left[ \left( \frac{r_s + d_s}{\lambda_s} \right) \left( e^{2\lambda_ss} - e^{\lambda_ss} \right) \right]^{1/2}. 
\end{align*}

\noindent Multiplying both sides by $\frac{1}{n} e^{-\lambda_m s}$ and integrating yields 

\begin{align*}
\frac{1}{n} \int_0^{u_n + \varepsilon} \mathbb{E} \left[ \left| X_s(s) - ne^{\lambda_ss} \right| \right] e^{-\lambda_m s} \, ds &\leq \frac{1}{n^{1/2}} \int_0^{u_n + \varepsilon} \left[ \left( \frac{r_s + d_s}{\lambda_s} \right) \left( e^{2\lambda_ss} - e^{\lambda_ss} \right) \right]^{1/2} e^{-\lambda_m s} \, ds\\
&=  \left( \frac{r_s + d_s}{-\lambda_sn} \right)^{1/2} \int_0^{u_n + \varepsilon} \left( 1 - e^{\lambda_ss} \right)^{1/2} e^{\lambda_ss/2} e^{-\lambda_m s} \, ds\\
&\leq  \left( \frac{r_s + d_s}{-\lambda_sn} \right)^{1/2} \int_0^{u_n + \varepsilon} e^{(\lambda_s/2 - \lambda_m )s} \, ds\\
&= \frac{1}{\lambda_s/2 - \lambda_m } \left( \frac{r_s + d_s}{-\lambda_sn} \right)^{1/2} \left( e^{(\lambda_s/2 - \lambda_m )(u_n + \varepsilon)} - 1 \right).
\end{align*}

\noindent Note that, since $u_n = \tilde{u}_n t_n = \frac{\tilde{u}_n}{-\lambda_s} \log n$, 

\begin{align*}
e^{(\lambda_s/2 - \lambda_m )(u_n + \varepsilon)} &= e^{(\lambda_s/2 - \lambda_m ) \frac{\tilde{u}_n}{-\lambda_s} \log n} e^{(\lambda_s/2 - \lambda_m ) \varepsilon}\\
&= n^{(\lambda_m /\lambda_s - 1/2) \tilde{u}_n} e^{(\lambda_s/2 - \lambda_m )\varepsilon},
\end{align*}

\noindent which converges to zero since $\tilde{u}_n \rightarrow \frac{-\lambda_s \alpha}{\lambda_m }$ as $n \rightarrow \infty$ by Proposition~\ref{prop:EstimateMut}.  Therefore, 

\begin{equation*}
\lim_{n \rightarrow \infty} \frac{1}{n} \int_0^{u_n+\varepsilon} \mathbb{E} \left[ \left| X_s(s) - ne^{\lambda_ss} \right| \right] e^{-\lambda_m s} \, ds = 0,
\end{equation*}

\noindent as desired.  Next, we will show the first term in (\ref{flucprob}) converges to zero as well.  

\begin{align*}
\MoveEqLeft[3] \mathbb{E} \left[ \left( n^{\alpha +\lambda_m  u_n^+(\varepsilon)/\lambda_s - 1} X_m(u_n + \varepsilon) - \frac{1}{n} \int_0^{u_n + \varepsilon} X_s(s) e^{-\lambda_m s} \, ds \right)^2 \right]\\
&\begin{multlined}= n^{2(\alpha + \lambda_m  u_n^+(\varepsilon)/\lambda_s - 1)} \mathbb{E} \left[ X_m(u_n + \varepsilon)^2 \right]\\
\qquad- 2 n^{\alpha + \lambda_m  u_n^+(\varepsilon)/\lambda_s - 2} \int_0^{u_n + \varepsilon} \mathbb{E} \left[ X_s(s) X_m(u_n + \varepsilon) \right] e^{-\lambda_m s} \, ds\\
+ \left( \frac{1}{n} \right)^2 \int_0^{u_n + \varepsilon} \int_0^{u_n + \varepsilon} \mathbb{E} \left[ X_s(s) X_s(y) \right] e^{-\lambda_m s} e^{-\lambda_m y} \, ds \, dy. \end{multlined} 
\end{align*}

\noindent From Lemma 1 in \cite{foo_dynamics_2013}, we know that 

\begin{align*}
\mathbb{E} \left[ X_m(u_n + \varepsilon)^2 \right] &= \left( \frac{1}{n^{\alpha}} \right)^2 \int_0^{u_n + \varepsilon} \int_0^{u_n + \varepsilon} \mathbb{E} \left[ X_s(s) X_s(y) \right] e^{\lambda_m (u_n + \varepsilon - s)} e^{\lambda_m (u_n + \varepsilon - y)} \, ds \, dy\\ 
&\hspace{4mm}+ \left( \frac{1}{n^{\alpha}} \right) \int_0^{u_n + \varepsilon} \mathbb{E} \left[ X_s(s) \right] \mathbb{E} \left[ \tilde{X}_m(u_n + \varepsilon - s)^2 \right] \, ds, \\
\mathbb{E} \left[ X_s(s) X_m(u_n + \varepsilon) \right] &= \left( \frac{1}{n^{\alpha}} \right) \int_0^{u_n + \varepsilon} \mathbb{E} \left[ X_s(y) X_s(s) \right] e^{\lambda_m (u_n + \varepsilon - y)} \, dy, 
\end{align*}

\noindent where $\tilde{X}_m$ is a binary branching process starting from size one with birth rate $r_m$ and death rate $d_m$.  Substituting these expressions into our equation yields 

\begin{align*}
\MoveEqLeft[3] \mathbb{E} \left[ \left( n^{\alpha +\lambda_m  u_n^+(\varepsilon)/\lambda_s - 1} X_m(u_n + \varepsilon) - \frac{1}{n} \int_0^{u_n + \varepsilon} X_s(s) e^{-\lambda_m s} \, ds \right)^2 \right]\\
&\begin{multlined}=  n^{2(\lambda_m  u_n^+(\varepsilon)/\lambda_s - 1)} \int_0^{u_n + \varepsilon} \int_0^{u_n + \varepsilon} \mathbb{E} \left[ X_s(s) X_s(y) \right] e^{\lambda_m (2u_n + 2\varepsilon - s - y)} \, ds \, dy \\
+  n^{\alpha + 2(\lambda_m  u_n^+(\varepsilon)/\lambda_s - 1)} \int_0^{u_n + \varepsilon} \mathbb{E} \left[ X_s(s) \right] \mathbb{E} \left[ \tilde{X}_m(u_n + \varepsilon - s)^2 \right] \, ds \\
- 2 n^{\lambda_m  u_n^+(\varepsilon)/\lambda_s - 2} \int_0^{u_n + \varepsilon} \int_0^{u_n + \varepsilon} \mathbb{E} \left[ X_s(s) X_s(y) \right] e^{\lambda_m (u_n + \varepsilon - s - y)} \, ds \, dy \\
+  n^{-2} \int_0^{u_n + \varepsilon} \int_0^{u_n + \varepsilon} \mathbb{E} \left[ X_s(s) X_s(y) \right] e^{-\lambda_m (s+y)} \, ds \, dy. \end{multlined} 
\end{align*}

\noindent Then, by the definition of $u_n^+(\varepsilon)$, 

\begin{align*}
\MoveEqLeft[3] \mathbb{E} \left[ \left( n^{\alpha +\lambda_m  u_n^+(\varepsilon)/\lambda_s - 1} X_m(u_n + \varepsilon) - \frac{1}{n} \int_0^{u_n + \varepsilon} X_s(s) e^{-\lambda_m s} \, ds \right)^2 \right]\\
&\begin{multlined}=  n^{-2 \left(\frac{\lambda_m (u_n + \varepsilon)}{\log n} + 1 \right)} \int_0^{u_n + \varepsilon} \int_0^{u_n + \varepsilon} \mathbb{E} \left[ X_s(s) X_s(y) \right] e^{\lambda_m (2u_n + 2\varepsilon - s - y)} \, ds \, dy \\
+  n^{\alpha - 2 \left(\frac{\lambda_m (u_n + \varepsilon)}{\log n} + 1 \right)} \int_0^{u_n + \varepsilon} \mathbb{E} \left[ X_s(s) \right] \mathbb{E} \left[ \tilde{X}_m(u_n + \varepsilon - s)^2 \right] \, ds \\
- 2 n^{-\frac{\lambda_m (u_n + \varepsilon)}{\log n} - 2} \int_0^{u_n + \varepsilon} \int_0^{u_n + \varepsilon} \mathbb{E} \left[ X_s(s) X_s(y) \right] e^{\lambda_m (u_n + \varepsilon - s - y)} \, ds \, dy \\
+  n^{-2} \int_0^{u_n + \varepsilon} \int_0^{u_n + \varepsilon} \mathbb{E} \left[ X_s(s) X_s(y) \right] e^{-\lambda_m (s+y)} \, ds \, dy\end{multlined} \\
&\begin{multlined}= \left( \frac{1}{n} \right)^2 \int_0^{u_n + \varepsilon} \int_0^{u_n + \varepsilon} \mathbb{E} \left[ X_s(s) X_s(y) \right] e^{-\lambda_m (s+y)} \, ds \, dy \\
\qquad +  n^{\alpha - 2} e^{-2\lambda_m (u_n + \varepsilon)} \int_0^{u_n + \varepsilon} \mathbb{E} \left[ X_s(s) \right] \mathbb{E} \left[ \tilde{X}_m(u_n + \varepsilon - s)^2 \right] \, ds \\
- 2 \left( \frac{1}{n} \right)^2 \int_0^{u_n + \varepsilon} \int_0^{u_n + \varepsilon} \mathbb{E} \left[ X_s(s) X_s(y) \right] e^{-\lambda_m (s+y)} \, ds \, dy \\
+ \left( \frac{1}{n} \right)^2 \int_0^{u_n + \varepsilon} \int_0^{u_n + \varepsilon} \mathbb{E} \left[ X_s(s) X_s(y) \right] e^{-\lambda_m (s+y)} \, ds \, dy \end{multlined} \\
&= n^{\alpha - 2} e^{-2 \lambda_m (u_n + \varepsilon)} \int_0^{u_n + \varepsilon} \mathbb{E} \left[ X_s(s) \right] \mathbb{E} \left[ \tilde{X}_m(u_n + \varepsilon - s)^2 \right] \, ds. 
\end{align*}

\noindent Note that 

\begin{align*}
\mathbb{E} \left[ X_s(t) \right] &= ne^{\lambda_st}, \\
\mathbb{E} \left[ \tilde{X}_m(t)^2 \right] &= \frac{2r_m}{\lambda_m } e^{2\lambda_m t} - \frac{r_m + d_m}{\lambda_m } e^{\lambda_m  t}. 
\end{align*}

\noindent Substituting these into the above expression yields 

\begin{align*}
\MoveEqLeft[3] \mathbb{E} \left[ \left( n^{\alpha +\lambda_m  u_n^+(\varepsilon)/\lambda_s-1} X_m(u_n + \varepsilon) - \frac{1}{n} \int_0^{u_n + \varepsilon} X_s(s) e^{-\lambda_m s} \, ds \right)^2 \right]\\
&= n^{\alpha - 2} e^{-2\lambda_m (u_n + \varepsilon)} \int_0^{u_n + \varepsilon} ne^{\lambda_ss} \left( \frac{2r_m}{\lambda_m } e^{2\lambda_m (u_n + \varepsilon - s)} - \frac{r_m + d_m}{\lambda_m } e^{\lambda_m (u_n + \varepsilon - s)} \right) \, ds \\
&= \frac{2r_m  n^{\alpha - 1}}{\lambda_m } \int_0^{u_n + \varepsilon} e^{(\lambda_s - 2\lambda_m )s} \, ds - \frac{ n^{\alpha - 1} (r_m + d_m)}{\lambda_m } e^{-\lambda_m (u_n + \varepsilon)} \int_0^{u_n + \varepsilon} e^{(\lambda_s - \lambda_m )s} \, ds \\
&= \frac{2r_m  n^{\alpha - 1}}{\lambda_m (\lambda_s - 2\lambda_m )} \left( e^{(\lambda_s - 2\lambda_m )(u_n + \varepsilon)} - 1 \right) - \frac{ n^{\alpha - 1} (r_m + d_m)}{\lambda_m (\lambda_s - \lambda_m )} \left( e^{(\lambda_s - 2\lambda_m )(u_n + \varepsilon)} - e^{-\lambda_m (u_n + \varepsilon)} \right).
\end{align*} 

\noindent Since $u_n = \tilde{u}_n t_n$ and $t_n = \frac{-1}{\lambda_s} \log n$, the expression above is equivalent to 

\begin{align*}
\begin{multlined} \frac{ n^{\alpha - 1}}{\lambda_m } \Biggl[ e^{(\lambda_s - 2\lambda_m )(\tilde{u}_n \frac{-1}{\lambda_s} \log n + \varepsilon)} \left( \frac{2r_m}{\lambda_s - 2\lambda_m } - \frac{r_m + d_m}{\lambda_s - \lambda_m } \right) \\
- \frac{2r_m}{\lambda_s - 2\lambda_m } + \frac{r_m + d_m}{\lambda_s - \lambda_m } e^{-\lambda_m (\tilde{u}_n \frac{-1}{\lambda_s} \log n + \varepsilon)} \Biggr]. \end{multlined} 
\end{align*}

\noindent Because $\tilde{u}_n \rightarrow \frac{-\lambda_s \alpha}{\lambda_m }$ as $n \rightarrow \infty$, $e^{(\lambda_s - 2\lambda_m )(\tilde{u}_n \frac{-1}{\lambda_s} \log n + \varepsilon)} \rightarrow 0$ and $e^{-\lambda_m (\tilde{u}_n \frac{-1}{\lambda_s} \log n + \varepsilon)} \rightarrow 0$.  Therefore, the entire expression above converges to zero since $\alpha < 1$, so we are done.  
\end{proof} 

\subsection{Proof of Lemma~\ref{NoPermRecurAmp}}\label{proof:NoPermRecurAmp}
\begin{proof}

Since $\mathbb{P} \left( \sum_{k=2}^M X_k(dt_n) - n \leq 0 \right) = 1 - \mathbb{P} \left( \sum_{k=2}^M X_k(dt_n) - n > 0 \right)$, it suffices to show 

\begin{equation*}
\lim_{n \rightarrow \infty} \mathbb{P} \left( \sum_{k=2}^M X_k(dt_n) - n > 0 \right) = 0.
\end{equation*}

\noindent Note that 

\begin{align*}
\MoveEqLeft[2] \mathbb{P} \left( \sum_{k=2}^M X_k(dt_n) - n > 0 \right) \\
&= \mathbb{P} \left( n^{\beta(M-2) + \lambda_Md/\lambda_2 - 1} \left( \sum_{k=2}^M X_k(dt_n) + \sum_{k=2}^M \phi_k(dt_n) - \sum_{k=2}^M \phi_k(dt_n) - n \right) > 0 \right) \\
&= \mathbb{P} \left( \sum_{k=2}^M \hat{A}_k(n) + \hat{A}_{\phi}(n) > 0 \right),
\end{align*}

\noindent where 

\begin{align*}
\hat{A}_k(n) &= n^{\beta(M-2) + \lambda_Md/\lambda_2 - 1} \left( X_k(dt_n) - \phi_k(dt_n) \right), \\
\hat{A}_{\phi}(n) &= n^{\beta(M-2) + \lambda_Md/\lambda_2 - 1} \left( \sum_{k=2}^M \phi_k(dt_n) - n \right).
\end{align*}

\noindent Note that 

\begin{align*}
\left| \hat{A}_k(n) \right| &= n^{\beta(M-2) + \lambda_Md/\lambda_2 - 1} \left| X_k(dt_n) - \phi_k(dt_n) \right| \\
&\leq \sup_{z \in [d, v_n^+(\varepsilon)]} n^{\beta(M-2) + \lambda_Mz/\lambda_2 - 1} \left| X_k(zt_n) - \phi_k(zt_n) \right|,
\end{align*}

\noindent which converges to zero in probability by Proposition~\ref{prop:FlucAmp}, below.  Now we just need to show that $\hat{A}_{\phi}(n)$ is negative in the large population limit.  By the definitions of $\phi_2$ and $\phi_k$, $k \geq 3$, 

\begin{align*}
\hat{A}_{\phi}(n) &= n^{\beta(M-2) + \lambda_Md/\lambda_2 - 1} \left( n^{1-d} + \sum_{k=3}^M \frac{ (-1)^k}{D^{k-2}} n^{1-(k-2)\beta} \tilde{S}_k(d) - n \right) \\
&= n^{\beta(M-2) + \lambda_Md/\lambda_2} \left( n^{-d} + \sum_{k=3}^M \frac{ (-1)^k}{D^{k-2}} n^{-\beta(k-2)} \tilde{S}_k(d) - 1 \right) \\
&\sim n^{\beta(M-2) + \lambda_Md/\lambda_2} \left( \sum_{k=3}^M \frac{ (-1)^k}{D^{k-2}} \cdot \frac{n^{-\beta(k-2) - \lambda_kd/\lambda_2}}{\tilde{P}_{k,k}} - 1 \right).
\end{align*}

\noindent For $3 \leq k < M$, we obviously have $k-2 < M-2$.  Since $d > -\lambda_2 \beta \frac{M-2}{\lambda_M - \lambda_2} = \frac{-\lambda_2 \beta}{D}$ by (\ref{restrict3}), this means that $Dd/\lambda_2 < -\beta$, and hence $\beta + Dd/\lambda_2 < 0$.  Therefore, for all $3 \leq k < M$,  

\begin{align*}
-(k-2)(\beta + Dd/\lambda_2) &< -(M-2)(\beta + Dd/\lambda_2) \\
- d - \beta(k-2) - (k-2)Dd/\lambda_2 &< -d - \beta(M-2) - (M-2)Dd/\lambda_2 \\
-\beta(k-2) - [\lambda_2 + (k-2)D]d/\lambda_2 &< -\beta(M-2) - [\lambda_2 + (M-2)D]d/\lambda_2 \\
-\beta(k-2) - \lambda_kd/\lambda_2 &< -\beta(M-2) - \lambda_Md/\lambda_2.
\end{align*}

\noindent This implies that 

\begin{align*}
\hat{A}_{\phi}(n) &\sim n^{\beta(M-2) + \lambda_Md/\lambda_2} \left( \frac{ (-1)^M}{D^{M-2} \tilde{P}_{M,M}} n^{-\beta(M-2) - \lambda_Md/\lambda_2} - 1 \right) \\
&= \frac{ (-1)^M}{D^{M-2} \tilde{P}_{M,M}} - n^{\beta(M-2) + \lambda_Md/\lambda_2}. 
\end{align*}

\noindent Since $d < \frac{-\lambda_2 \beta(M-2)}{\lambda_M}$ by $(\ref{restrict3})$, we have that $\lambda_M d/\lambda_2 > -\beta(M-2)$, and hence $n^{\beta(M-2) + \lambda_Md/\lambda_2} \rightarrow \infty$ as $n \rightarrow \infty$.  Therefore, $\hat{A}_{\phi}(n)$ is definitely negative in the large population limit.  Putting this all together, we have our desired result: $\mathbb{P} \left( \sum_{k=2}^M \hat{A}_k(n) + \hat{A}_{\phi}(n) > 0 \right) \rightarrow 0$ as $n \rightarrow \infty$. 
\end{proof} 

\subsection{Proof of Proposition~\ref{prop:EstimateAmp}}\label{proof:EstimateAmp} 

\begin{proof}

Let $\hat{f}_n(z) = \sum_{k=2}^M \phi_k(z t_n) - n$.  Taking the first derivative of $\phi_2$ with respect to $z$, we get 

\begin{equation*}
\frac{d}{dz} \phi_2(z t_n) = -n^{1-z} \log n.
\end{equation*}

\noindent Taking a second derivative yields 

\begin{align*}
\frac{d^2}{dz^2} \phi_2(z t_n) &= (\log n)^2 n^{1-z} \\
&> 0.
\end{align*}

\noindent Hence $\phi_2(z t_n)$ is concave up everywhere.  Now for $k>2$, we have 

\begin{equation*}
\phi_k(z t_n) = \frac{ (-1)^k}{D^{k-2}} n^{1-(k-2)\beta} \tilde{S}_k(z).
\end{equation*}

\noindent Taking the first derivative of this equation with respect to $z$, we get 

\begin{equation*}
\frac{d}{dz} \phi_k(z t_n) = \frac{ (-1)^k}{D^{k-2}} n^{1 - (k-2)\beta} \log n \left( -\frac{\lambda_i}{\lambda_2} \right) \tilde{S}_k(z).
\end{equation*}

\noindent Taking a second derivative yields 

\begin{align*}
\frac{d^2}{dz^2} \phi_k(zt_n) &= \frac{ (-1)^k}{D^{k-2}} n^{1-(k-2)\beta} (\log n)^2 \left( \frac{\lambda_i}{\lambda_2} \right)^2 \tilde{S}_k(z) \\
&\sim \frac{ (-1)^k}{D^{k-2}} n^{1-(k-2)\beta} (\log n)^2 \frac{(\lambda_k/\lambda_2)^2 n^{-\lambda_kz/\lambda_2}}{\tilde{P}_{k,k}} \\
&> 0.
\end{align*}

\noindent So $\phi_k(z t_n)$ is concave up in the large population limit for all $k \geq 2$.  Therefore, $\sum_{k=2}^M \phi_k(z t_n)$ is concave up since a sum of concave up functions is always concave up.  Hence $\hat{f}_n(z)$ is concave up as well.  Since $\hat{f}_n(z)$ is clearly differentiable everywhere, this implies that either $\hat{f}_n(z)$ is monotonically increasing everywhere, or $\hat{f}_n(z)$ is decreasing on the first part of its domain and then increasing on the rest of its domain.  Note that $\hat{f}_n(0) = 0$.  We also have that 

\begin{align*}
\hat{f}_n(b_n) &= n^{1-b_n} + \sum_{k=3}^M \frac{ (-1)^k}{D^{k-2}} n^{1-(k-2)\beta} \tilde{S}_k(b_n) - n \\
&\begin{multlined}= n \Biggl[ n^{\frac{\lambda_2}{\lambda_M} \beta(M-2)} n^{\frac{1}{\lambda_M t_n} \log \left[ \frac{ (-1)^M}{D^{M-2}\tilde{P}_{M,M}} \left( \frac{\lambda_2 - \lambda_M}{\lambda_2} \right) \right]} \\
+ \sum_{k=3}^M \frac{ (-1)^k}{D^{k-2}} \sum_{i=2}^k \frac{n^{-\beta(k-2)} n^{\frac{\lambda_i}{\lambda_M} \beta(M-2)} n^{\frac{\lambda_i}{\lambda_M \lambda_2 t_n} \log \left[ \frac{ (-1)^M}{D^{M-2}\tilde{P}_{M,M}} \left( \frac{\lambda_2 - \lambda_M}{\lambda_2} \right) \right]}}{\tilde{P}_{i,k}} - 1 \Biggr] \end{multlined} \\
&\sim n \left[ \sum_{k=3}^M \frac{ (-1)^k}{D^{k-2}} \cdot \frac{n^{\beta[ \lambda_k(M-2)/\lambda_M - (k-2)]} \left[ \frac{ (-1)^M}{D^{M-2}\tilde{P}_{M,M}} \left( \frac{\lambda_2 - \lambda_M}{\lambda_2} \right) \right]^{-\lambda_k/\lambda_M}}{\tilde{P}_{k,k}} - 1 \right] \\
&\begin{multlined}= n \Biggl[ \sum_{k=3}^{M-1} \frac{ (-1)^k}{D^{k-2}} \cdot \frac{n^{\beta[ \lambda_k(M-2)/\lambda_M - (k-2)]} \left[ \frac{ (-1)^M}{D^{M-2}\tilde{P}_{M,M}} \left( \frac{\lambda_2 - \lambda_M}{\lambda_2} \right) \right]^{-\lambda_k/\lambda_M}}{\tilde{P}_{k,k}} \\
+ \frac{\lambda_2}{\lambda_2 - \lambda_M} - 1 \Biggr]. \end{multlined} 
\end{align*}

\noindent Note that 

\begin{align*}
\beta \left[ \frac{\lambda_k(M-2)}{\lambda_M} - (k-2) \right] &= \beta \frac{ \left[ \lambda_2 + (k-2) \frac{\lambda_M - \lambda_2}{M-2} \right] (M-2) - \left[ \lambda_2 + (M-2) \frac{\lambda_M - \lambda_2}{M-2} \right] (k-2)}{\lambda_M} \\
&= \beta \frac{ \lambda_2(M-2) + (k-2)(\lambda_M - \lambda_2) - \lambda_2(k-2) - (k-2)(\lambda_M - \lambda_2)}{\lambda_M} \\
&= \beta \frac{\lambda_2}{\lambda_M} (M-k), 
\end{align*}

\noindent so we can substitute this to get 

\begin{align*}
\hat{f}_n(b_n) &\sim n \Biggl[ \sum_{k=3}^{M-1} \frac{ (-1)^k \left[ \frac{ (-1)^M}{D^{M-2}\tilde{P}_{M,M}} \left( \frac{\lambda_2 - \lambda_M}{\lambda_2} \right) \right]^{-\lambda_k/\lambda_M}}{D^{k-2} \tilde{P}_{k,k}} n^{\beta \lambda_2(M-k)/\lambda_M} + \frac{\lambda_2}{\lambda_2 - \lambda_M} - 1 \Biggr] \\
&\sim n \left( \frac{\lambda_2}{\lambda_2 - \lambda_M} - 1 \right) \\
&< 0. 
\end{align*}

\noindent So $\hat{f}_n(b_n) < 0$ in the large population limit.  Similarly, 

\begin{align*}
\hat{f}_n(B_n) &= n^{1-B_n} + \sum_{k=3}^M \frac{ (-1)^k}{D^{k-2}} n^{1-(k-2)\beta} \tilde{S}_k(B_n) - n \\
&\begin{multlined}= n \Biggl[ n^{\frac{\lambda_2}{\lambda_M} \beta(M-2)} n^{\frac{1}{\lambda_M t_n} \log \left[ \frac{ (-1)^M}{D^{M-2}\tilde{P}_{M,M}} \right]} \\
+ \sum_{k=3}^M \frac{ (-1)^k}{D^{k-2}} \sum_{i=2}^k \frac{n^{-\beta(k-2)} n^{\frac{\lambda_i}{\lambda_M} \beta(M-2)} n^{\frac{\lambda_i}{\lambda_M \lambda_2 t_n} \log \left[ \frac{ (-1)^M}{D^{M-2}\tilde{P}_{M,M}} \right]}}{\tilde{P}_{i,k}} - 1 \Biggr] \end{multlined} \\
&\sim n \left[ \sum_{k=3}^M \frac{ (-1)^k}{D^{k-2}} \cdot \frac{n^{\beta[ \lambda_k(M-2)/\lambda_M - (k-2)]} \left[ \frac{ (-1)^M}{D^{M-2}\tilde{P}_{M,M}} \right]^{-\lambda_k/\lambda_M}}{\tilde{P}_{k,k}} - 1 \right] \\
&= n \left[ \sum_{k=3}^{M-1} \frac{ (-1)^k \left[ \frac{ (-1)^M}{D^{M-2}\tilde{P}_{M,M}} \right]^{-\lambda_k/\lambda_M}}{D^{k-2} \tilde{P}_{k,k}} n^{\beta \lambda_2(M-k)/\lambda_M} \right] \\
&> 0. 
\end{align*}

\noindent So $\hat{f}_n(B_n) > 0$ in the large population limit.  Therefore, we know that $\hat{f}_n(z)$ must be monotonically decreasing on the first part of its domain and monotonically increasing on the rest of its domain.  This implies that there exists one and only one positive solution $\tilde{v}_n$ to the equation $\hat{f}_n(z) = 0$.  And since $\hat{f}_n(b_n) < 0$ and $\hat{f}_n(B_n) > 0$ in the large population limit, we must have $b_n < \tilde{v}_n < B_n$.  Lastly, because $\lim_{n \rightarrow \infty} b_n = \lim_{n \rightarrow \infty} B_n = -\frac{\lambda_2}{\lambda_M}\beta(M-2)$, the solution $\tilde{v}_n \rightarrow -\frac{\lambda_2}{\lambda_M}\beta(M-2)$ as $n \rightarrow \infty$ as well. 
\end{proof}

\subsection{Proof of Theorem~\ref{LLNAmp}}\label{proof:LLNAmp} 

\begin{proof}

In order to show the desired result, we must show 

\begin{equation*}
\lim_{n \rightarrow \infty} \mathbb{P} (\omega_n > v_n + \varepsilon) + \lim_{n \rightarrow \infty} \mathbb{P} (\omega_n < v_n - \varepsilon) = 0.
\end{equation*}

\noindent Let's start by proving that $\mathbb{P} (\omega_n < v_n - \varepsilon) \rightarrow 0$ as $n \rightarrow \infty$.  Note that 

\begin{align*}
\MoveEqLeft[2] \mathbb{P} (\omega_n < v_n - \varepsilon) \\
&= \mathbb{P} \left( \frac{\omega_n}{t_n} < v_n^-(\varepsilon) \right) \\
&\leq \mathbb{P} \left( \sup_{z \in [d, v_n^-(\varepsilon)]} \left( \sum_{k=2}^M X_k(zt_n) - n \right) > 0 \right) \\
&= \mathbb{P} \left( \sup_{z \in [d, v_n^-(\varepsilon)]} n^{\beta(M-2) + \lambda_Mz/\lambda_2 - 1} \left( \sum_{k=2}^M X_k(zt_n) + \sum_{k=2}^M \phi_k(zt_n) - \sum_{k=2}^M \phi_k(zt_n) - n \right) > 0 \right) \\
&\leq \mathbb{P} \left( \sum_{k=2}^M \hat{B}_k(n, \varepsilon) + \hat{B}_{\phi}(n, \varepsilon) > 0 \right),
\end{align*}

\noindent where 

\begin{align*}
\hat{B}_k(n, \varepsilon) &= \sup_{z \in [d, v_n^-(\varepsilon)]} n^{\beta(M-2) + \lambda_Mz/\lambda_2 - 1} \left( X_k(zt_n) - \phi_k(zt_n) \right), \\
\hat{B}_{\phi}(n, \varepsilon) &= \sup_{z \in [d, v_n^-(\varepsilon)]} n^{\beta(M-2) + \lambda_Mz/\lambda_2 - 1} \left( \sum_{k=2}^M \phi_k(zt_n) - n \right). 
\end{align*}

\noindent Note that 

\begin{align*}
\MoveEqLeft[2] \sup_{z \in [d, v_n^-(\varepsilon)]} \left| n^{\beta(M-2) + \lambda_Mz/\lambda_2 - 1} \left( X_k(zt_n) - \phi_k(zt_n) \right) \right| \\
&\leq \sup_{z \in [d, v_n^+(\varepsilon)]} n^{\beta(M-2) + \lambda_Mz/\lambda_2 - 1} \left| X_k(zt_n) - \phi_k(zt_n) \right|,
\end{align*}

\noindent which converges to zero in probability by Proposition~\ref{prop:FlucAmp}.  Now we just need to show that $\hat{B}_{\phi}(n, \varepsilon)$ is negative in the large population limit.  Let $\hat{g}_n(z) = n^{\beta(M-2) + \lambda_M z/\lambda_2 - 1} \left( \sum_{k=2}^M \phi_k(z t_n) - n \right)$.  Using the definitions of $\phi_2$ and $\phi_k$ for $k>2$, we see that 

\begin{align*}
\hat{g}_n(z) &= n^{\beta(M-2) + \lambda_Mz/\lambda_2 - 1} \left( n^{1-z} + \sum_{k=3}^M \frac{ (-1)^k}{D^{k-2}} n^{1-(k-2)\beta} \tilde{S}_k(z) - n \right) \\
&= n^{\beta(M-2) + z(\lambda_M/\lambda_2 - 1)} + \sum_{k=3}^M \frac{ (-1)^k}{D^{k-2}} \sum_{i=2}^k \frac{n^{\beta(M-k) + (\lambda_M - \lambda_i)z/\lambda_2}}{\tilde{P}_{i,k}} - n^{\beta(M-2) + \lambda_Mz/\lambda_2} \\
&= n^{(M-2)(\beta + Dz /\lambda_2)} + \sum_{k=3}^M \frac{ (-1)^k}{D^{k-2}} \sum_{i=2}^k \frac{n^{\beta(M-k) + (M-i)Dz/\lambda_2}}{\tilde{P}_{i,k}} - n^{\beta(M-2) + \lambda_Mz/\lambda_2}. 
\end{align*}

\noindent Taking the derivative with respect to $z$ yields 

\begin{align*}
\hat{g}_n^{\prime}(z) &= n^{(M-2)(\beta + Dz /\lambda_2)} \log n \cdot \frac{(M-2)D}{\lambda_2} \\
&\phantom{==}+ \sum_{k=3}^M \frac{ (-1)^k}{D^{k-2}} \sum_{i=2}^k \frac{n^{\beta(M-k) + (M-i)Dz/\lambda_2} \log n \cdot \frac{(M-i)D}{\lambda_2}}{\tilde{P}_{i,k}} \\
&\phantom{==}- n^{\beta(M-2)}\log n \cdot \frac{\lambda_M}{\lambda_2} n^{\lambda_Mz/\lambda_2}
\end{align*}
Since $\lambda_2 < 0$ we know that the inner sum of the second term will be asymptotically dominated by the $i = k$ term as $n \to \infty$. Thus,
\begin{align*}
\hat{g}_n^{\prime}(z) &\sim  \frac{\log n}{\lambda_2} \left[ (M-2)D \cdot n^{(M-2)(\beta + Dz /\lambda_2)}+ \sum_{k=3}^M \frac{ (-1)^k (M-k)}{D^{k-3} \tilde{P}_{k,k}} n^{(M-k)(\beta + Dz /\lambda_2)} - \lambda_M n^{\beta(M-2)+\lambda_Mz/\lambda_2} \right],
\end{align*}

\noindent Let $z \in [d, v_n^{-}(\varepsilon)]$. Since $z \ge d$ we have that $z > -\lambda_2 \beta \frac{M-2}{\lambda_M - \lambda_2} = -\lambda_2 \beta/D$ by (\ref{restrict3}).  Therefore, $Dz /\lambda_2 < -\beta$, and hence $\beta + Dz /\lambda_2 < 0$. So $(M-k)(\beta + Dz /\lambda_2) \le 0$ for $2 \le k \le M$. On the other hand, since $z \le v_n^{-}(\varepsilon) < -\frac{\lambda_2}{\lambda_M} \beta (M-2)$ for sufficiently large $n$ by Proposition~\ref{prop:EstimateAmp}, we have that $\lambda_M z/\lambda_2 > - \beta(M-2)$, and hence $\beta(M-2) + \lambda_M z /\lambda_2 > 0$. Together, this implies that when $z \in [d, v_n^{-}(\varepsilon)]$,

\begin{align*}
\hat{g}_n^{\prime}(z) &\sim \frac{\log n}{\lambda_2} \left[ - \lambda_M n^{\beta(M-2)+\lambda_Mz/\lambda_2}\right] \\
&= \frac{\lambda_M \log n}{-\lambda_2} n^{\beta(M-2)+\lambda_Mz/\lambda_2}\\
&> 0 \text{ for sufficiently large $n$}. 
\end{align*}

\noindent Hence $\hat{g}_n(z)$ is monotonically increasing on the interval $[d, v_n^-(\varepsilon)]$.  Therefore, we may rewrite $\hat{B}_{\phi}(n, \varepsilon)$ as 

\begin{equation*}
\hat{B}_{\phi}(n, \varepsilon) = n^{\beta(M-2) + \lambda_M v_n^-(\varepsilon)/\lambda_2 - 1} \left( \sum_{k=2}^M \phi_k(v_n^-(\varepsilon) t_n) - n \right). 
\end{equation*}

\noindent Then by the definitions of $\phi_2$ and $\phi_k$ for $k>2$, we have 

\begin{align*}
\hat{B}_{\phi}(n, \varepsilon) &= n^{\beta(M-2) + \lambda_M v_n^-(\varepsilon)/\lambda_2 - 1} \left( n^{1-v_n^-(\varepsilon)} + \sum_{k=3}^M \frac{ (-1)^k}{D^{k-2}} n^{1-(k-2)\beta} \tilde{S}_k(v_n^-(\varepsilon)) - n \right) \\
&= \Biggl( n^{(M-2)(\beta + D v_n^-(\varepsilon)/\lambda_2)} \\
&\phantom{==}+ n^{\beta(M-2) + \lambda_M v_n^-(\varepsilon)/\lambda_2 - 1} \sum_{k=3}^M \frac{ (-1)^k}{D^{k-2}} n^{1-(k-2)\beta} \sum_{i=2}^k \frac{n^{-v_n^-(\varepsilon) - (i-2)D v_n^-(\varepsilon)/\lambda_2}}{\tilde{P}_{i,k}}\\
&\phantom{==}-n^{\beta(M-2) + \lambda_M v_n^-(\varepsilon)/\lambda_2} \Biggr) 
\end{align*}
Similar to the previous calculations, since $\lambda_2 < 0$ we know that the inner sum of the second term will be dominated by the $i = k$ term as $n \to \infty$. Thus, 
\begin{align*}
\hat{B}_{\phi}(n, \varepsilon)&\sim   \Biggl( n^{(M-2)(\beta + D v_n^-(\varepsilon)/\lambda_2)} \\
&\phantom{==}+ n^{\beta(M-2) + \lambda_M v_n^-(\varepsilon)/\lambda_2 - 1} 
\sum_{k=3}^M \frac{ (-1)^k}{D^{k-2}\tilde{P}_{k,k}} n^{1-(k-2)\beta-v_n^-(\varepsilon) - (k-2)D v_n^-(\varepsilon)/\lambda_2} \\
&\phantom{==}-n^{\beta(M-2) + \lambda_M v_n^-(\varepsilon)/\lambda_2} \Biggr) \\
&= \Biggl( n^{(M-2)(\beta + D v_n^-(\varepsilon)/\lambda_2)} \\
&\phantom{==}+ n^{(M-2)(\beta + D v_n^-(\varepsilon)/\lambda_2)} 
\sum_{k=3}^M \frac{ (-1)^k}{D^{k-2}\tilde{P}_{k,k}} n^{-(k-2)(\beta+ D v_n^-(\varepsilon)/\lambda_2)} \\
&\phantom{==}-n^{\beta(M-2) + \lambda_M v_n^-(\varepsilon)/\lambda_2} \Biggr) \\
&= \Biggl( n^{(M-2)(\beta + D v_n^-(\varepsilon)/\lambda_2)} +
\sum_{k=3}^M \frac{ (-1)^k}{D^{k-2}\tilde{P}_{k,k}} n^{(M-k)(\beta+ D v_n^-(\varepsilon)/\lambda_2)}-n^{\beta(M-2) + \lambda_M v_n^-(\varepsilon)/\lambda_2} \Biggr) 
\end{align*}

\noindent Since $v_n^-(\varepsilon) > d$, we have that $v_n^-(\varepsilon) > -\lambda_2 \beta \frac{M-2}{\lambda_M - \lambda_2} = -\lambda_2 \beta/D$ by (\ref{restrict3}).  Therefore, $Dv_n^-(\varepsilon)/\lambda_2 < -\alpha$, and hence $\beta + Dv_n^-(\varepsilon)/\lambda_2 < 0$.  This implies that $(M-k)(\beta+ D v_n^-(\varepsilon)/\lambda_2) < 0$ for $2 \le k \le M$. Moreover, since $v_n^-(\varepsilon) < \frac{-\lambda_2}{\lambda_M}\beta(M-2)$ for sufficiently large $n$ by Proposition~\ref{prop:EstimateAmp}, we have that $\frac{\lambda_M}{\lambda_2} v_n^-\varepsilon > - \beta(M-2)$ and hence $\beta(M-2) +\frac{\lambda_M}{\lambda_2} v_n^-\varepsilon > 0 $. Thus, 
\[
\hat{B}_{\phi}(n, \varepsilon)\sim -n^{\beta(M-2) + \lambda_M v_n^-(\varepsilon)/\lambda_2}.
\]

\noindent So $\hat{B}_{\phi}(n, \varepsilon)$ is definitely negative in the large population limit.  Putting this all together, we have that $\mathbb{P} \left( \sum_{k=2}^M \hat{B}_k(n, \varepsilon) + \hat{B}_{\phi}(n, \varepsilon) > 0 \right) \rightarrow 0$ as $n \rightarrow \infty$.  Therefore, we have shown $\mathbb{P} (\omega_n < v_n - \varepsilon) \rightarrow 0$ as $n \rightarrow \infty$.  The proof that $\mathbb{P}(\omega_n > v_n + \varepsilon) \rightarrow 0$ as $n \rightarrow \infty$ follows using a similar argument.  
\end{proof}

\subsection{Proof of Proposition~\ref{prop:FlucAmp}}\label{proof:FlucAmp}

\begin{proof}

Let's start by proving the $k=2$ case.  That is, we want to show that 

\begin{equation*}
\lim_{n \rightarrow \infty} \mathbb{P} \left( \sup_{z \in [d, v_n^+(\varepsilon)]} n^{\beta(M-2) + \lambda_M z/\lambda_2 - 1} \left| X_2(zt_n) - \phi_2(zt_n) \right| > \delta \right) = 0.
\end{equation*}

\noindent Because a branching process normalized by its mean is a martingale, we know that 
\[Z_2(z) = n^{\beta(M-2) + z - 1} \left(X_2(zt_n) - \phi_2(zt_n) \right)\]
is a martingale in $z$.  To prove our desired result for $k=2$, we need to show 

\begin{equation*}
\lim_{n \rightarrow \infty} \mathbb{P} \left( \sup_{z \in [d, v_n^+(\varepsilon)]} n^{(\lambda_M - \lambda_2)z/\lambda_2} |Z_2(z)| > \delta \right) = 0,
\end{equation*}

\noindent or, equivalently, 

\begin{equation*}
\lim_{n \rightarrow \infty} \mathbb{P} \left( \sup_{z \in [d, v_n^+(\varepsilon)]} n^{(M-2)Dz/\lambda_2} |Z_2(z)| > \delta \right) = 0.
\end{equation*}

\noindent Note that 

\begin{align*}
\sup_{z \in [d, v_n^+(\varepsilon)]} n^{(M - 2)Dz/\lambda_2} |Z_2(z)| &\leq \sup_{z \in [d, v_n^+(\varepsilon)]} n^{(M - 2)Dz/\lambda_2} \cdot \sup_{z \in [d, v_n^+(\varepsilon)]} |Z_2(z)| \\
&= n^{(M - 2)Dd/\lambda_2} \cdot \sup_{z \in [d, v_n^+(\varepsilon)]} |Z_2(z)|.
\end{align*}

\noindent Therefore, we have that 

\begin{align*}
\MoveEqLeft[2] \mathbb{P} \left( \sup_{z \in [d, v_n^+(\varepsilon)]} n^{(M-2)Dz/\lambda_2} |Z_2(z)| > \delta \right) \\
&\leq \mathbb{P} \left( n^{(M-2)Dd/\lambda_2} \cdot \sup_{z \in [d, v_n^+(\varepsilon)]} |Z_2(z)| > \delta \right) \\
&= \mathbb{P} \left( \sup_{z \in [d, v_n^+(\varepsilon)]} |Z_2(z)| > \delta \cdot n^{(2-M)Dd/\lambda_2} \right) \\
&\leq \frac{1}{\delta} n^{(M-2)Dd/\lambda_2} \cdot \mathbb{E} \left[ \left| Z_2(v_n^+(\varepsilon)) \right| \right] \text{ by Doob's Martingale Inequality} \\
&= \frac{1}{\delta} n^{(M-2)Dd/\lambda_2} \cdot \mathbb{E} \left[ \left| n^{\beta(M-2) + v_n^+(\varepsilon) - 1} (X_2(v_n^+(\varepsilon) t_n) - \phi_2(v_n^+(\varepsilon) t_n)) \right| \right] \\
&= \frac{1}{\delta} n^{\beta(M-2) + (M-2)Dd/\lambda_2 + v_n^+(\varepsilon) - 1} \cdot \mathbb{E} \left[ \left| X_2(v_n^+(\varepsilon) t_n) - \phi_2(v_n^+(\varepsilon) t_n) \right| \right].
\end{align*}

\noindent Clearly, 

\begin{align*}
\left| X_2(v_n^+(\varepsilon) t_n) - \phi_2(v_n^+(\varepsilon) t_n) \right| &\leq \left| X_2(v_n^+(\varepsilon) t_n) \right| + \left| \phi_2(v_n^+(\varepsilon) t_n) \right| \\
&= X_2(v_n^+(\varepsilon) t_n) + \phi_2(v_n^+(\varepsilon) t_n).
\end{align*}

\noindent This implies that 

\begin{align*}
\mathbb{E} \left[ \left| X_2(v_n^+(\varepsilon) t_n) - \phi_2(v_n^+(\varepsilon) t_n) \right| \right] &\leq \mathbb{E} \left[ X_2(v_n^+(\varepsilon) t_n) + \phi_2(v_n^+(\varepsilon) t_n) \right] \\
&= 2 \phi_2(v_n^+(\varepsilon) t_n).
\end{align*}

\noindent And hence 

\begin{align*}
\mathbb{P} \left( \sup_{z \in [d, v_n^+(\varepsilon)]} n^{(M - 2)Dz/\lambda_2} |Z_2(z)| > \delta \right) &\leq \frac{2}{\delta} n^{\beta(M-2) + (M - 2)Dd/\lambda_2 + v_n^+(\varepsilon) - 1} n^{1-v_n^+(\varepsilon)} \\
&= \frac{2}{\delta} n^{(M-2)(\beta + Dd/\lambda_2)}. 
\end{align*}

\noindent So it suffices to show that $\beta + Dd/\lambda_2 < 0$.  From (\ref{restrict3}), we have that 

\begin{equation*}
d > -\lambda_2 \beta \frac{M-2}{\lambda_M - \lambda_2} = \frac{-\lambda_2 \beta}{m},
\end{equation*}

\noindent which implies that $Dd/\lambda_2 < -\beta$, so we are done with the $k=2$ case.  

Next, let's prove the desired result for $k>2$.  Note that 

\begin{align*}
\phi_k(zt_n) &= \int_0^{zt_n} n^{-\beta} \phi_{k-1}(s) e^{\lambda_k(zt_n - s)} \, ds \\
&= n^{-\beta -\lambda_k z / \lambda_2} \int_0^{zt_n} \phi_{k-1}(s) e^{-\lambda_ks} \, ds. 
\end{align*}

\noindent Therefore, 

\begin{align*}
\MoveEqLeft[2] n^{\beta(M-2) + \lambda_M z/\lambda_2 - 1} \left( X_k(zt_n) - \phi_k(zt_n) \right) \\
&= n^{\beta(M-2) + \lambda_M z/\lambda_2 - 1} X_k(zt_n) - n^{\beta(M-3) + (\lambda_M - \lambda_k)z/\lambda_2 - 1} \int_0^{zt_n} \phi_{k-1}(s) e^{-\lambda_k s} \, ds \\
&= n^{\beta(M-2) + \lambda_M z/\lambda_2 - 1} X_k(zt_n) - n^{\beta(M-3) + (M-k)Dz/\lambda_2 - 1} \int_0^{zt_n} \phi_{k-1}(s) e^{-\lambda_k s} \, ds \\
&\begin{multlined}= n^{\beta(M-2) + \lambda_M z/\lambda_2 - 1} X_k(zt_n) - n^{\beta(M-3) + (M-k)Dz/\lambda_2 - 1} \int_0^{zt_n} X_{k-1}(s) e^{-\lambda_k s} \, ds \\
+ n^{\beta(M-3) + (M-k)Dz/\lambda_2 - 1} \int_0^{zt_n} \left( X_{k-1}(s) - \phi_{k-1}(s) \right) e^{-\lambda_k s} \, ds. \end{multlined}
\end{align*}

\noindent Taking the absolute value of both sides and using the triangle inequality, we get 

\begin{align*}
\MoveEqLeft[2] n^{\beta(M-2) + \lambda_M z/\lambda_2 - 1} \left| X_k(zt_n) - \phi_k(zt_n) \right| \\
&\begin{multlined}\leq \left| n^{\beta(M-2) + \lambda_Mz/\lambda_2 - 1} X_k(zt_n) - n^{\beta(M-3) + (M-k)Dz/\lambda_2 - 1} \int_0^{zt_n} X_{k-1}(s) e^{-\lambda_k s} \, ds \right| \\
+ n^{\beta(M-3) + (M-k)Dz/\lambda_2 - 1} \int_0^{zt_n} \left| X_{k-1}(s) - \phi_{k-1}(s) \right| e^{-\lambda_k s} \, ds. \end{multlined}
\end{align*}

\noindent This implies that 

\begin{align*}
\MoveEqLeft[2] \sup_{z \in [d, v_n^+(\varepsilon)]} n^{\beta(M-2) + \lambda_M z/\lambda_2 - 1} \left| X_k(zt_n) - \phi_k(zt_n) \right| \\
&\begin{multlined}\leq \sup_{z \in [d, v_n^+(\varepsilon)]} \left| n^{\beta(M-2) + \lambda_Mz/\lambda_2 - 1} X_k(zt_n) - n^{\beta(M-3) + (M-k)Dz/\lambda_2 - 1} \int_0^{zt_n} X_{k-1}(s) e^{-\lambda_ks} \, ds \right| \\
+ \sup_{z \in [d, v_n^+(\varepsilon)]} n^{\beta(M-3) + (M-k)Dz/\lambda_2 - 1} \int_0^{zt_n} \left| X_{k-1}(s) - \phi_{k-1}(s) \right| e^{-\lambda_ks} \, ds. \end{multlined} 
\end{align*}

\noindent Hence 

\begin{align}
\MoveEqLeft[2] \mathbb{P} \left( \sup_{z \in [d, v_n^+(\varepsilon)]} n^{\beta(M-2) + \lambda_M z/\lambda_2 - 1} \left| X_k(zt_n) - \phi_k(zt_n) \right| > \delta \right) \nonumber \\
&\leq \mathbb{P} \left( \sup_{z \in [d, v_n^+(\varepsilon)]} \left| n^{\beta(M-2) + \lambda_Mz/\lambda_2 - 1} X_k(zt_n) - n^{\beta(M-3) + (M-k)Dz/\lambda_2 - 1} \int_0^{zt_n} X_{k-1}(s) e^{-\lambda_ks} \, ds \right| > \delta/2 \right) \\
&\hspace{4mm}+ \mathbb{P} \left( \sup_{z \in [d, v_n^+(\varepsilon)]} n^{\beta(M-3) + (M-k)Dz/\lambda_2 - 1} \int_0^{zt_n} \left| X_{k-1}(s) - \phi_{k-1}(s) \right| e^{-\lambda_ks} \, ds > \delta/2 \right). \label{bigguy}
\end{align}

\noindent As a reminder, our goal is to show that (\ref{bigguy}) converges to 0 as $n \rightarrow \infty$.  

Let's start with the second term.  Note that 

\begin{align*}
\MoveEqLeft[2] \sup_{z \in [d, v_n^+(\varepsilon)]} n^{\beta(M-3) + (M-k)Dz/\lambda_2 - 1} \int_0^{zt_n} \left| X_{k-1}(s) - \phi_{k-1}(s) \right| e^{-\lambda_ks} \, ds \\
&\leq \sup_{z \in [d, v_n^+(\varepsilon)]} n^{\beta(M-3) + (M-k)Dz/\lambda_2 - 1} \sup_{z \in [d, v_n^+(\varepsilon)]} \int_0^{zt_n} \left| X_{k-1}(s) - \phi_{k-1}(s) \right| e^{-\lambda_ks} \, ds \\
&= n^{\beta(M-3) + (M-k)Dd/\lambda_2 - 1} \int_0^{v_n^+(\varepsilon)t_n} \left| X_{k-1}(s) - \phi_{k-1}(s) \right| e^{-\lambda_ks} \, ds.
\end{align*}

\noindent So, in order to show the second term in (\ref{bigguy}) converges to 0, we must show 

\begin{equation*}
\lim_{n \rightarrow \infty} \mathbb{P} \left( n^{\beta(M-3) + (M-k)Dd/\lambda_2 - 1} \int_0^{v_n^+(\varepsilon) t_n} \left| X_{k-1}(s) - \phi_{k-1}(s) \right| e^{-\lambda_k s} \, ds > \delta/2 \right) = 0.
\end{equation*}

\noindent Since convergence in mean implies convergence in probability (by Markov's Inequality), it suffices to show that 

\begin{equation*}
\lim_{n \rightarrow \infty} \mathbb{E} \left( n^{\beta(M-3) + (M-k)Dd/\lambda_2 - 1} \int_0^{v_n^+(\varepsilon) t_n} \left| X_{k-1}(s) - \phi_{k-1}(s) \right| e^{-\lambda_k s} \, ds \right) = 0
\end{equation*}

\noindent or, equivalently, 

\begin{equation*}
\lim_{n \rightarrow \infty} n^{\beta(M-3) + (M-k)Dd/\lambda_2 - 1} \int_0^{v_n + \varepsilon} \mathbb{E} \left[ \left| X_{k-1}(s) - \phi_{k-1}(s) \right| \right] e^{-\lambda_k s} \, ds = 0.
\end{equation*}

\noindent Note that 

\begin{align*}
\left| X_{k-1}(s) - \phi_{k-1}(s) \right| &\leq \left| X_{k-1}(s) \right| + \left| \phi_{k-1}(s) \right| \text{ by the triangle inequality}\\
&= X_{k-1}(s) + \phi_{k-1}(s). 
\end{align*}

\noindent Taking the mean of both sides yields 

\begin{align*}
\mathbb{E} \left[ \left| X_{k-1}(s) - \phi_{k-1}(s) \right| \right] &\leq \mathbb{E} \left[ X_{k-1}(s) + \phi_{k-1}(s) \right] \\
&= 2 \phi_{k-1}(s).
\end{align*}

\noindent Multiplying both sides by $n^{\beta(M-3) + (M-k)Dd/\lambda_2 - 1} e^{-\lambda_k s}$ and integrating gives us the following inequality: 

\begin{equation*}
\begin{multlined}
n^{\beta(M-3) + (M-k)Dd/\lambda_2 - 1} \int_0^{v_n + \varepsilon} \mathbb{E} \left[ \left| X_{k-1}(s) - \phi_{k-1}(s) \right| \right] e^{-\lambda_k s} \, ds \\
\leq n^{\beta(M-3) + (M-k)Dd/\lambda_2 - 1} \int_0^{v_n + \varepsilon} 2 \phi_{k-1}(s) e^{-\lambda_k s} \, ds.
\end{multlined} 
\end{equation*}

\noindent So it suffices to show that 

\begin{equation*}
\lim_{n \rightarrow \infty} n^{\beta(M-3) + (M-k)Dd/\lambda_2 - 1} \int_0^{v_n + \varepsilon} 2 \phi_{k-1}(s) e^{-\lambda_k s} \, ds = 0
\end{equation*}

\noindent or, equivalently, 

\begin{equation*}
\lim_{n \rightarrow \infty} n^{\beta(M-3) + (M-k)Dd/\lambda_2 - 1} \int_0^{v_n + \varepsilon} \mathbb{E} \left[ X_{k-1}(s) \right] e^{-\lambda_k s} \, ds = 0.
\end{equation*}

\noindent Using the definition of $\mathbb{E} \left[ X_{k-1}(s) \right]$, we see that 

\begin{align*}
\MoveEqLeft[2] n^{\beta(M-3) + (M-k)Dd/\lambda_2 - 1} \int_0^{v_n + \varepsilon} \mathbb{E} \left[ X_{k-1}(s) \right] e^{-\lambda_k s} \, ds \\
&= n^{\beta(M-3) + (M-k)Dd/\lambda_2 - 1} \int_0^{v_n + \varepsilon} \left[  n^{1-(k-3)\beta} (-1)^{k-1} S_{k-1}(s) \right] e^{-\lambda_k s} \, ds \\
&=  n^{\beta(M-k) + (M-k)Dd/\lambda_2} (-1)^{k-1} \int_0^{v_n + \varepsilon} \sum_{i=2}^{k-1} \frac{e^{(\lambda_i - \lambda_k)s}}{P_{i,k-1}} \, ds \\
&= \left( \frac{1}{m} \right)^{k-3} n^{(M-k)(\beta + Dd/\lambda_2)} (-1)^{k-1} \sum_{i=2}^{k-1} \int_0^{v_n + \varepsilon} \frac{e^{(i-k)Ds}}{\tilde{P}_{i,k-1}} \, ds \\
&= \frac{1}{D^{k-2}} n^{(M-k)(\beta + Dd/\lambda_2)} (-1)^k \sum_{i=2}^{k-1} \frac{1}{\tilde{P}_{i,k}} \left( e^{(i-k)D(v_n + \varepsilon)} - 1 \right).
\end{align*}

\noindent Since $v_n = \tilde{v}_n t_n = -\frac{\tilde{v}_n}{\lambda_2} \log n$, 

\begin{align*}
e^{(i-k)D(v_n + \varepsilon)} &= e^{(k-i)D \frac{\tilde{v}_n}{\lambda_2} \log n} e^{(i-k)D \varepsilon} \\
&= n^{(k-i)D \tilde{v}_n / \lambda_2} e^{(i-k)D\varepsilon}.
\end{align*}

\noindent So we want to show that 

\begin{equation*}
\lim_{n \rightarrow \infty} \frac{1}{D^{k-2}} n^{(M-k)(\beta + Dd/\lambda_2)} (-1)^k \sum_{i=2}^{k-1} \frac{1}{\tilde{P}_{i,k}} \left( n^{(k-i)D \tilde{v}_n / \lambda_2} e^{(i-k)D\varepsilon} - 1 \right) = 0
\end{equation*}

\noindent or, equivalently, 

\begin{equation*}
\lim_{n \rightarrow \infty} n^{(M-k)(\beta + Dd/\lambda_2)} \sum_{i=2}^{k-1} \frac{1}{\tilde{P}_{i,k}} \left( n^{(k-i)D \tilde{v}_n / \lambda_2} e^{(i-k)D\varepsilon} - 1 \right) = 0.
\end{equation*}

\noindent Since $\tilde{v}_n \rightarrow - \frac{\lambda_2}{\lambda_M} \beta (M-2)$ as $n \rightarrow \infty$, we have that $n^{(k-i)D \tilde{v}_n /\lambda_2} \rightarrow 0$ and hence $n^{(k-i)D \tilde{v}_n /\lambda_2} e^{(i-k)D\varepsilon} - 1 \rightarrow -1$.  Therefore, this problem simplifies to showing that $n^{(M-k)(\beta + Dd/\lambda_2)} \rightarrow 0$ as $n \rightarrow \infty$.  But we have already seen that $\beta + Dd/\lambda_2 < 0$ in the $k=2$ case, so we have the desired result.  

Now that we have proved the second term in (\ref{bigguy}) converges to 0 as $n \rightarrow \infty$, all we have left is to show that the first term converges to 0 as well.  By Lemma 1 in \cite{durrett_evolution_2010}, we know that 

\begin{equation*}
e^{-\lambda_k t} X_k(t) - \int_0^t n^{-\beta} e^{-\lambda_k s} X_{k-1}(s) \, ds
\end{equation*}

\noindent is a martingale.  Setting $t = zt_n = -\frac{z}{\lambda_2} \log n$, we get that 

\begin{equation*}
n^{\lambda_k z/\lambda_2} X_k(zt_n) - n^{-\beta} \int_0^{zt_n} e^{-\lambda_k s} X_{k-1} (s) \, ds 
\end{equation*}

\noindent is a martingale in $z$.  Since linear combinations of martingales are also martingales, 

\begin{equation*}
n^{\lambda_k z/\lambda_2 + \beta(M-2) + (\lambda_M - \lambda_k) z/\lambda_2 - 1} X_k(zt_n) - n^{-\beta + \beta(M-2) + (\lambda_M - \lambda_k) z/\lambda_2 - 1} \int_0^{zt_n} X_{k-1}(s) e^{-\lambda_k s} \, ds 
\end{equation*}

\noindent is also a martingale in $z$.  The expression above simplifies to 

\begin{equation*}
n^{\beta(M-2) + \lambda_M z/\lambda_2 - 1} X_k(zt_n) - n^{\beta(M-3) + (M-k)D z/\lambda_2 - 1} \int_0^{zt_n} X_{k-1}(s) e^{-\lambda_k s} \, ds.
\end{equation*}

\noindent Therefore, 

\begin{equation*}
\left| n^{\beta(M-2) + \lambda_M z/\lambda_2 - 1} X_k(zt_n) - n^{\beta(M-3) + (M-k)D z/\lambda_2 - 1} \int_0^{zt_n} X_{k-1}(s) e^{-\lambda_k s} \, ds \right| 
\end{equation*}

\noindent is a non-negative submartingale in $z$, so we can apply Doob's Martingale Inequality to get 

\begin{align*}
\MoveEqLeft[2] \mathbb{P} \left( \sup_{z \in [d, v_n^+(\varepsilon)]} \left| n^{\beta(M-2) + \lambda_Mz/\lambda_2 - 1} X_k(zt_n) - n^{\beta(M-3) + (M-k)Dz/\lambda_2 - 1} \int_0^{zt_n} X_{k-1}(s) e^{-\lambda_ks} \, ds \right| > \delta/2 \right) \\
&\begin{multlined}\leq \frac{4}{\delta^2} \cdot \mathbb{E} \Biggl[ \Biggl( n^{\beta(M-2) + \lambda_M v_n^+(\varepsilon)/\lambda_2 - 1} X_k(v_n + \varepsilon) \\
- n^{\beta(M-3) + (M-k)D v_n^+(\varepsilon)/\lambda_2 - 1} \int_0^{v_n + \varepsilon} X_{k-1} (s) e^{-\lambda_k s} \, ds \Biggr)^2 \Biggr]. \end{multlined} \\
\end{align*}

\noindent Therefore, it suffices to show that 

\begin{equation*}
\lim_{n \rightarrow \infty} \mathbb{E} \left[ \left( n^{\beta(M-2) + \lambda_M v_n^+(\varepsilon)/\lambda_2 - 1} X_k(v_n + \varepsilon) - n^{\beta(M-3) + (M-k)D v_n^+(\varepsilon)/\lambda_2 - 1} \int_0^{v_n + \varepsilon} X_{k-1} (s) e^{-\lambda_k s} \, ds \right)^2 \right] = 0.
\end{equation*}

\noindent We can expand the quantity above as follows: 

\begin{align*}
\MoveEqLeft[2] \mathbb{E} \left[ \left( n^{\beta(M-2) + \lambda_M v_n^+(\varepsilon)/\lambda_2 - 1} X_k(v_n + \varepsilon) - n^{\beta(M-3) + (M-k)D v_n^+(\varepsilon)/\lambda_2 - 1} \int_0^{v_n + \varepsilon} X_{k-1} (s) e^{-\lambda_k s} \, ds \right)^2 \right] \\
&\begin{multlined}= n^{2\beta(M-2) + 2\lambda_M v_n^+(\varepsilon)/\lambda_2 - 2} \mathbb{E} \left[ X_k(v_n + \varepsilon)^2 \right] \\
- 2 n^{\beta(2M-5) + (2\lambda_M - \lambda_k) v_n^+(\varepsilon)/\lambda_2 - 2} \int_0^{v_n + \varepsilon} \mathbb{E} \left[ X_{k-1}(s) X_k(v_n + \varepsilon) \right] e^{-\lambda_k s} \, ds \\
+  n^{2\beta(M-3) + 2(\lambda_M - \lambda_k) v_n^+(\varepsilon)/\lambda_2 - 2} \int_0^{v_n + \varepsilon} \int_0^{v_n + \varepsilon} \mathbb{E} \left[ X_{k-1}(s) X_{k-1}(y) \right] e^{-\lambda_k s} e^{-\lambda_k y} \, ds \, dy. \end{multlined} 
\end{align*} 

\noindent By making a slight modification to the proof of Lemma 1 in \cite{foo_dynamics_2013}, we have that 

\begin{align*}
\mathbb{E} \left[ X_k(v_n + \varepsilon)^2 \right] &= (n^{-\beta})^2 \int_0^{v_n + \varepsilon} \int_0^{v_n + \varepsilon} \mathbb{E} \left[ X_{k-1}(s) X_{k-1}(y) \right] e^{\lambda_k(v_n + \varepsilon - s)} e^{\lambda_k(v_n + \varepsilon - y)} \, ds \, dy \\
&\hspace{4mm}+ n^{-\beta} \int_0^{v_n + \varepsilon} \mathbb{E} \left[ X_{k-1}(s) \right] \mathbb{E} \left[ \tilde{X}_k(v_n + \varepsilon - s)^2 \right] \, ds \\
\mathbb{E} \left[ X_{k-1}(s) X_k(v_n + \varepsilon) \right] &= n^{-\beta} \int_0^{v_n + \varepsilon} \mathbb{E} \left[ X_{k-1}(s) X_{k-1}(y) \right] e^{-\lambda_k(v_n + \varepsilon - y)} \, dy
\end{align*}

\noindent where $\tilde{X}_k$ is a binary branching process starting from size one with birth rate $r_k$ and death rate $d_k$.  Substituting these expressions into our equation yields 

\begin{align*}
\MoveEqLeft[2] \mathbb{E} \left[ \left( n^{\beta(M-2) + \lambda_M v_n^+(\varepsilon)/\lambda_2 - 1} X_k(v_n + \varepsilon) - n^{\beta(M-3) + (M-k)D v_n^+(\varepsilon)/\lambda_2 - 1} \int_0^{v_n + \varepsilon} X_{k-1} (s) e^{-\lambda_k s} \, ds \right)^2 \right] \\
&\begin{multlined}=  n^{2\beta(M-3) + 2\lambda_M v_n^+(\varepsilon)/\lambda_2 - 2} \int_0^{v_n + \varepsilon} \int_0^{v_n + \varepsilon} \mathbb{E} \left[ X_{k-1}(s) X_{k-1}(y) \right] e^{\lambda_k(2v_n + 2\varepsilon - s - y)} \, ds \, dy \\
+ n^{\beta(2M-5) + 2\lambda_M v_n^+(\varepsilon)/\lambda_2 - 2} \int_0^{v_n + \varepsilon} \mathbb{E} \left[ X_{k-1}(s) \right] \mathbb{E} \left[ \tilde{X}_k(v_n + \varepsilon - s)^2 \right] \, ds \\
- 2 n^{2\beta(M-3) + (2\lambda_M - \lambda_k) v_n^+(\varepsilon)/\lambda_2 - 2} \int_0^{v_n + \varepsilon} \int_0^{v_n + \varepsilon} \mathbb{E} \left[ X_{k-1}(s) X_{k-1}(y) \right] e^{\lambda_k(v_n + \varepsilon - s - y)} \, ds \, dy \\
+  n^{2\beta(M-3) + 2(\lambda_M - \lambda_k) v_n^+(\varepsilon)/\lambda_2 - 2} \int_0^{v_n + \varepsilon} \int_0^{v_n + \varepsilon} \mathbb{E} \left[ X_{k-1}(s) X_{k-1}(y) \right] e^{-\lambda_k(s+y)} \, ds \, dy. \end{multlined} 
\end{align*} 

\noindent Then, by definition of $v_n^+(\varepsilon)$, 

\begin{align*}
\MoveEqLeft[2] \mathbb{E} \left[ \left( n^{\beta(M-2) + \lambda_M v_n^+(\varepsilon)/\lambda_2 - 1} X_k(v_n + \varepsilon) - n^{\beta(M-3) + (M-k)D v_n^+(\varepsilon)/\lambda_2 - 1} \int_0^{v_n + \varepsilon} X_{k-1} (s) e^{-\lambda_k s} \, ds \right)^2 \right] \\
&\begin{multlined}=  n^{2\beta(M-3) - 2} e^{-2\lambda_M(v_n + \varepsilon)} \int_0^{v_n + \varepsilon} \int_0^{v_n + \varepsilon} \mathbb{E} \left[ X_{k-1}(s) X_{k-1}(y) \right] e^{\lambda_k(2v_n + 2\varepsilon - s - y)} \, ds \, dy \\
+ n^{\beta(2M-5) - 2} e^{-2\lambda_M(v_n + \varepsilon)} \int_0^{v_n + \varepsilon} \mathbb{E} \left[ X_{k-1}(s) \right] \mathbb{E} \left[ \tilde{X}_k(v_n + \varepsilon - s)^2 \right] \, ds \\ 
- 2  n^{2\beta(M-3) - 2} e^{-(2\lambda_M - \lambda_k)(v_n + \varepsilon)} \int_0^{v_n + \varepsilon} \int_0^{v_n + \varepsilon} \mathbb{E} \left[ X_{k-1}(s) X_{k-1}(y) \right] e^{\lambda_k(v_n + \varepsilon - s - y)} \, ds \, dy \\
+  n^{2\beta(M-3) - 2} e^{-2(\lambda_M - \lambda_k)(v_n + \varepsilon)} \int_0^{v_n + \varepsilon} \int_0^{v_n + \varepsilon} \mathbb{E} \left[ X_{k-1}(s) X_{k-1}(y) \right] e^{-\lambda_k(s+y)} \, ds \, dy \end{multlined} \\
&\begin{multlined}=  n^{2\beta(M-3) - 2} e^{2(\lambda_k - \lambda_M)(v_n + \varepsilon)} \int_0^{v_n + \varepsilon} \int_0^{v_n + \varepsilon} \mathbb{E} \left[ X_{k-1}(s) X_{k-1}(y) \right] e^{-\lambda_k(s+y)} \, ds \, dy \\
+ n^{\beta(2M-5) - 2} e^{-2\lambda_M(v_n + \varepsilon)} \int_0^{v_n + \varepsilon} \mathbb{E} \left[ X_{k-1}(s) \right] \mathbb{E} \left[ \tilde{X}_k(v_n + \varepsilon - s)^2 \right] \, ds \\
- 2 n^{2\beta(M-3) - 2} e^{2(\lambda_k - \lambda_M)(v_n + \varepsilon)} \int_0^{v_n + \varepsilon} \int_0^{v_n + \varepsilon} \mathbb{E} \left[ X_{k-1}(s) X_{k-1}(y) \right] e^{-\lambda_k(s+y)} \, ds \, dy \\
+  n^{2\beta(M-3) - 2} e^{2(\lambda_k - \lambda_M)(v_n + \varepsilon)} \int_0^{v_n + \varepsilon} \int_0^{v_n + \varepsilon} \mathbb{E} \left[ X_{k-1}(s) X_{k-1}(y) \right] e^{-\lambda_k(s+y)} \, ds \, dy \end{multlined} \\
&= n^{\beta(2M-5) -2} e^{-2\lambda_M(v_n + \varepsilon)} \int_0^{v_n + \varepsilon} \mathbb{E} \left[ X_{k-1}(s) \right] \mathbb{E} \left[ \tilde{X}_k(v_n + \varepsilon - s)^2 \right] \, ds. 
\end{align*} 

\noindent Note that 

\begin{align*}
\mathbb{E} \left[ X_{k-1}(s) \right] &=  n^{1-(k-3)\beta} (-1)^{k-1} S_{k-1}(s), \\
\mathbb{E} \left[ \tilde{X}_k(v_n + \varepsilon - s)^2 \right] &= \frac{2r_k}{\lambda_k} e^{2\lambda_k(v_n + \varepsilon - s)} - \frac{r_k + d_k}{\lambda_k} e^{\lambda_k(v_n + \varepsilon - s)}. 
\end{align*}

\noindent Substituting these into the above expression yields 

\begin{align*}
\MoveEqLeft[2] \mathbb{E} \left[ \left( n^{\beta(M-2) + \lambda_M v_n^+(\varepsilon)/\lambda_2 - 1} X_k(v_n + \varepsilon) - n^{\beta(M-3) + (M-k)D v_n^+(\varepsilon)/\lambda_2 - 1} \int_0^{v_n + \varepsilon} X_{k-1} (s) e^{-\lambda_k s} \, ds \right)^2 \right] \\
&\begin{multlined}= n^{\beta(2M-5) - 2} e^{-2\lambda_M(v_n + \varepsilon)} \cdot \\
\int_0^{v_n + \varepsilon}  n^{1-(k-3)\beta} (-1)^{k-1} S_{k-1}(s) \left( \frac{2r_k}{\lambda_k} e^{2\lambda_k(v_n + \varepsilon - s)} - \frac{r_k + d_k}{\lambda_k} e^{\lambda_k(v_n + \varepsilon - s)} \right) \, ds \end{multlined} \\
&\begin{multlined}=  n^{\beta(2M-k-2)-1} e^{-2\lambda_M(v_n + \varepsilon)} (-1)^{k-1} \Biggl[ \frac{2r_k}{\lambda_k} e^{2\lambda_k(v_n + \varepsilon)} \int_0^{v_n + \varepsilon} \sum_{i=2}^{k-1} \frac{e^{(\lambda_i - 2\lambda_k)s}}{P_{i,k-1}} \, ds \\
- \frac{r_k + d_k}{\lambda_k} e^{\lambda_k(v_n + \varepsilon)} \int_0^{v_n + \varepsilon} \sum_{i=2}^{k-1} \frac{e^{(\lambda_i - \lambda_k)s}}{P_{i,k-1}} \, ds \Biggr] \end{multlined} \\
&\begin{multlined}=  n^{\beta(2M-k-2)-1} (-1)^{k-1} e^{-2\lambda_M(v_n + \varepsilon)} \frac{1}{\lambda_k} \cdot \\
\sum_{i=2}^{k-1} \left[ \frac{2r_k \left( e^{\lambda_i(v_n + \varepsilon)} - e^{2\lambda_k(v_n + \varepsilon)} \right)}{(\lambda_i - 2\lambda_k) P_{i,k-1}} - \frac{(r_k + d_k) \left( e^{\lambda_i(v_n + \varepsilon)} - e^{\lambda_k(v_n + \varepsilon)} \right)}{(\lambda_i - \lambda_k) P_{i,k-1}} \right] \end{multlined} \\
&\begin{multlined}=  n^{\beta(2M-k-2)-1} (-1)^{k-1} e^{-2\lambda_M(v_n + \varepsilon)} \frac{1}{\lambda_k} \cdot \\
\sum_{i=2}^{k-1} \frac{2r_k(\lambda_i - \lambda_k) \left( e^{\lambda_i(v_n + \varepsilon)} - e^{2\lambda_k(v_n + \varepsilon)} \right) - (r_k + d_k)(\lambda_i - 2\lambda_k) \left( e^{\lambda_i(v_n + \varepsilon)} - e^{\lambda_k(v_n + \varepsilon)} \right)}{(\lambda_i - \lambda_k)(\lambda_i - 2\lambda_k) P_{i,k-1}}. \end{multlined} 
\end{align*}

\noindent Now note that 

\begin{align*}
\MoveEqLeft[2] 2r_k(\lambda_i - \lambda_k) \left( e^{\lambda_i(v_n + \varepsilon)} - e^{2\lambda_k(v_n + \varepsilon)} \right) - (r_k + d_k)(\lambda_i - 2\lambda_k) \left( e^{\lambda_i(v_n + \varepsilon)} - e^{\lambda_k(v_n + \varepsilon)} \right) \\
&= \lambda_k \left( \lambda_i + 2d_k \right) e^{\lambda_i(v_n + \varepsilon)} + \left( r_k + d_k \right) \left( \lambda_i - 2\lambda_k \right) e^{\lambda_k(v_n + \varepsilon)} + 2r_k \left(\lambda_k - \lambda_i \right) e^{2\lambda_k(v_n + \varepsilon)}. 
\end{align*}

\noindent Substituting this back in, we get 

\begin{align*}
\MoveEqLeft[2] \mathbb{E} \left[ \left( n^{\beta(M-2) + \lambda_M v_n^+(\varepsilon)/\lambda_2 - 1} X_k(v_n + \varepsilon) - n^{\beta(M-3) + (M-k)D v_n^+(\varepsilon)/\lambda_2 - 1} \int_0^{v_n + \varepsilon} X_{k-1} (s) e^{-\lambda_k s} \, ds \right)^2 \right] \\
&\begin{multlined}=  n^{\beta(2M-k-2)-1} (-1)^{k-1} e^{-2\lambda_M(v_n + \varepsilon)} \frac{1}{\lambda_k} \cdot \\
\sum_{i=2}^{k-1} \frac{\lambda_k \left( \lambda_i + 2d_k \right) e^{\lambda_i(v_n + \varepsilon)} + \left( r_k + d_k \right) \left( \lambda_i - 2\lambda_k \right) e^{\lambda_k(v_n + \varepsilon)} + 2r_k \left(\lambda_k - \lambda_i \right) e^{2\lambda_k(v_n + \varepsilon)}}{(\lambda_i - \lambda_k)(\lambda_i - 2\lambda_k) P_{i,k-1}} \end{multlined} \\
&\begin{multlined}\sim \frac{ n^{\beta(2M-k-2) -1} (-1)^{k-1} e^{-2\lambda_M(v_n + \varepsilon)}}{\lambda_k(\lambda_{k-1} - \lambda_k)(\lambda_{k-1} - 2\lambda_k) P_{k-1,k-1}} \cdot \\
\left[ (r_k + d_k)(\lambda_{k-1} - 2\lambda_k) e^{\lambda_k(v_n + \varepsilon)} + 2r_k (\lambda_k - \lambda_{k-1}) e^{2\lambda_k(v_n + \varepsilon)} \right]. \end{multlined} 
\end{align*}

\noindent Since $v_n = \tilde{v}_n t_n = -\frac{\tilde{v}_n}{\lambda_2} \log n$,

\begin{align*}
\MoveEqLeft[2] \mathbb{E} \left[ \left( n^{\beta(M-2) + \lambda_M v_n^+(\varepsilon)/\lambda_2 - 1} X_k(v_n + \varepsilon) - n^{\beta(M-3) + (M-k)D v_n^+(\varepsilon)/\lambda_2 - 1} \int_0^{v_n + \varepsilon} X_{k-1} (s) e^{-\lambda_k s} \, ds \right)^2 \right] \\
&\begin{multlined}\sim \frac{ n^{\beta(2M-k-2)-1} (-1)^{k-1} e^{-2\lambda_M \varepsilon} e^{\frac{2\lambda_M \tilde{v}_n}{\lambda_2} \log n}}{\lambda_k (\lambda_{k-1} - \lambda_k) (\lambda_{k-1} - 2\lambda_k) P_{k-1,k-1}} \cdot \\
\left[ (r_k + d_k) (\lambda_{k-1} - 2\lambda_k) e^{\lambda_k \varepsilon} e^{\frac{\lambda_k \tilde{v}_n}{-\lambda_2} \log n} + 2r_k (\lambda_k - \lambda_{k-1}) e^{2\lambda_k \varepsilon} e^{\frac{2\lambda_k \tilde{v}_n}{-\lambda_2} \log n} \right] \end{multlined} \\
&\begin{multlined}= \frac{ (-1)^{k-1} e^{-2\lambda_M \varepsilon}}{\lambda_k(\lambda_{k-1} - \lambda_k)(\lambda_{k-1} - 2\lambda_k) P_{k-1,k-1}} \cdot \\
\hspace{4mm} \Bigl[ (r_k + d_k) (\lambda_{k-1} - 2\lambda_k) e^{\lambda_k \varepsilon} n^{\beta(2M-k-2) - 1 + (2\lambda_M - \lambda_k) \tilde{v}_n/\lambda_2} \\
+ 2r_k (\lambda_k - \lambda_{k-1}) e^{2\lambda_k \varepsilon} n^{\beta(2M-k-2) - 1 + 2(\lambda_M - \lambda_k) \tilde{v}_n/\lambda_2} \Bigr]. \end{multlined} 
\end{align*}

\noindent Note that 

\begin{align*}
\frac{2\lambda_M - \lambda_k}{\lambda_2} &= \frac{2(\lambda_2 + (M-2)D) - (\lambda_2 + (k-2)D)}{\lambda_2} \\
&= 1 + \frac{D(2M-k-2)}{\lambda_2}. 
\end{align*}

\noindent Therefore, 

\begin{align*}
\lim_{n \rightarrow \infty} n^{\beta(2M-k-2) - 1 + (2\lambda_M - \lambda_k) \tilde{v}_n/\lambda_2} &= \lim_{n \rightarrow \infty} n^{\beta(2M-k-2) - 1 + \tilde{v}_n + \frac{D(2M-k-2)}{\lambda_2} \tilde{v}_n} \\
&= \lim_{n \rightarrow \infty} e^{ \left[ \beta(2M-k-2) - 1 + \tilde{v}_n + \frac{D(2M-k-2)}{\lambda_2} \tilde{v}_n \right] \log n}. 
\end{align*}

\noindent Then, since 

\begin{align*}
\MoveEqLeft[3] \lim_{n \rightarrow \infty} \left[ \beta(2M-k-2) - 1 + \tilde{v}_n + \frac{D(2M-k-2)}{\lambda_2} \tilde{v}_n \right] \\
&= \beta(2M-k-2) - 1 - \frac{\lambda_2}{\lambda_M} \beta (M-2) - \frac{D(2M-k-2)}{\lambda_2} \frac{\lambda_2}{\lambda_M} \beta(M-2) \text{ by Proposition~\ref{prop:EstimateAmp}} \\ 
&= \beta(2M-k-2) \left( 1 - \frac{M-2}{\lambda_M} \frac{\lambda_M - \lambda_2}{M-2} \right) - 1 - \frac{\lambda_2}{\lambda_M} \beta(M-2) \\
&= \frac{\lambda_2}{\lambda_M} \beta(M-k) - 1 \\
&< 0,
\end{align*}

\noindent we have that 

\begin{equation*}
\lim_{n \rightarrow \infty} n^{\beta(2M-k-2) - 1 + (2\lambda_M - \lambda_k) \tilde{v}_n/\lambda_2} = 0. 
\end{equation*}

\noindent Similarly, note that 

\begin{align*}
2 \frac{\lambda_M - \lambda_k}{\lambda_2} &= 2 \frac{(\lambda_2 + (M-2)D) - (\lambda_2 + (k-2)D)}{\lambda_2} \\
&= 2D \frac{M-k}{\lambda_2}. 
\end{align*}

\noindent Therefore, 

\begin{align*}
\lim_{n \rightarrow \infty} n^{\beta(2M-k-2) - 1 + 2(\lambda_M - \lambda_k) \tilde{v}_n/\lambda_2} &= \lim_{n \rightarrow \infty} n^{\beta(2M-k-2) - 1 + 2D \frac{M-k}{\lambda_2} \tilde{v}_n} \\
&= \lim_{n \rightarrow \infty} e^{ \left[ \beta(2M-k-2) - 1 + 2D \frac{M-k}{\lambda_2} \tilde{v}_n \right] \log n}. 
\end{align*}

\noindent We also have that 

\begin{align*}
\MoveEqLeft[3] \lim_{n \rightarrow \infty} \left[ \beta(2M-k-2) - 1 + 2D \frac{M-k}{\lambda_2} \tilde{v}_n \right] \\
&= \beta(2M-k-2) - 1 - 2D \frac{M-k}{\lambda_2} \frac{\lambda_2}{\lambda_M} \beta(M-2) \text{ by Proposition~\ref{prop:EstimateAmp}} \\
&= \beta(M-2) + \beta(M-k) - 1 - 2\beta \left( 1 - \frac{\lambda_2}{\lambda_M} \right) (M-k) \\
&= \beta(M-2) - 1 + \beta(M-k) \left( 2 \frac{\lambda_2}{\lambda_M} - 1 \right).  
\end{align*}

\noindent Since $\beta(M-k) (2\lambda_2/\lambda_M - 1) < 0$ and $\beta < 1/(M-2)$, the above limit is negative, and hence 

\begin{equation*}
\lim_{n \rightarrow \infty} n^{\beta(2M-k-2) - 1 + 2(\lambda_M - \lambda_k) \tilde{v}_n/\lambda_2} = 0. 
\end{equation*}

\noindent So we are done.  
\end{proof} 

\newpage
\bibliography{references.bib}

\end{document}